\documentclass[traditabstract]{aa} 
\usepackage{xcolor}
\usepackage{url}
\usepackage{graphicx}
\usepackage{longtable,ltcaption}
\usepackage{rotating}
\usepackage{lscape}
\usepackage{sidecap}
\usepackage{amsmath} 
\usepackage{subcaption}
\usepackage{slantsc}

\begin{document}

\title{A spectroscopic and kinematic survey of fast hot subdwarfs}

\author{S.~Geier \inst{1}
   \and U.~Heber \inst{2}
   \and A.~Irrgang \inst{2}
   \and M.~Dorsch \inst{1,2}
   \and A.~Bastian \inst{1}
   \and P.~Neunteufel \inst{3}
   \and T.~Kupfer \inst{4,5,6}
   \and S.~Bloemen \inst{7}
   \and S.~Kreuzer \inst{2}
   \and L.~M\"oller \inst{2}
   \and M.~Schindewolf \inst{2}
   \and D.~Schneider \inst{2}
   \and E.~Ziegerer \inst{2}
   \and I.~Pelisoli \inst{8,1}
   \and V.~Schaffenroth \inst{9,1}
   \and B.~N.~Barlow \inst{10}
   \and R.~Raddi \inst{11,2}   
   \and S.~J.~Geier \inst{12,13}
   \and N.~Reindl \inst{14,1}
   \and T.~Rauch \inst{15}   
   \and P.~Nemeth \inst{16,17,2}
   \and B.~T.~G\"ansicke \inst{8}
   }

\offprints{S.\,Geier,\\ \email{sgeier@astro.physik.uni-potsdam.de}}

\institute{Institut f\"ur Physik und Astronomie, Universit\"at Potsdam, Haus 28, Karl-Liebknecht-Str. 24/25, 14476 Potsdam-Golm, Germany
\and Dr.~Karl~Remeis-Observatory \& ECAP, Astronomical Institute, Friedrich-Alexander University Erlangen-Nuremberg, Sternwartstr.~7, 96049 Bamberg, Germany
\and Max Planck Institut für Astrophysik, Karl-Schwarzschild-Straße 1, 85748 Garching bei München, Germany
\and Hamburger Sternwarte, University of Hamburg, Gojenbergsweg 112, 21029 Hamburg, Germany
\and Texas Tech University, Department of Physics \& Astronomy, Box 41051, 79409, Lubbock, TX, USA
\and Kavli Institute for Theoretical Physics, University of California, Santa Barbara, CA, 93106, USA
\and Department of Astrophysics/IMAPP, Radboud University Nijmegen, P.O. Box 9010, Nijmegen 6500 GL, The Netherlands
\and Department of Physics, University of Warwick, Conventry CV4 7AL, UK
\and Th\"uringer Landessternwarte Tautenburg, Sternwarte 5, 07778 Tautenburg, Germany
\and Department of Physics, High Point University, One University Parkway, High Point, NC, 27268, USA
\and Departament de Física, Universitat Politecnica de Catalunya, c/Esteve Terrades 5, E-08860 Castelldefels, Spain
\and Gran Telescopio Canarias (GRANTECAN), Cuesta de San Jose s/n, 38712 Brena Baja, La Palma, Spain
\and Instituto de Astrofisica de Canarias, Via Lactea s/n, 38205 La Laguna, Tenerife, Spain
\and Landessternwarte Heidelberg, Zentrum für Astronomie, Ruprecht-Karls-Universität, K\"onigstuhl 12, 69117 Heidelberg, Germany
\and Institute for Astronomy and Astrophysics, Kepler Center for Astro and Particle Physics, Eberhard Karls University, Sand 1, 72076 T\"ubingen, Germany
\and Astronomical Institute of the Czech Academy of Sciences, CZ 251\,65, Ond\v{r}ejov, Czech Republic
\and Astroserver.org, F\H{o} t\'er 1, HU 8533 Malomsok, Hungary}

\date{Received \ Accepted}

\abstract{Hot subdwarfs (sdO/B) are the stripped helium cores of red giants formed by binary interactions. Close hot subdwarf binaries with massive white dwarf companions have been proposed as possible progenitors of thermonuclear supernovae type Ia (SN\,Ia). If the supernova is triggered by stable mass transfer from the helium star, the companion should survive the explosion and should be accelerated to high velocities. The hypervelocity star US\,708 is regarded as the prototype for such an ejected companion. To find more of those objects we conducted an extensive spectroscopic survey. Candidates for such fast stars have been selected from the spectroscopic database of the Sloan Digital Sky Survey (SDSS) and several ground-based proper motion surveys. Follow-up spectroscopy has been obtained with several 4m- to 10m-class telescopes. Combining the results from quantitative spectroscopic analyses with space-based astrometry from \textit{Gaia} Early Data Release 3 (EDR3) we determined the atmospheric and kinematic parameters of 53 fast hot subdwarf stars. None of these stars is unbound to the Galaxy, although some have Galactic restframe velocities close to the Galactic escape velocity. 21 stars are apparently single objects, which crossed the Galactic disc within their lifetimes in the sdO/B stage and could be regarded as potential candidates for the SN Ia ejection scenario. However, the properties of the full sample are more consistent with a pure old Galactic halo population. We therefore conclude that the fast sdO/B stars we found are likely to be extreme halo stars.

}

\authorrunning{Geier et al.}
\titlerunning{Fast hot subdwarf stars}

\maketitle

\section{Introduction \label{sec:intro}}

Hot subdwarf stars (sdO/Bs) have been discovered as a prominent population of faint blue stars at high Galactic latitudes, but were subsequently found in all Galactic populations (e.\ g.\ Greenstein \& Sargent \cite{greenstein74}; Luo et al. \cite{luo20}). They are much smaller and of lower mass than hot main sequence stars of similar spectral types and most of them are likely core helium-burning extreme horizontal branch (EHB) stars (Heber et al.\ \cite{heber86}) with very thin hydrogen envelopes. The state of the art in hot subdwarf research has been reviewed by Heber (\cite{heber09,heber16}). 

The origin of sdO/B stars is still not fully understood, because they can only be formed if the progenitor loses its envelope almost entirely after passing the red-giant branch. While single-star scenarios are discussed, binary interactions are more likely (Pelisoli et al.\ \cite{pelisoli20}). Systematic surveys for radial velocity (RV) variable stars revealed that about one third of the hot subdwarf stars are members of short-period, single-lined binaries (Maxted et al.\ \cite{maxted01}; see Geier et al.\ \cite{geier22} for a review) with white dwarf (WD), M-type main sequence or brown dwarf companions (Schaffenroth et al. \cite{schaffenroth22} and references therein). Those are formed after a common envelope phase, during which the companion becomes completely immersed in the red-giant envelope. Stable mass transfer to a main sequence companion in a wide binary and mergers involving He-WDs have been proposed as possible formation channels as well (Han et al. \cite{han02,han03}; Chen et al. \cite{chen13}; Zhang \& Jeffery \cite{zhang12}). Double-lined hot subdwarf binaries  with FGK-type main-sequence companions have indeed been discovered to be wide (e.\ g.\ Vos et al.\ \cite{vos18}). Merger candidates among the single sdO/Bs have also been found (e.g. Dorsch et al. \cite{dorsch22}; Werner et al. \cite{werner22}).

Since most sdB stars are regarded as the progeny of low-mass stars, they can belong to all Galactic populations, either the young thin disc, the older thick disc, or even the Galactic halo. Differences between these populations in metallicity and age are predicted to significantly affect hot subdwarf formation (e.g. Han \cite{han08}; Vos et al. \cite{vos20}). Because the abundances in sdB atmospheres are altered by diffusion processes (e. g. Geier \cite{geier13a}), their population membership cannot be deduced from their abundance patterns, but only from their kinematic properties. While disc stars orbit the Galactic centre with moderate velocities and eccentricities, halo stars can have retrograde, very eccentric and highly inclined trajectories with respect to the Galactic plane. The kinematic method has been used to determine the population membership of rather bright samples of field sdBs (de Boer et al. \cite{deboer97}; Altmann et al. \cite{altmann04}; Kawka et al. \cite{kawka15}; Heber \cite{heber16}; Martin et al. \cite{martin17}; Bobylev \& Bajkova \cite{bobylev19}; Luo et al. \cite{luo20,luo21}). However, the early studies were limited by the poor quality of the tangential velocities and the lack of information about the variability of their radial velocities. While \textit{Gaia} measurements have remedied the first issue, the second one is still a drawback even for the most recent studies. Previous studies concluded that most field sdO/Bs are members of the Galactic disc. The $<10\%$ candidates for halo sdO/Bs can have very high Galactic restframe velocities close to the escape velocity of the Galaxy. 

However, there are alternative ways to accelerate hot subdwarf stars to (very) high velocities, which have been studied intensively after the discovery of the hypervelocity star (HVS) US\,708, which is a He-rich sdO (He-sdO) with a velocity high enough to be unbound to the Galaxy, thereby excluding an origin in the bound halo population (Hirsch et al. \cite{hirsch05}). Perets (\cite{perets09}) proposed the disruption of a hierarchical triple system by the supermassive black hole in the centre of the Galaxy (a variant of the slingshot scenario proposed by Hills \cite{hills88}) and the subsequent merger of an ejected close helium white dwarf (He-WD) binary as a possible, yet quite complicated scenario for US\,708. Geier et al. (\cite{geier15a}), however, could exclude a Galactic centre origin of this star, calling for a different ejection scenario.

Possible alternatives are related to the supernova ejection scenario (Blaauw \cite{blaauw61}), where a close binary is disrupted by the explosion of one component as supernova and the companion is ejected. In principle, core-collapse supernovae can eject He-stars (Renzo et al. \cite{renzo19}; Tauris \cite{tauris15}), but much higher velocities are possible if thermonuclear supernovae are involved. Close hot subdwarf binaries with massive WD companions turned out to be candidates for the progenitors of such thermonuclear supernovae of type Ia (Maxted et al. \cite{maxted00}; Geier et al. \cite{geier07}; Pelisoli et al.\ \cite{pelisoli21}) for the double-degenerate merger scenario (Iben \& Tutukov \cite{iben84}; Webbink \cite{webbink84}). Shen et al. (\cite{shen18}) proposed a scenario according to which the WD companion survives the SN Ia (see also Pakmor et al. \cite{pakmor21}) and discovered several candidates with extremely high velocities ($1000-3000\,{\rm km\,s^{-1}}$). These objects are more luminous than normal WDs, resemble hot subdwarfs of lower temperature, and might remain in such a puffed-up non-equilibrium state for quite some time after the explosion (Bauer et al. \cite{bauer19}; Liu et al. \cite{liu21}). 

Also the exploding WDs themselves might survive the explosion and end up as fast moving objects with very peculiar abundance patterns (e.\ g. Jordan et al. \cite{jordan12}; Bravo et al. \cite{bravo16}). Several fast candidates with such  very peculiar surface compositions have been discovered (Vennes et al. \cite{vennes17}; Raddi et al. \cite{raddi18a,raddi18b,raddi19}; G\"ansicke et al. \cite{gaensicke20}; El-Badry et al. \cite{elbadry23}; Igoshev et al. \cite{igoshev23}; Scholz \cite{scholz24}; Werner et al. \cite{werner24}). Most of these fast objects also have luminosities similar to hot subdwarfs, but significantly lower temperatures than US\,708. A possible evolutionary connection has been proposed by Shen et al. (\cite{shen18}), according to which US\,708 might represent the state after the puffed-up phase when the star is moving back to the WD cooling tracks (Bauer et al. \cite{bauer19}). 

Hot subdwarfs are also good candidates for ejected companions in the single-degenerate SN Ia scenario, in which stable mass transfer from a donor star triggers the explosion of a WD (Whelan \& Iben \cite{whelan73}; Nomoto \cite{nomoto82}). The WD might either explode when reaching the Chandrasekhar mass or by the detonation of a helium layer on top of a C/O WD core (Woosley \& Weaver \cite{woosley94}; Fink et al. \cite{fink10}). The hot subdwarf companions ejected in such a scenario can reach velocities of several hundred ${\rm km\,s^{-1}}$ and exceed the escape velocity of the Galaxy (Justham et al. \cite{justham09}; Wang \& Han \cite{wang09}; Geier et al. \cite{geier13}; Neunteufel \cite{neunteufel20}; Neunteufel et al. \cite{neunteufel19,neunteufel21,neunteufel22}). Most of the predicted properties of those ejected He-star companions are consistent with the properties of US\,708 (Geier et al. \cite{geier15a}).

Several sdO/B binary progenitor candidates for this scenario with massive WD companions are known (Mereghetti et al. \cite{mereghetti09,mereghetti21}; Vennes et al. \cite{vennes12}; Geier et al. \cite{geier13}; Kupfer et al. \cite{kupfer22}) as well as several other extremely close sdO/B+WD systems with lower mass companions  (Kupfer et al. \cite{kupfer17a,kupfer17b}; Luo et al. \cite{luo24}), some of them even found in a state of ongoing mass transfer (Kupfer et al. \cite{kupfer20a,kupfer20b}). Although those binaries do not fulfill all the criteria for SN\,Ia progenitor candidates, they provide evidence for a rich population of interacting sdB+WD systems. In contrast to that, no additional candidate for an ejected hot subdwarf companion similar to US\,708 has been discovered yet. 

The search for both the close binary progenitors and the surviving ejected companions of thermonuclear supernovae was one of the key science goals of a long-term survey project, which started in 2009 and will be briefly summarised in the following sections. In this paper, we present the final results of the spectroscopic and kinematic survey of fast hot subdwarfs. 

\section{Previous results from the spectroscopic survey}

A dedicated survey was performed to find hot subdwarf stars with massive compact companions, such as massive white dwarfs ($>$1.0\,$M_{\rm \odot}$), neutron stars or stellar mass black holes: the ``Massive Unseen Companions to Hot Faint Underluminous Stars from SDSS'' project, or short MUCHFUSS. Hot subdwarf stars were selected from the Sloan Digital Sky Survey (SDSS, e.g. Ahn et al. \cite{ahn12}) by colour, followed by visual inspection of the spectra. Hot subdwarf stars with high radial velocity variations were selected as candidates for follow-up spectroscopy to derive the radial velocity curves and the binary mass functions of the systems (see Geier et al. \cite{geier11a}; Geier \cite{geier15b}). 

Hot subdwarfs with extreme space velocities were studied in the Hyper-MUCHFUSS spin-off project. The initial survey was restricted to sdO/B stars with constant RVs exceeding $100\,{\rm km\,s^{-1}}$. Atmospheric parameters, spectroscopic distances and RVs were obtained from SDSS spectra. Proper motions were determined using position measurements from ground-based catalogues (Tillich et al. \cite{tillich11}). Twelve high-velocity sdO/Bs were discovered, either originating from the central bulge region of the Galaxy or the outer Galactic disc. While eleven of these stars were found to be bound to the Galaxy, one star was proposed to be unbound. 

\begin{table*}[t!]
\caption{Summary of the follow-up observations in the course of the MUCHFUSS project.}
\label{tab:runs}
\begin{center}
\begin{tabular}{lllcll} \hline\hline
\noalign{\smallskip}
Date & Nights & Telescope\,\&\,Instrument & Resolution [$\lambda$/$\Delta\lambda$] &  Coverage [\AA]  &  Observer\\ \hline
\noalign{\smallskip}
2015/08/20 -- 2015/08/22 & 3   & ING-WHT+ISIS      &  2000  &  3440 -- 5270     &  T. K. \\
2015/09/04 -- 2015/09/06 & 3   & CAHA-3.5m+TWIN    &  4000  &  3460 -- 5630     &  M.S., S.K. \\
2016/12/27 -- 2016/12/31 & 3   & ING-WHT+ISIS      &  2000  &  3440 -- 5270     &  S. B. \\
2017/04/18 -- 2017/04/21 & 4   & ESO-NTT+EFOSC2    &   600  &  3270 -- 5240     &  S. K. \\
2017/07/24 -- 2017/07/26 & 3   & ING-WHT+ISIS      &  2000  &  3440 -- 5270     &  D.S., S.G. \\
2018/01/03 -- 2018/01/05 & 3   & ING-WHT+ISIS      &  2000  &  3440 -- 5270     &  S. B. \\
2015,2016,2019           & 3.5 & ESO-VLT+X-shooter  &  4000  &  3000 -- 10000    &  Service \\
2017/2018                & 2.5 & GTC+OSIRIS        &  1000  &  3600 -- 7500     &  Service \\
2018/05/15               & 1   & Keck+ESI          &  7000  &  4000 -- 10000    &  T. K. \\
\noalign{\smallskip}
\hline
\noalign{\smallskip}
                         &     & Palomar-5m+DBSP$^{\rm a}$   &  1500   &  3800 -- 5700     &  T. K. \\
                         &     & Keck+LRIS$^{\rm a}$         &   900   &  3200 -- 5600     &  T. K. \\
\hline
\end{tabular}
\tablefoot{
$^{\rm a}$ Observations obtained as backup during other projects.
The spectra were reduced with different standard software packages such as IRAF and MIDAS. The LRIS, ESI and X-shooter spectra were reduced using pipelines provided by the Mauna Kea observatory and ESO, respectively.}
\end{center}
\end{table*}

However, a more detailed analysis of this latter star PB\,3877 revealed it to be a composite binary consisting of an sdB and a K-type main-sequence companion with an orbital period of the order of hundreds of days. Furthermore, a bound orbit could not be excluded any more. Since such a wide binary is quite fragile, any acceleration scenario requiring close binary interactions or close encounters with massive objects could be excluded, leaving a primordial or accreted halo origin as the only viable scenario for this star (Nemeth et al. \cite{nemeth16}). 

In contrast to that, a detailed follow-up study of US\,708 (Geier et al. \cite{geier15a}) based on improved spectroscopy and astrometry revealed that this star is moving even faster than estimated before (Hirsch et al. \cite{hirsch05}). With a Galactic restframe velocity $\sim 1000\,{\rm km\,s^{-1}}$ it turned out to be the fastest unbound star in the Galaxy known by then (see also Neunteufel \cite{neunteufel20}). In addition, an origin close to the Galactic center could be excluded and with that the Hills ejection mechanism. Furthermore, the spectrum of US\,708 shows significant rotational broadening of the spectral lines quite untypical for a  He-sdO, but perfectly consistent with a prior phase of tidal spin-up in an ultracompact binary consisting of a low-mass hot subdwarf with $\sim0.3\,M_{\rm \odot}$ and a WD with more than $1.0\,M_{\rm \odot}$. These results are consistent with US\,708 being the ejected companion of a thermonuclear SN.

Finally, detailed analyses of four more fast hot subdwarf candidates were performed: two from the sample of Tillich et al. (\cite{tillich11}) and two more discovered in the MUCHFUSS survey (Geier et al. \cite{geier15b}). All of these stars are bound to the Galaxy and three can be traced back to the Galactic center or the inner disc. The He-sdO SDSS J205030.39$-$061957.8 turned out to be very similar to US\,708 regarding its spectroscopic parameters. It originates from the outskirts of the Galactic disc, but is also on a bound Galactic orbit (Ziegerer et al. \cite{ziegerer17}). 

\section{Revisiting and extending the sample}
\label{sec:extended}

The results of the \textit{Gaia} mission are a game changer, because it provides astrometry of unprecedented precision allowing for detailed kinematic studies. Large surveys have provided UV, optical and IR photometry, which allow us to construct detailed spectral energy distributions.
Spectral data bases (e.g. SDSS, LAMOST) provide multi-epoch spectroscopy. This wealth of new information motivated us to revisit and extend the sample of fast hot subdwarfs. We present the extension of the sample in Sect.~\ref{sec:extended}.
The quantitative spectral analysis is presented in Sect.~\ref{sec:homogeneous}. 
Multi-epoch spectra are used to check for radial velocity variability in Sect. \ref{sec:radial_velocity}. Spectral energy distributions are constructed and spectroscopic distances are derived in Sect.~\ref{sec:sed} and then combined with \textit{Gaia} proper motions (PMs) to carry out a kinematic study (Sect.~\ref{sec:kinematic}).

\subsection{Target selection}\label{selection}

To substantially increase the sample of high velocity sdO/Bs and to find unbound objects similar to the prototype US\,708, we performed an extended and more systematic survey, including archival spectroscopic and ground-based astrometric data. Our aim was to select candidates with high Galactic restframe velocities, for which we wanted to obtain spectroscopic follow-up observations for detailed analyses. This part of the project has been conducted mostly between 2015 and 2018 with the data and methods available to us at that time. 

Starting point were again the regular data releases of the Sloan Digital Sky Survey (up to DR14), including newly classified hot subdwarfs (Geier et al. \cite{geier15b}; Kepler et al. \cite{kepler16,kepler19}). To further increase the input sample, we compiled a comprehensive catalogue of more than $5600$ spectroscopically identified hot subdwarfs from the literature (Geier et al. \cite{geier17}). 

For more than $1900$ sdO/Bs the spectroscopically determined fundamental parameters effective temperature $T_{\rm eff}$, surface gravity $\log{g}$, helium abundance $\log{n(\rm He)/n(\rm H)}$ and radial velocity (RV) have been published in the literature. To improve our target selection, we performed quantitative spectral analyses of both the SDSS and the follow-up spectra of the most promising candidates for fast sdO/Bs. The method is described in Geier et al. (\cite{geier11b}). Spectroscopic distances have been determined as described in Ramspeck et al. (\cite{ramspeck01}) assuming the canonical mass of $0.47\,{\rm M_{\odot}}$ for the subdwarfs.   

\begin{figure*}[t!]
\begin{center}
	\resizebox{9cm}{!}{\includegraphics{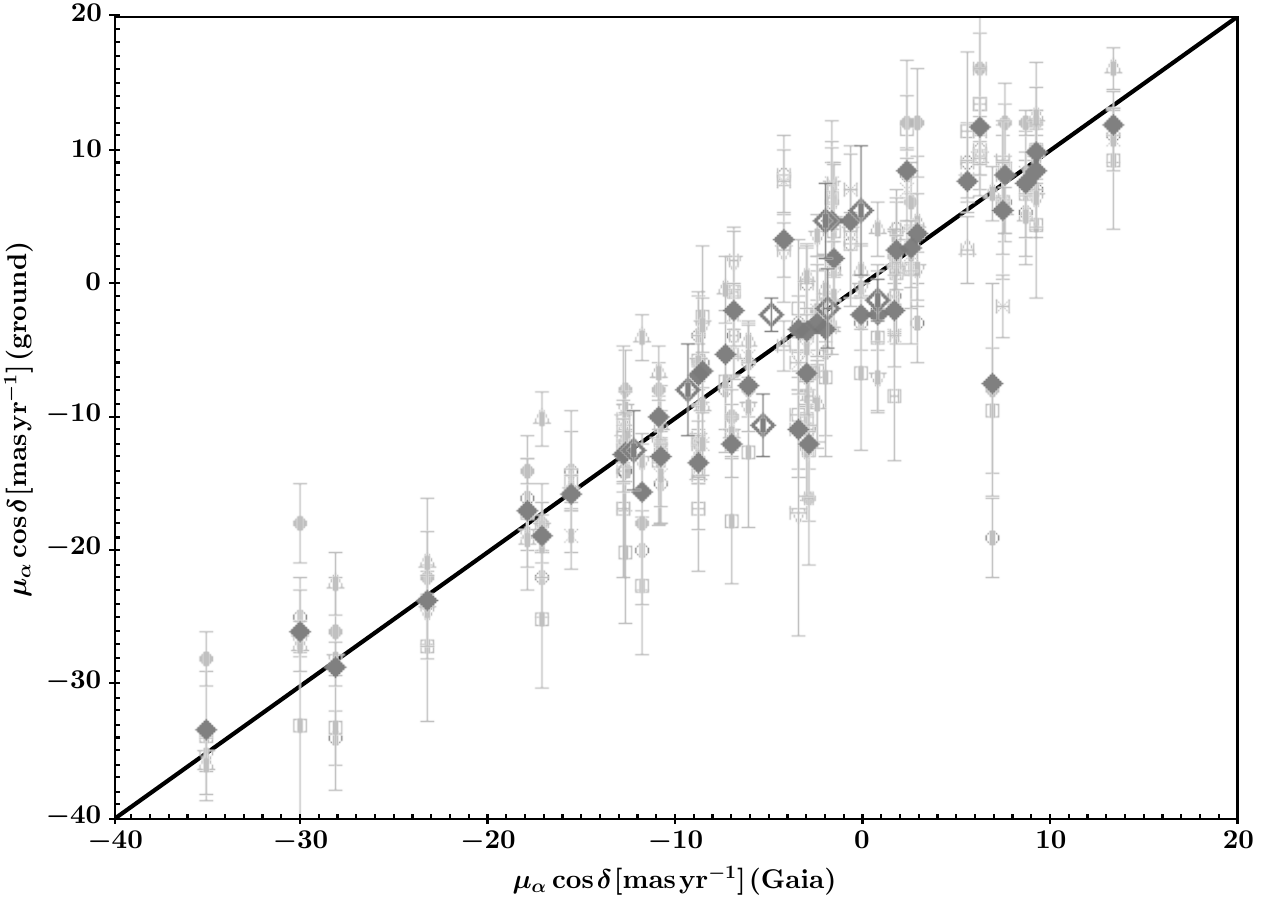}}
	\resizebox{9cm}{!}{\includegraphics{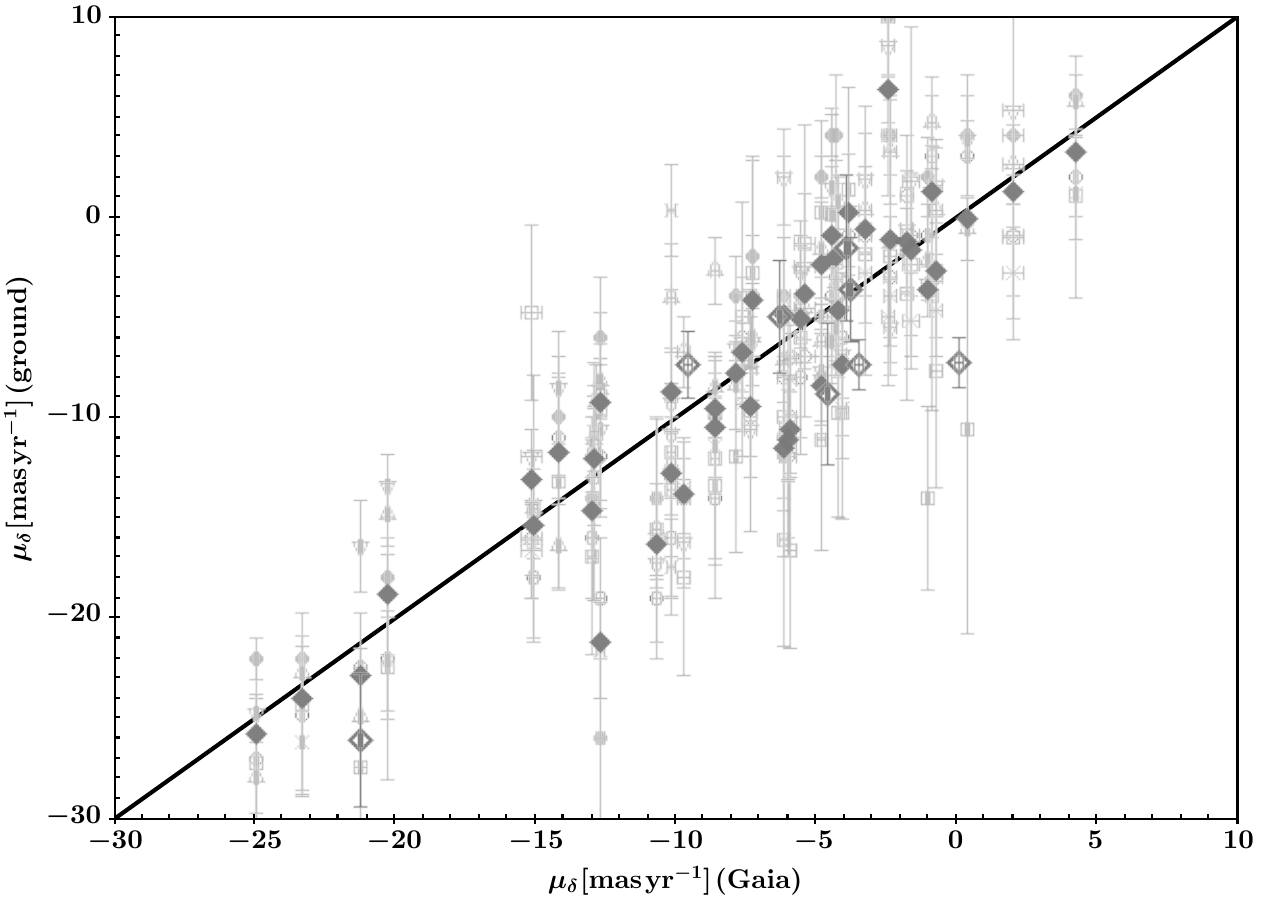}}
\end{center} 
\caption{Proper motion measurements in right ascension (left panel) and declination (right panel). \textit{Gaia} proper motions from Early Data Release 3 (EDR3, \textit{Gaia} collaboration \cite{gaia21}) are plotted against ground based measurements from SDSS\,DR9 (Ahn et al. \cite{ahn12}, open circles), PPMXL catalog of positions and proper motions on the ICRS (Roeser et al. \cite{roeser10}, open squares), Whole-Sky USNO-B1.0 Catalog (Monet et al. \cite{monet03}, open diamonds), Absolute Proper motions Outside the Plane catalog (APOP, Qi et al. \cite{qi15}, open downward triangles), Gaia-PS1-SDSS (GPS1, Tian et al. \cite{tian17}, open upward triangles), and Hot Stuff for One Year (HSOY, Altmann et al. \cite{altmann17}, crosses). PMs from the Fourth U.S. Naval Observatory CCD Astrograph Catalog (UCAC4, Zacharias et al. \cite{zacharias13}) have been used as well, but are not plotted here, because only few stars have measurements in this catalogue. The averages of the ground-based values from those surveys are plotted as filled diamonds. The ground-based measurements by Tillich et al. (\cite{tillich11}) and Ziegerer et al. (\cite{ziegerer17}) are plotted as open dark grey diamonds.}
\label{fig:pmcomp}
\end{figure*}

Because our survey started well before the \textit{Gaia} era, ground-based as well as hybrid ground- and space-based PMs have been taken from several catalogues (see Fig.~\ref{fig:pmcomp}). Since Galactic field sdO/B stars have typical distances of the order of several hundred pc to a few kpc, their PMs are quite small and often close to the detection limits of the ground-based surveys. Extreme care has to be taken using such PMs (see Ziegerer et al. \cite{ziegerer15}), which is why we applied a very conservative approach and only selected stars where all the independent PM measurements from the literature (each stars has about $4-5$) were significant and consistent with each other. Only in a few cases, where the majority of the other measurements was consistent and the data quality high, we allowed for one outlier at most. 

We calculated the Galactic velocity components and the total Galactic restframe velocities by combining the RV, PM and spectroscopic distance measurements. For the spectroscopic follow-up we finally selected the candidates with Galactic restframe velocities $v_{\rm grf}>300\,{\rm km\,s^{-1}}$. 

\subsection{Spectroscopic follow-up observations}

A follow-up campaign was conducted to obtain spectra of higher quality and to improve the accuracy of the derived spectroscopic parameters. Second epoch spectra are also needed to identify RV variable stars, since the close binary fraction of sdBs is very high. Reliable kinematic parameters can only be derived for stars with constant RV, and only single hot subdwarfs can be candidates for the ejected companions of thermonuclear supernovae. 

The campaign has been conducted from 2015 to 2018 in dedicated observing runs with 4m- to 10m telescopes equipped with medium-resolution spectrographs. In total more than 26 nights of observing time have been granted for this project (see Table~\ref{tab:runs}). 

Additional spectra were retrieved from the SDSS DR17 (taken with the original SDSS spectrograph as well as with the BOSS spectrograph, Abdurro'uf et al. \cite{sdssiv}) and LAMOST DR8 data bases.
An overview of the spectroscopic data used for the spectroscopic analysis is given in Table \ref{tab:spec_overview}.

\subsection{Cleaning the sample}

From the initial sample of fast sdO/Bs candidates we have removed many false-positives as observational data was successively added. This happened for different reasons outlined in the following.

\paragraph{Stars with erroneous parameters:}

Over the last decade, several new ground-based proper motion catalogues were published (see Sect.~\ref{selection}) and continually used to check the consistency of the individual measurements. Many stars turned out to have inconsistent PMs and were therefore removed from the sample. Finally, the highest quality, space-based PMs provided by \textit{Gaia} allowed us to eliminate the remaining candidates with erroneous measurements. 

Fig.~\ref{fig:pmcomp} shows a comparison of the ground-based proper motions of our final sample from the most relevant surveys with the proper motions from the \textit{Gaia} mission. The results are, except for some outliers, in general consistent within the uncertainties confirming our ground-based target selection. 

The spectroscopic follow-up allowed for an improved determination of the atmospheric parameters with respect to the ones determined from the SDSS spectra alone. In many cases this had a significant impact on the spectroscopic distances, which had previously often been overestimated. Many candidates turned out to be much closer than estimated and their Galactic velocity components are thus much slower. One of those stars, the He-sdO SDSS\,J141812.51$-$024426.8 has recently been reanalysed by Werner et al. (\cite{werner22}).

\paragraph{Composites and blends:}%\label{sec:misclass2} label unreferenced

Spectra of higher resolution and quality revealed spectroscopic signatures of cool MS companions in several of our programme stars, which initially ranked among our top candidates for fast sdO/Bs. The high space velocities of those stars are likely caused by inaccurate atmospheric parameters, which were determined by fitting single-star model spectra and led to an overestimation of the spectroscopic distances. However, we also found a composite system with a confirmed high space velocity in our survey (Nemeth et al. \cite{nemeth16}). 

To identify other composite binaries or blended sources we performed an analysis of the spectral energy distributions (SEDs) of all the stars from our sample, using a $\chi^{2}$-minimization method to match filter-averaged magnitudes calculated from model spectra to observed magnitudes drawn automatically from photometric catalogues in the literature (see Heber et al. \cite{heber18} for a more detailed description and Table~\ref{tab:photo_surveys} for a list of the photometric catalogues used). The method allows to fit two stellar components and can be used to determine the effective temperatures of both stars, the surface ratio and the angular diameter of the primary in composite binaries (e.\ g.\ Dorsch et al.\ \cite{dorsch21}). Since the SEDs include photometry up to the infrared regime, this method is very well suited to detect cool companions, even if they are too faint to show significant spectroscopic features. In addition, comparing the surface ratio with theoretical predictions for the radii of hot subdwarfs and cool companions of the determined temperatures, significant mismatches allow us to identify physically unassociated blends.

In this study, we only present the analysis of the sdO/Bs with SEDs indicative of single stars, which do not show any additional IR-excess. The analysis of the composite binaries will be presented in another paper of this series (Heber et al. in prep.).

\paragraph{Proto-He-WDs:} %\label{sec:misclass3} label unreferenced
Another potential reason for misclassification as a very fast star might be the underlying assumption that the stars are in the (post-)EHB phase and that they all have the same mass of $0.47\,M_{\rm \odot}$. It is known that red giant branch stars can be stripped via binary interactions even before they ignite core helium-burning (e.g. Iben \& Tutukov \cite{iben86}; Benvenuto \& De Vito \cite{benvenuto04}). Those objects subsequently cool down and evolve to become He-WDs. Stars in this proto-He-WD stage can have similar atmospheric parameters as sdB or blue horizontal branch (BHB) stars, but have significantly lower masses ($\sim0.2-0.3\,M_{\rm \odot}$), often smaller radii and therefore also lower luminosities (e.g. Heber et al. \cite{heber03}; Kawka et al. \cite{kawka15}). 

That such objects can be easily misclassified as much more luminous and therefore distant stars is impressively demonstrated by the putative top hypervelocity BHB candidate SDSS\,J160429.12+100002.2. Detailed spectroscopic follow-up revealed that it is a proto-He-WD in a close binary, which is orbited by a very likely substellar companion (Irrgang et al. \cite{irrgang21}), drastically reducing its estimated space velocity. 

Although these proto-He-WDs appear to be much rarer than sdO/Bs in the (post-)EHB phases (especially at the hot end of the EHB, where their evolutionary times are one to two orders of magnitude shorter), our selection for apparently large space velocities may be biased towards such low-mass objects. We therefore conclude that the relative fraction of proto-He-WDs in our sample may be higher than in the field population. 

\paragraph{Runaway main-sequence star:}

Finally, the B-type star PG\,1610+062, an object originally suspected to be an RV-variable hot subdwarf rather than a fast moving one (Geier et al. \cite{geier15b}), eventually turned out to be a runaway main sequence star with an exceptionally high ejection velocity from the disc, which challenges current ejection scenarios (Irrgang et al. \cite{irrgang19,irrgang21}).

\section{Homogeneous spectral  analysis}\label{sec:homogeneous}

The finally selected objects showing diverse spectral types are classified, analysed and checked for RV variability in the following sections. 

\begin{figure*}
\begin{center}
\includegraphics[width=0.8\textwidth]{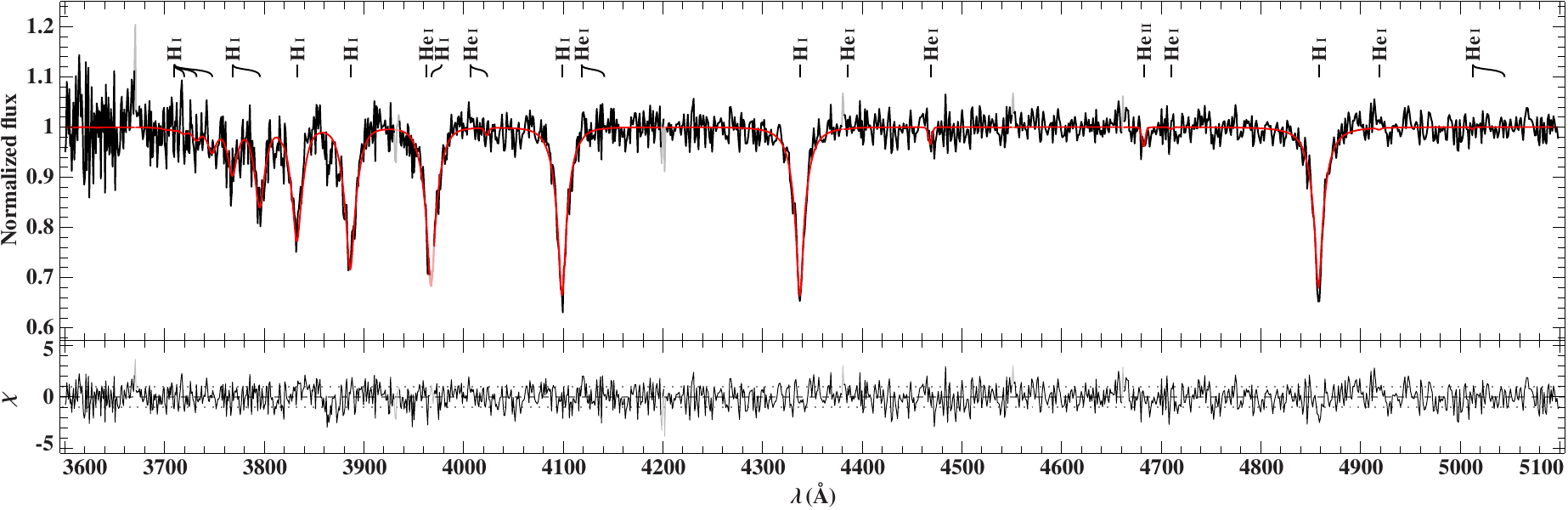}
\includegraphics[width=0.8\textwidth]{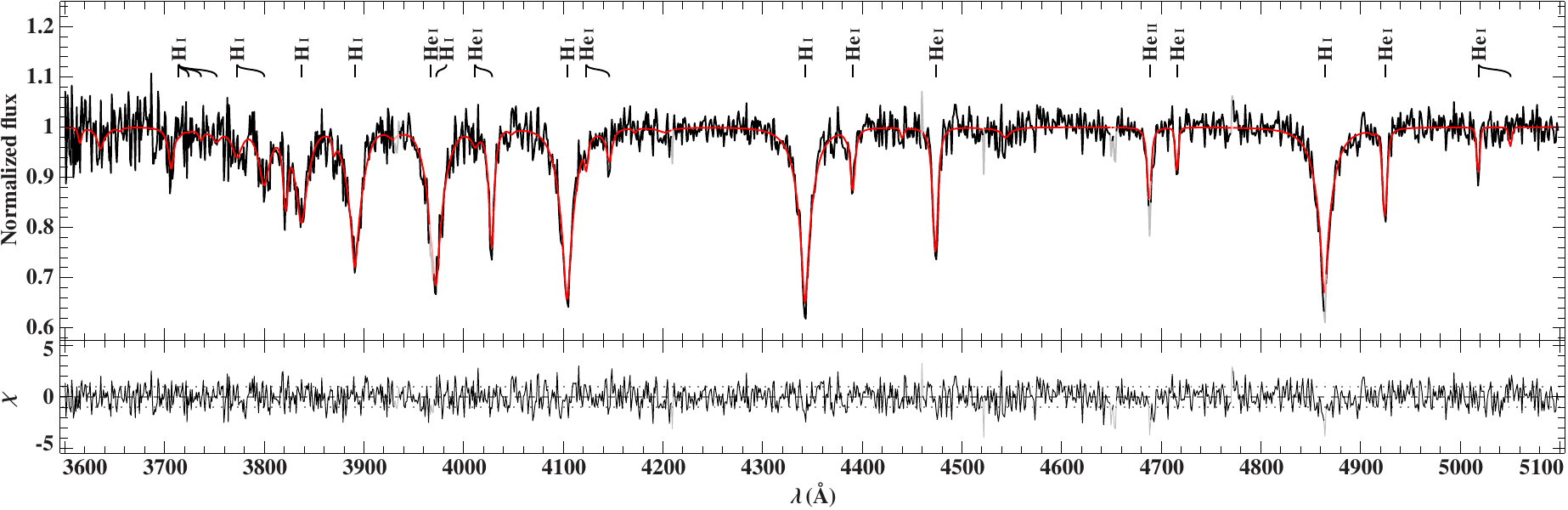}
\includegraphics[width=0.8\textwidth]{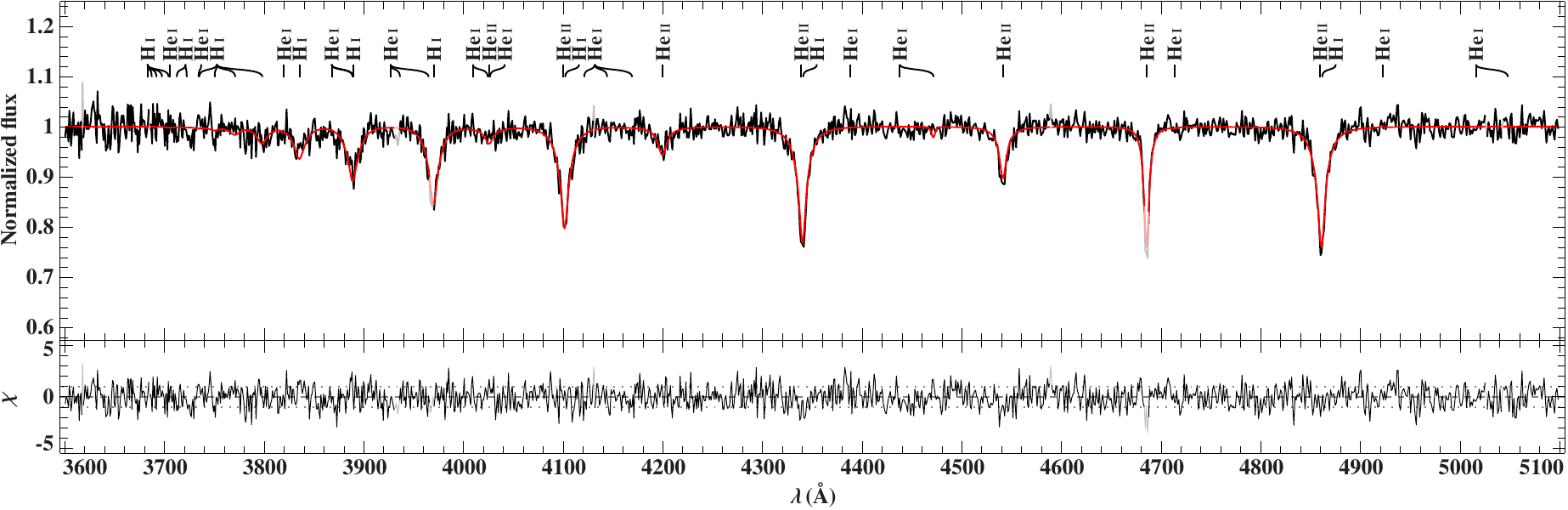}
\caption{Excerpts of BOSS spectral fits of He-poor hot subdwarfs. From top to bottom: SDSSJ124248.89+133632.6 (sdB), SDSSJ124310.58+343358.4 (sdOB), SDSSJ124819.08+035003.2 (sdO). In the upper panels the best-fit synthetic spectrum (red) is shown along with the observed one. The lower panels give residual $\chi$, whereby the long-dashed horizontal lines mark the zero values and the short-dashed horizontal lines deviations in terms of $1\sigma$. Parts of the spectra, which have been excluded from the fits, are marked in grey.}
\label{fig:specfig1}
\end{center}
\end{figure*}

\begin{figure*}
\begin{center}
\includegraphics[width=0.8\textwidth]{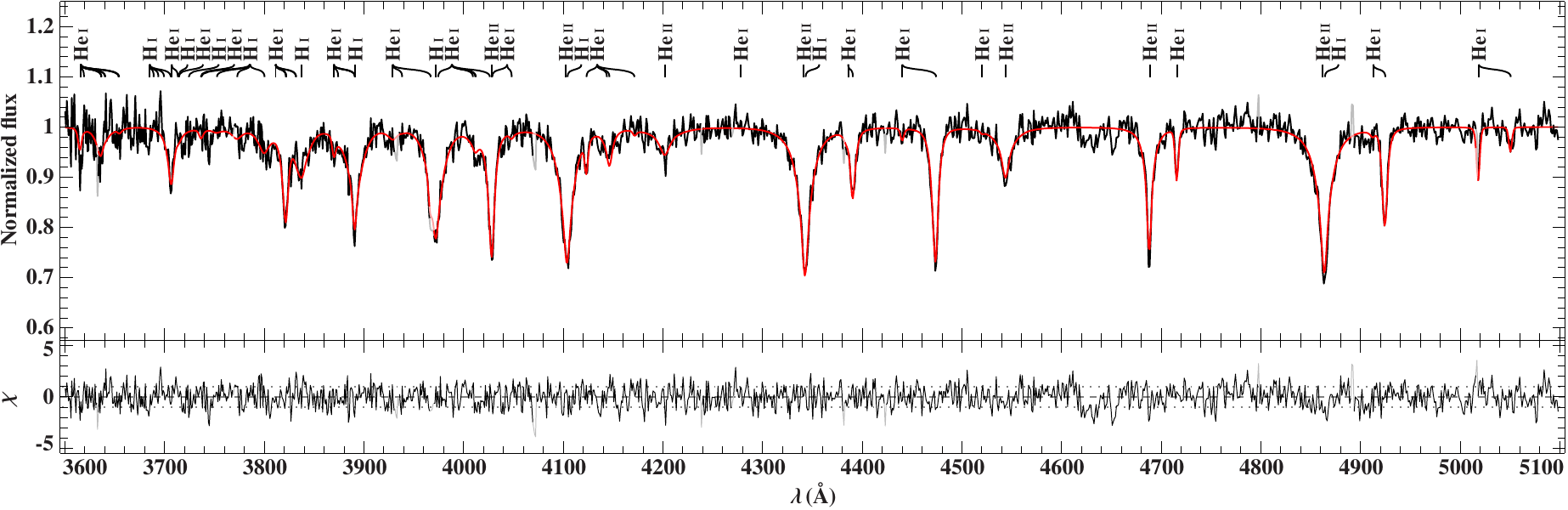}
\includegraphics[width=0.8\textwidth]{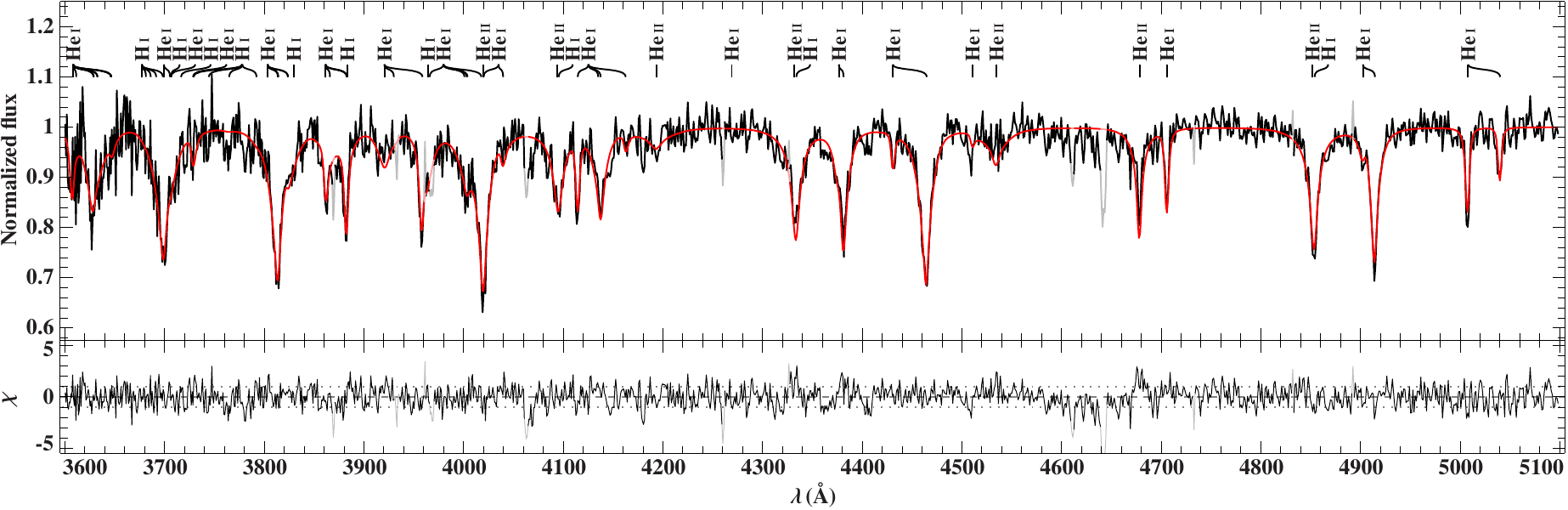}
\includegraphics[width=0.8\textwidth]{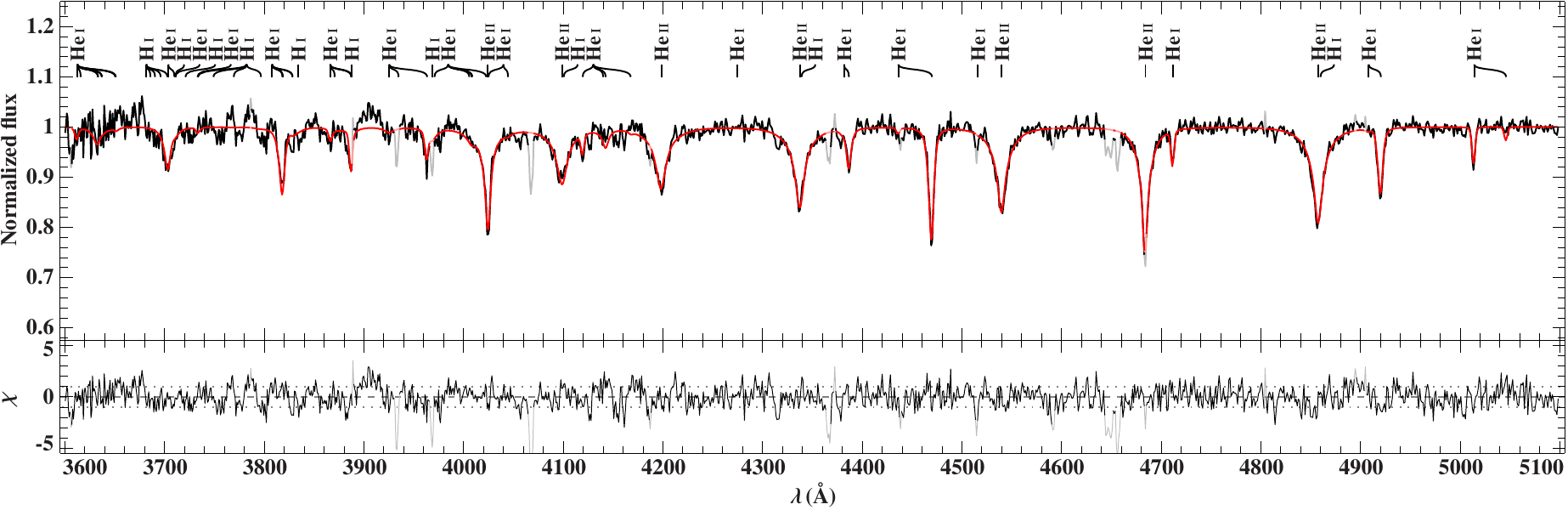}
\caption{Same as Fig. \ref{fig:specfig1}, but for He-rich hot subdwarfs. From top to bottom: SDSSJ120352.24+235343.3 (iHe-sdOB), SDSSJ222515.34-011156.8 (eHe-sdOB), SDSSJ080833.76+180221.8 (eHe-sdO).
}
\label{fig:specfig2}
\end{center}
\end{figure*}

%ISIS Spectra
\begin{figure*}
\begin{center}
\includegraphics[width=0.8\textwidth]{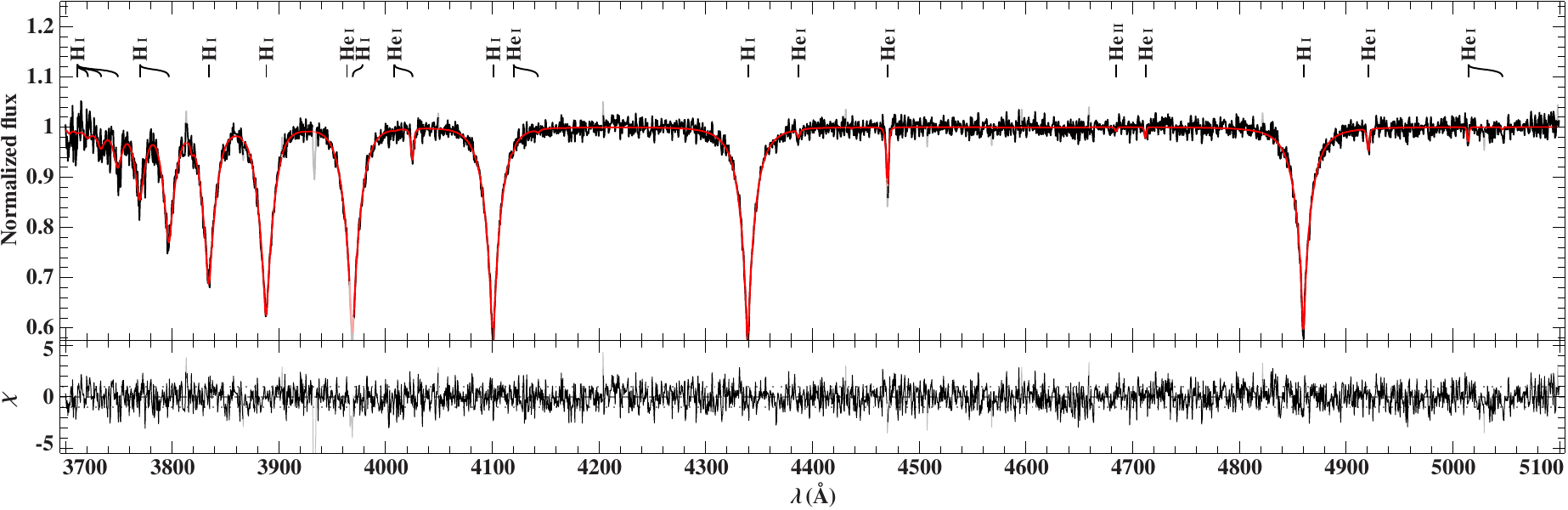}
\includegraphics[width=0.8\textwidth]{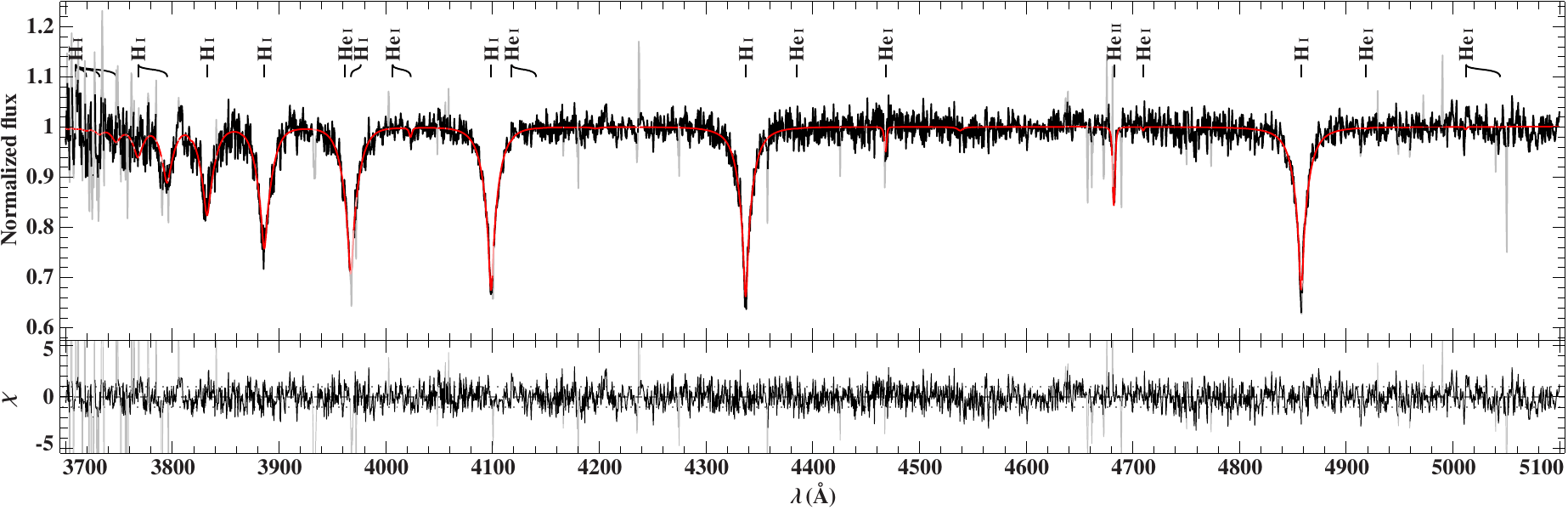}
\includegraphics[width=0.8\textwidth]{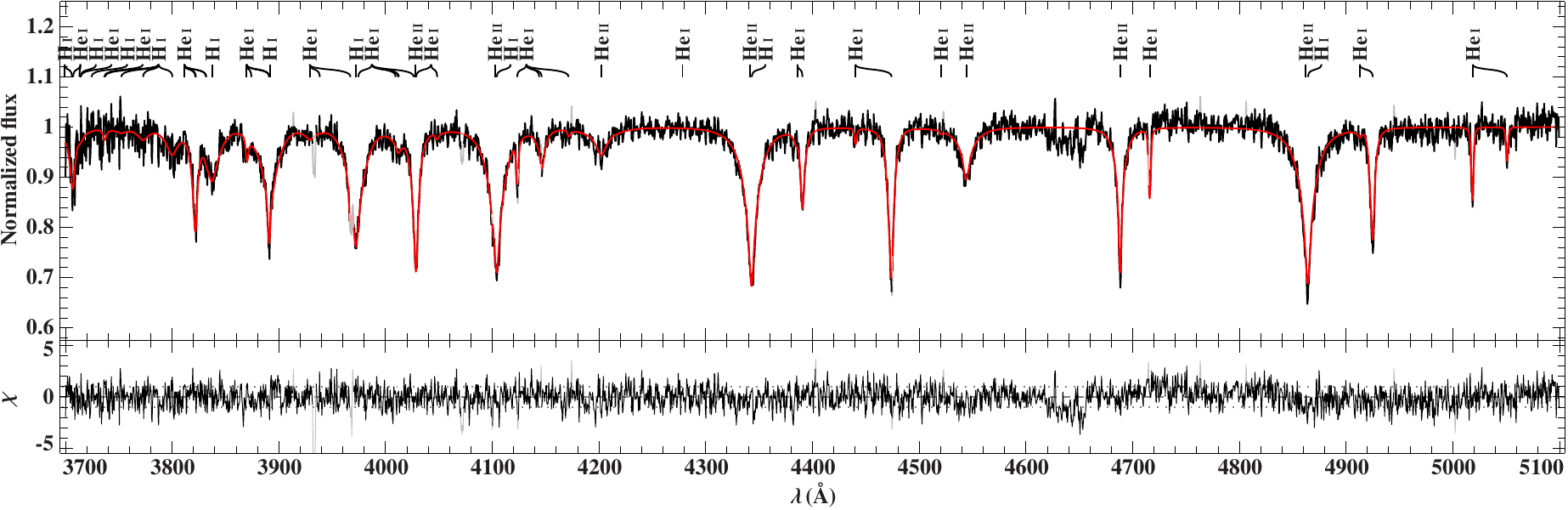}
\caption{Same as Fig. \ref{fig:specfig1}, but showing fits of the WHT/ISIS spectra of hot subdwarfs. From top to bottom: SDSSJ130543.97+115840.8 (sdB), SDSSJ221728.35+121642.6 (sdOB), SDSSJ120352.24+235343.3 (iHe-sdOB).}
\label{fig:specfig3}
\end{center}
\end{figure*}

\subsection{Spectral classification}\label{sec:spec_class}

A spectral classification scheme has been established by Drilling et al. (\cite{drilling13}) and was applied e.\ g.\ by Lei et al. (\cite{lei19}) and  Jeffery et al. (\cite{jeffery21}). 
This classification scheme adds a third dimension to the MK scheme by defining 40 helium classes to address the diverse helium line strengths in the optical spectra. We refrain from applying this scheme to our sample, but condense it into just nine spectral sub-classes. 
We can easily distinguish hydrogen dominated from helium dominated spectra. Helium lines may or may not be present for the H-dominated spectral types. In sdB stars He\,$\textsc{i}$ lines may be present but no He\,$\textsc{ii}$ lines. Somewhat hotter H strong-lined stars are classified as sdOB if they display He\,$\textsc{ii}$ 4686\,\AA\ in addition to He\,$\textsc{i}$ lines.  In sdO stars 
the Balmer lines dominate the spectrum but He\,$\textsc{ii}$ lines are stronger than the He\,$\textsc{i}$ lines, the latter may even be absent in the hottest.

The helium strong spectral types are more delicate to classify because both temperature and helium abundance have a strong impact on the spectral appearance. We may distinguish extremely He strong-lined hot subdwarfs (He-sdB, He-sdOB and He-sdO) from intermediate ones by the presence or absence of any hydrogen Balmer lines. In the former case they will be classified as intermediate (iHe), in the latter as extreme helium (eHe) subdwarfs. The relative strength of He\,$\textsc{i}$ and He\,$\textsc{ii}$ lines varies with temperature. While no He\,$\textsc{ii}$ lines are detectable in He-sdBs, He\,$\textsc{ii}$ lines are stronger than He\,$\textsc{i}$ lines in He-sdOs. In He-sdOBs the strongest lines are from He\,$\textsc{i}$, but He\,$\textsc{ii}$ is detectable and may reach similar strength. 

Hence we distinguish three H strong-lined classes (sdB, sdOB, and sdO) and six He strong-lined classes (iHe-sdB, iHe-sdOB, iHe-sdO, eHe-sdB, eHe-sdOB, and eHe-sdO). In Table \ref{tab:spec} we give the spectral classifications of all programme stars\footnote{J174211.75+643009.9 turns out to be a BHB star}. In Fig.\ \ref{fig:specfig1} we show examples for H strong-lined types, while Fig. \ref{fig:specfig2}  shows examples of He dominated spectra.  

The final sample of fast hot subdwarfs comprises 53 stars. We classified 29 sdBs ($55\%$), five sdOBs ($9\%$), five sdOs ($9\%$), seven iHe-sdOBs ($13\%$), two iHe-sdOs ($4\%$), two eHe-sdOBs ($4\%$), two eHe-sdO ($4\%$), and one BHB ($2\%$). There are no He-sdB stars in the sample. 

\subsection{Model atmospheres}\label{sec:model_atmospheres}

The chemical composition of sdB stars is known to be non-solar and varies considerably from star to star (Pereira \cite{pereira11}; Naslim et al. \cite{naslim13}; Geier \cite{geier13a}). Because individual abundances for the programme stars are not available, we used the average abundance pattern from Pereira (\cite{pereira11}) as reported by Naslim et al. (\cite{naslim13}) for the model atmosphere calculations. 

Model atmospheres and synthetic spectra have been calculated in a hybrid local thermodynamic equilibrium (LTE)/non-LTE (NLTE) approach with a suite of numerical codes. The ATLAS\,12 code (Kurucz \cite{kurucz96}) was used to calculate the temperature and density stratification of the
atmosphere in LTE with the most recent Kurucz line lists\footnote{\url{http://kurucz.harvard.edu/linelists/gfnew/} as of October 8, 2017.} to incorporate the line blanketing effect. The opacity sampling technique implemented in ATLAS\,12 allows non-solar chemical abundances to be used without pre-computing opacity distribution functions.

Once the model atmosphere calculations had been converged, detailed hydrogen and helium model atoms were set up to calculate occupation numbers for all relevant levels of hydrogen, He\,\textsc{i}, and He\,\textsc{ii}.
The DETAIL code (Giddings \cite{giddings81}; 
Butler \& Giddings \cite{butler85}) numerically solves the coupled equations for radiative transfer and statistical equilibrium. The emerging H/He spectrum was then computed with the SURFACE code (Butler \& Giddings \cite{butler85}) using detailed line broadening tables for hydrogen and helium transitions (see Irrgang et al. \cite{irrgang18b} for more details). 

We calculated extensive grids covering the entire parameter range of hot subluminous stars. The backbone grid covers effective temperatures from 9000\,K to 55\,000\,K, surface gravities log\,$g$ = 4.6 to 6.6 and helium abundances from log\,$n$(He) = $-$5.05 to $-$0.041. Extensions to high $T_{\rm eff}$ (75\,000\,K for log\,$g$ = 5.2 to 6.6) and high helium abundances up to log\,n(He)=-0.001 (for $T_{\rm eff}$ =25\,000 to 55\,000\,K) 
were computed to cover the parameters for all subtypes. 

\subsection{Quantitative spectral analyses}\label{sec:spec_analysis}

We replaced the previously used selective $\chi^{2}$ minimisation fitting procedure (Napiwotzki \cite{napiwotzki99}; Hirsch \cite{hirsch09}) by a global one developed by Irrgang et al. (\cite{irrgang14}). This means that we consider the whole useful spectral range, usually 3600\,\AA\ to 6700\,\AA, rather than pre-selecting spectral lines, remove artifacts and simultaneously fit the entire multi-parameter space (effective temperature, surface gravity, helium abundance, projected rotational velocity, and radial velocity). Available spectra of a star were simultaneously fit to determine the atmospheric parameters, projected rotation velocity and radial velocities. Example fits of BOSS spectra are shown (blue spectral parts  only) in Fig. \ref{fig:specfig1} and Fig. \ref{fig:specfig2}. Fits of WHT/ISIS spectra are shown in Fig. \ref{fig:specfig3}. Reduced $\chi$ is plotted as well to demonstrate the quality of fit. The resulting atmospheric parameters are listed in Table \ref{tab:spec}.

\begin{SCfigure*}
\centering
\includegraphics[width=0.63\textwidth]{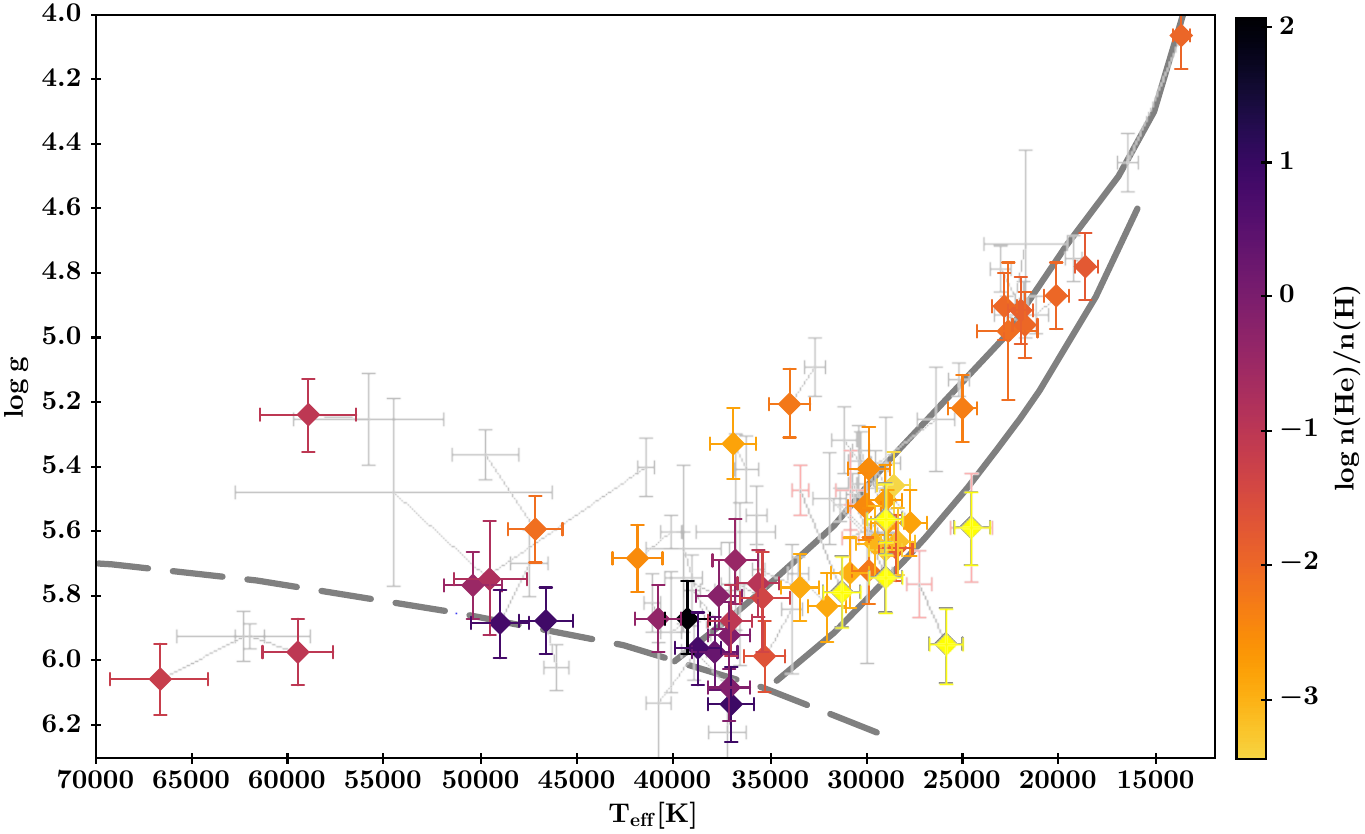} 
\caption{$T_{\rm eff}-\log{g}$ diagram of the sample of fast stars. The colour scales with the helium abundance from light orange to dark purple. The light grey symbols mark the results of the preliminary spectroscopic analysis on which the target selection was based. They are connected with the corresponding values from the new homogeneous analysis. The zero-age and terminal-age EHBs for a subsolar metallicity of $-1.48$ (solid grey lines) have been interpolated from evolutionary tracks by Dorman et al. (\cite{dorman93}). The helium main sequence (dashed line) is taken from Paczynski (\cite{paczynki71}).}
\label{spectro_kiel}
\end{SCfigure*}

\begin{SCfigure*}
\includegraphics[width=0.63\textwidth]{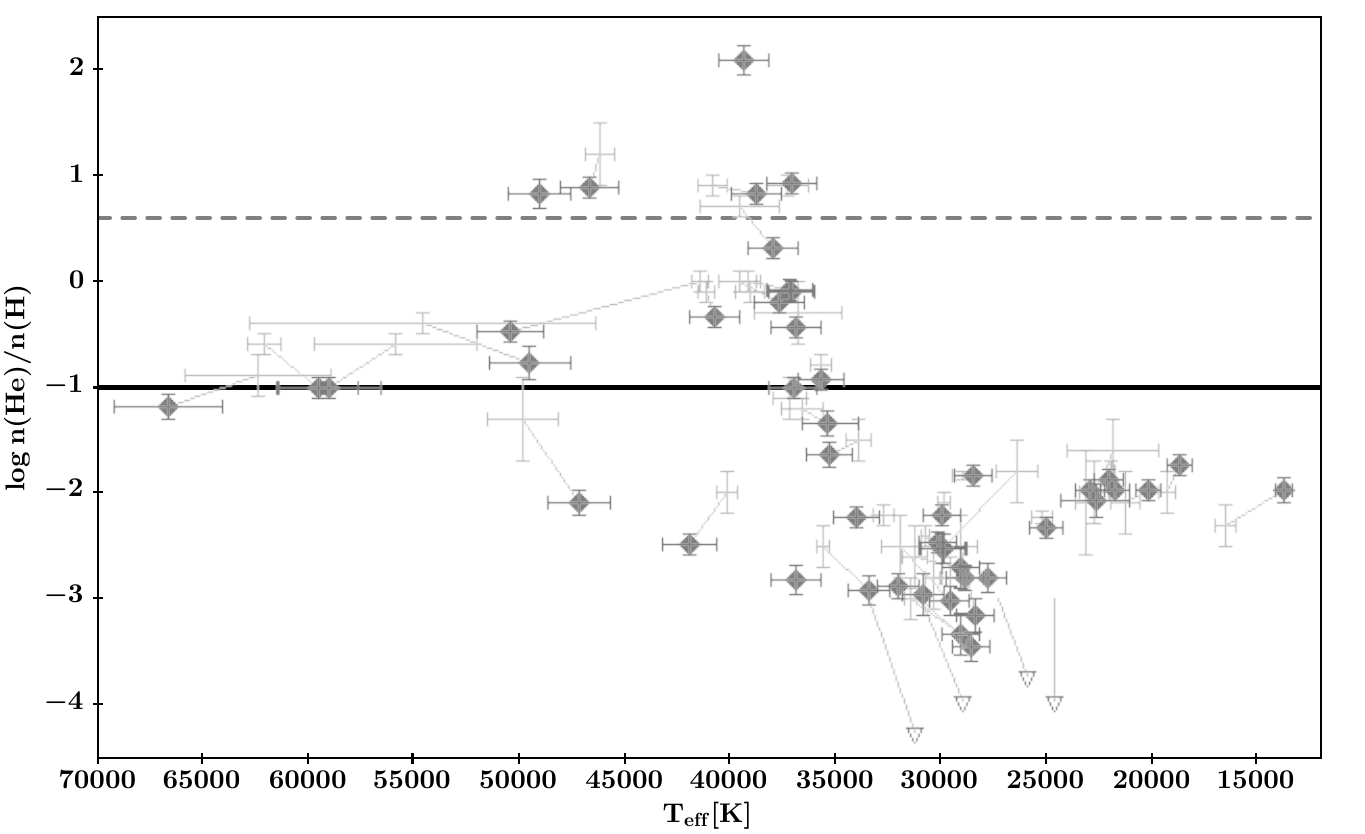} 
\caption{$T_{\rm eff}-\log{n({\rm He})/n({\rm H})}$ diagram of the sample compared to the preliminary analysis (see Fig.~\ref{spectro_kiel}). Solar helium abundance is marked by the solid horizontal line, while the dotted line marks the transition between intermediate and extreme helium abundance.}
\label{spectro_he}
\end{SCfigure*}

The $T_{\rm eff}-\log{g}$ and $T_{\rm eff}-\log{n({\rm He})/n({\rm H})}$ diagrams (Fig.\ \ref{spectro_kiel} and \ref{spectro_he}) generally resemble the corresponding diagrams of the field population (reviewed by Heber \cite{heber16}). Furthermore, it can be seen that the overall distribution of parameters determined in our new homogeneous analysis matches the predicted location close to the EHB better than the preliminary analysis on which the target selection was based.

\begin{table*}
\centering
\renewcommand{\arraystretch}{1.13}
%\begin{longtable}{llllllllll}
\caption{\label{tab:spec} Classifications and spectroscopic parameters. Spectral types are defined in section \ref{sec:spec_class}. Effective temperature, surface gravity, and helium abundance are listed with their 1$\sigma$ uncertainties. The mean radial velocity and its standard deviation, the $\log{p}$ value to indicate radial velocity variability, and the number of spectra used are listed as well. The last column lists the category as defined in Sect. \ref{sec:radial_velocity}.}
\begin{tabular}{llllllllll}
\hline\hline
\noalign{\smallskip}
Name               &  Class      &  $T_{\rm eff}$          & $\log{g}$                 & $\log{n(\mathrm{He}/\mathrm{H})}$
& RV$_\mathrm{average}$         & $\log{p}$ & N$_{\rm spec}$ & category \\
                   &             &  [K]                    &                           &                            & ${\rm [km\,s^{-1}]}$ &         &              & \\
\noalign{\smallskip}
\hline
\noalign{\smallskip}
J022422.21+000313.5 & iHe-sdOB   & 37159$_{-67  }^{+69  }$ & 5.923$_{-0.012}^{+0.012}$ & -0.084$_{-0.009}^{+0.009}$ &  -218 $\pm$2         & -0.24   &  46          & 2 \\
J080833.76+180221.8 & eHe-sdO    & 46626$_{-83  }^{+88  }$ & 5.878$_{-0.018}^{+0.023}$ & 0.8778$_{-0.043}^{+0.038}$ &  -93  $\pm$2         & -1.72   &  13          & 1 \\
J082802.03+404008.9 & sdO        & 47109$_{-321 }^{+321 }$ & 5.597$_{-0.022}^{+0.019}$ & -2.097$_{-0.045}^{+0.044}$ &  -175 $\pm$3         & -0.51   &  13          & 1 \\
J084556.85+135211.3 & sdB        & 24621$_{-564 }^{+388 }$ & 5.591$_{-0.045}^{+0.061}$ & $<$-3.96                   &  109  $\pm$7         & -0.50   &  3           & 4 \\
J090252.99+073533.9 & eHe-sdOB   & 39319$_{-218 }^{+166 }$ & 5.870$_{-0.056}^{+0.041}$ & 2.0840$_{-0.097}^{+0.105}$ &  -135 $\pm$5         & -0.03   &  4           & 4 \\
J091512.06+191114.6 & iHe-sdOB   & 37613$_{-202 }^{+220 }$ & 5.800$_{-0.037}^{+0.032}$ & -0.195$_{-0.026}^{+0.027}$ &  -57  $\pm$5         & -0.41   &  9           & 2 \\
J094850.47+551631.6 & sdB        & 35292$_{-246 }^{+268 }$ & 5.988$_{-0.037}^{+0.049}$ & -1.632$_{-0.050}^{+0.053}$ &  -117 $\pm$7         & -0.44   &  5           & 1 \\
J102057.16+013751.2 & sdB        & 28545$_{-142 }^{+99  }$ & 5.458$_{-0.022}^{+0.017}$ & -3.440$_{-0.093}^{+0.093}$ &  232  $\pm$3         & -0.09   &  7           & 1 \\
J102439.43+383917.9 & sdB        & 29921$_{-58  }^{+68  }$ & 5.726$_{-0.010}^{+0.011}$ & -2.214$_{-0.016}^{+0.016}$ &  47   $\pm$3         & -5.87   &  9           & 1 \\
J103810.94+253204.8 & sdB        & 22883$_{-56  }^{+58  }$ & 4.905$_{-0.003}^{+0.006}$ & -1.972$_{-0.012}^{+0.012}$ &  208  $\pm$2         & -0.08   &  7           & 1 \\
J120352.24+235343.3 & iHe-sdOB   & 40736$_{-34  }^{+35  }$ & 5.870$_{-0.009}^{+0.009}$ & -0.349$_{-0.006}^{+0.006}$ &  177  $\pm$2         & -0.72   &  11          & 1 \\
J120521.48+224702.2 & iHe-sdO    & 50349$_{-339 }^{+346 }$ & 5.767$_{-0.034}^{+0.033}$ & -0.470$_{-0.029}^{+0.029}$ &  146  $\pm$4         & -1.56   &  12          & 2 \\
J121703.12+454539.3 & iHe-sdOB   & 37019$_{-248 }^{+278 }$ & 6.137$_{-0.053}^{+0.053}$ & 0.9197$_{-0.038}^{+0.041}$ &  -377 $\pm$5         & -0.55   &  3           & 4 \\
J123137.56+074621.7 & sdB        & 25035$_{-31  }^{+31  }$ & 5.222$_{-0.005}^{+0.005}$ & -2.332$_{-0.012}^{+0.012}$ &  464  $\pm$2         & -1.14   &  8           & 3 \\
J123428.30+262757.9 & sdB        & 30836$_{-296 }^{+266 }$ & 5.731$_{-0.043}^{+0.043}$ & -2.953$_{-0.372}^{+0.176}$ &  64   $\pm$9         & -0.17   &  7           & 2 \\
J123953.52+062853.0 & sdB        & 20125$_{-186 }^{+115 }$ & 4.872$_{-0.022}^{+0.018}$ & -1.974$_{-0.025}^{+0.025}$ &  92   $\pm$3         & -1.26   &  7           & 2 \\
J124248.89+133632.6 & sdOB       & 36847$_{-430 }^{+383 }$ & 5.329$_{-0.045}^{+0.039}$ & -2.812$_{-0.091}^{+0.091}$ &  -170 $\pm$4         & -0.40   &  7           & 1 \\
J124310.58+343358.4 & sdOB   & 35643$_{-133 }^{+131 }$ & 5.762$_{-0.005}^{+0.005}$ & -0.928$_{-0.014}^{+0.014}$ &  166  $\pm$3         & -0.79   &  8           & 2 \\
J124819.08+035003.2 & sdO        & 59518$_{-409 }^{+418 }$ & 5.974$_{-0.016}^{+0.015}$ & -1.009$_{-0.004}^{+0.007}$ &  9    $\pm$3         & -0.23   &  9           & 1 \\
J130543.97+115840.8 & sdB        & 30119$_{-55  }^{+34  }$ & 5.525$_{-0.005}^{+0.007}$ & -2.458$_{-0.012}^{+0.013}$ &  -76  $\pm$2         & -0.15   &  10          & 1 \\
J133135.41+020919.8 & sdB %sdOB
& 33390$_{-212 }^{+178 }$ & 5.775$_{-0.028}^{+0.026}$ & -2.908$_{-0.078}^{+0.081}$ &  -24  $\pm$4         & -0.30   &  5           & 4 \\
J133417.10+173850.7 & sdB %sdOB
& 28961$_{-245 }^{+353 }$ & 5.565$_{-0.045}^{+0.046}$ & $<$-3.36                   &  -89  $\pm$13        & -1.48   &  3           & 4 \\
J135651.26+155810.4 & sdB        & 22667$_{-1407}^{+1519}$ & 4.980$_{-0.185}^{+0.189}$ & -2.078$_{-0.183}^{+0.123}$ &  206  $\pm$16        & -0.06   &  3           & 4 \\
J140532.34+410626.1 & iHe-sdOB   & 36823$_{-524 }^{+520 }$ & 5.693$_{-0.085}^{+0.092}$ & -0.437$_{-0.069}^{+0.041}$ &  -168 $\pm$9         & -0.82   &  4           & 2 \\
J143127.88+014416.2 & sdB        & 29057$_{-266 }^{+301 }$ & 5.502$_{-0.042}^{+0.041}$ & -2.703$_{-0.171}^{+0.165}$ &  -47  $\pm$8         & -0.44   &  7           & 2 \\
J143258.05+011857.9 & sdB        & 27734$_{-136 }^{+135 }$ & 5.574$_{-0.022}^{+0.022}$ & -2.808$_{-0.103}^{+0.099}$ &  29   $\pm$5         & -0.26   &  7           & 2 \\
J144209.90+105733.9 & sdB        & 28891$_{-181 }^{+152 }$ & 5.576$_{-0.029}^{+0.028}$ & -2.801$_{-0.071}^{+0.071}$ &  24   $\pm$4         & -0.10   &  7           & 2 \\
J145141.40+090645.1 & sdB        & 18659$_{-205 }^{+130 }$ & 4.783$_{-0.028}^{+0.031}$ & -1.740$_{-0.024}^{+0.023}$ &  -36  $\pm$3         & -1.65   &  5           & 3 \\
J145930.70+175846.1 & sdB        & 25906$_{-478 }^{+398 }$ & 5.948$_{-0.050}^{+0.065}$ & $<$-3.72                   &  -89  $\pm$10        & -0.13   &  4           & 2 \\
J150222.35+320220.9 & sdB        & 28998$_{-118 }^{+136 }$ & 5.661$_{-0.027}^{+0.019}$ & -3.328$_{-0.171}^{+0.164}$ &  -95  $\pm$4         & -0.18   &  7           & 2 \\
J151248.61+042205.5 & sdB        & 31235$_{-258 }^{+213 }$ & 5.787$_{-0.043}^{+0.038}$ & $<$-4.25                   &  -165 $\pm$7         & -0.22   &  7           & 3 \\
J153419.42+372557.2 & iHe-sdOB   & 37899$_{-241 }^{+350 }$ & 5.973$_{-0.052}^{+0.055}$ & 0.3123$_{-0.036}^{+0.038}$ &  90   $\pm$7         & -0.81   &  6           & 2 \\
J154958.29+043820.1 & sdOB
%iHe-sdOB  
& 36968$_{-233 }^{+252 }$ & 5.876$_{-0.047}^{+0.043}$ & -1.010$_{-0.026}^{+0.017}$ &  -143 $\pm$6         & -0.28   &  7           & 2 \\
J161143.29+554044.3 & sdO 
%iHe-sdO    
& 58973$_{-1746}^{+1672}$ & 5.243$_{-0.043}^{+0.054}$ & -1.009$_{-0.051}^{+0.037}$ &  -442 $\pm$6         & -0.39   &  4           & 4 \\
J163213.05+205124.0 & sdB        & 28420$_{-27  }^{+27  }$ & 5.654$_{-0.003}^{+0.004}$ & -1.842$_{-0.006}^{+0.006}$ &  -243 $\pm$1         & -0.12   &  20          & 1 \\
J164419.44+452326.7 & sdB        & 31991$_{-100 }^{+100 }$ & 5.833$_{-0.003}^{+0.039}$ & -2.881$_{-0.081}^{+0.079}$ &  -310 $\pm$4         & -0.91   &  7           & 2 \\
J164853.26+121703.0 & sdB        & 29870$_{-555 }^{+539 }$ & 5.408$_{-0.079}^{+0.074}$ & -2.522$_{-0.142}^{+0.090}$ &  -74  $\pm$7         & -1.24   &  5           & 4 \\
J165924.75+273244.4 & iHe-sdO    & 49493$_{-1269}^{+1118}$ & 5.750$_{-0.150}^{+0.142}$ & -0.775$_{-0.098}^{+0.122}$ &  -396 $\pm$15        & -0.61   &  6           & 2 \\
J170256.38+241757.9 & sdB        & 29887$_{-553 }^{+534 }$ & 5.408$_{-0.078}^{+0.074}$ & -2.522$_{-0.142}^{+0.089}$ &  -32  $\pm$6         & -0.42   &  9           & 2 \\
J171533.84+365214.8 & sdB        & 21769$_{-201 }^{+215 }$ & 4.966$_{-0.027}^{+0.026}$ & -1.972$_{-0.028}^{+0.026}$ &  160  $\pm$2         & -0.86   &  12          & 2 \\
J172736.02+361706.3 & sdB        & 28973$_{-238 }^{+220 }$ & 5.742$_{-0.039}^{+0.039}$ & $<$-3.95                   &  -95  $\pm$9         & -0.36   &  4           & 4 \\
J174211.75+643009.9$^a$ & BHB        & 13716$_{-100 }^{+79  }$ & 4.068$_{-0.025}^{+0.023}$ & -1.966$_{-0.057}^{+0.060}$ &  -501 $\pm$3         & -0.35   &  4           & 3 \\
J180313.45+234000.1 & sdOB       & 33945$_{-359 }^{+218 }$ & 5.205$_{-0.048}^{+0.034}$ & -2.221$_{-0.060}^{+0.041}$ &  -56  $\pm$3         & -4.43   &  5           & 1 \\
J184832.52+181540.0 & sdB        & 29549$_{-205 }^{+353 }$ & 5.639$_{-0.053}^{+0.034}$ & -3.008$_{-0.127}^{+0.099}$ &  -168 $\pm$6         & -0.16   &  10          & 1 \\
J204358.55-065025.8 & iHe-sdOB   & 37092$_{-178 }^{+132 }$ & 6.084$_{-0.021}^{+0.026}$ & -0.106$_{-0.018}^{+0.019}$ &  -316 $\pm$4         & -2.64   &  10          & 1 \\
J205030.39-061957.8 & eHe-sdO    & 48981$_{-254 }^{+294 }$ & 5.886$_{-0.031}^{+0.047}$ & 0.8096$_{-0.083}^{+0.094}$ &  -497 $\pm$6         & -0.19   &  8           & 1 \\
J210907.28+103640.6 & sdB        & 30846$_{-1321}^{+1198}$ & 5.902$_{-0.192}^{+0.162}$ & -2.000$_{-0.192}^{+0.162}$ &  -60  $\pm$27        & -1.09   &  2           & 4 \\
J212300.31+043453.0 & sdOB       & 35355$_{-970 }^{+603 }$ & 5.807$_{-0.104}^{+0.132}$ & -1.352$_{-0.095}^{+0.071}$ &  -389 $\pm$12        & -0.11   &  7           & 2 \\
J212449.22+061956.4 & sdO        & 66602$_{-1555}^{+1683}$ & 6.058$_{-0.039}^{+0.037}$ & -1.186$_{-0.042}^{+0.044}$ &  -276 $\pm$5         & -1.42   &  9           & 1 \\
J215648.71+003620.7 & sdB        & 28397$_{-171 }^{+150 }$ & 5.636$_{-0.028}^{+0.029}$ & -3.158$_{-0.143}^{+0.129}$ &  -170 $\pm$4         & -0.46   &  10          & 2 \\
J220759.08+204505.9 & sdB        & 22019$_{-81  }^{+61  }$ & 4.920$_{-0.015}^{+0.003}$ & -1.875$_{-0.015}^{+0.015}$ &  -339 $\pm$2         & -0.30   &  14          & 2 \\
J221728.38+121643.8 &  sdO
%sdB        
& 41843$_{-156 }^{+139 }$ & 5.685$_{-0.019}^{+0.016}$ & -2.482$_{-0.033}^{+0.033}$ &  -226 $\pm$3         & -0.18   &  6           & 3 \\
J222515.34-011156.8 & eHe-sdOB   & 38721$_{-255 }^{+261 }$ & 5.959$_{-0.041}^{+0.058}$ & 0.8159$_{-0.033}^{+0.035}$ &  -447 $\pm$4         & -1.02   &  13          & 2 \\
\noalign{\smallskip}                                         
\hline\hline                                       
\end{tabular} 
\tablefoot{The uncertainties in the table are statistical uncertainties only. Systematic uncertainties of $\pm3\%$ in $T_{\rm eff}$ and $\pm0.1$ in $\log{g}$ %and $\log{ \frac{n(\mathrm{He})}{n(\mathrm{H})}}$ 
should be added in quadrature for a more realistic estimate of the uncertainties.\\
$^a$Atmospheric parameters derived with a tailored grid for BHB stars.}  
\end{table*}

\subsection{Multi-epoch radial velocities}\label{sec:radial_velocity}

Radial velocities of all the stars have been measured along with the atmospheric parameters from the available medium-resolution spectra. The low-resolution spectra obtained with GTC/OSIRIS are not suited to measure sufficiently accurate RVs. 
We also encountered problems with spectra taken during our observing run with WHT/ISIS in July 2017 (see Table~\ref{tab:runs}), where the seeing reached down to the astonishingly small value of $0.3\,{\rm arcsec}$. Since the smallest slit provided by ISIS has $0.6\,{\rm arcsec}$ and the telescope was not defocused to match the slit width, significant movement of the stars within the slit lead to erroneous RVs, which had to be discarded. 

Because a large fraction of the sdB stars is known to be radial velocity variable, multi-epoch measurements are needed to identify the variables in the sample. Therefore, we made use of the individual observations as well as the combined SDSS spectra from the final data release of the fourth phase of the Sloan Digital Sky Survey  (SDSS-IV, DR17, Abdurro'uf et al. \cite{sdssiv})  as well as from our follow-up spectroscopy. 

To estimate the probability that a star is variable in radial velocity we calculate the statistical $p$-value as described by Maxted et al. (\cite{maxted01}). Assuming $\chi^2$ statistics we determine the goodness of fit of a constant radial velocity to the observed radial velocities. Then the $p$-value is the probability to obtain the observed $\chi^2$ value or higher from random fluctuations of a constant value. To allow for the systematic errors, we added an external error in quadrature to all radial velocity uncertainties prior to calculating statistics. For spectra taken with the SDSS and BOSS spectrographs Yanny et al. (\cite{yanny09}) and Bolton et al. (\cite{bolton12}) estimate systematic radial velocity uncertainties of $4\,{\rm km\,s^{-1}}$ and $4.5\,{\rm km\,s^{-1}}$, respectively. Here we adopted $5\,{\rm km\,s^{-1}}$ for all spectra, irrespective of their spectral resolution. If $\log{p}<-4$, we consider the variations not to be caused by random fluctuation, but the star to be a radial-velocity variable close binary (Maxted et al. \cite{maxted01}).

For all stars multi-epoch observations are available and the number of individual measurements as well as the resulting p-values are given in Table~\ref{tab:spec}. The time spans between the individual RVs is quite different and we, therefore, divided the sample in four categories. Objects in the best category number 1 have several spectra from SDSS, which have been taken at distinct epochs separated by one day or more. In addition, category 1 stars have follow-up observations taken with different instruments usually several years after the SDSS epochs. Stars of category number 2 have SDSS spectra, only, but again taken at several distinct epochs separated by at least one day. Stars of category number 3 only have individual SDSS spectra, which have been taken back-to-back usually within less than one hour, but their longer term RV variability can be checked with follow-up spectra taken with different instruments. Finally. the least constraining category number 4 has only individual SDSS spectra taken back-to-back within less than one hour. Since the typical orbital periods of sdBs are ranging from hours to days, category 4 stars can only probe the closest binary systems with the shortest periods. Non-detections of RV variations in category 4 stars are therefore not suited to exclude close binarity in general.

\begin{figure}[t!]
\begin{center}
\includegraphics[width=\columnwidth]{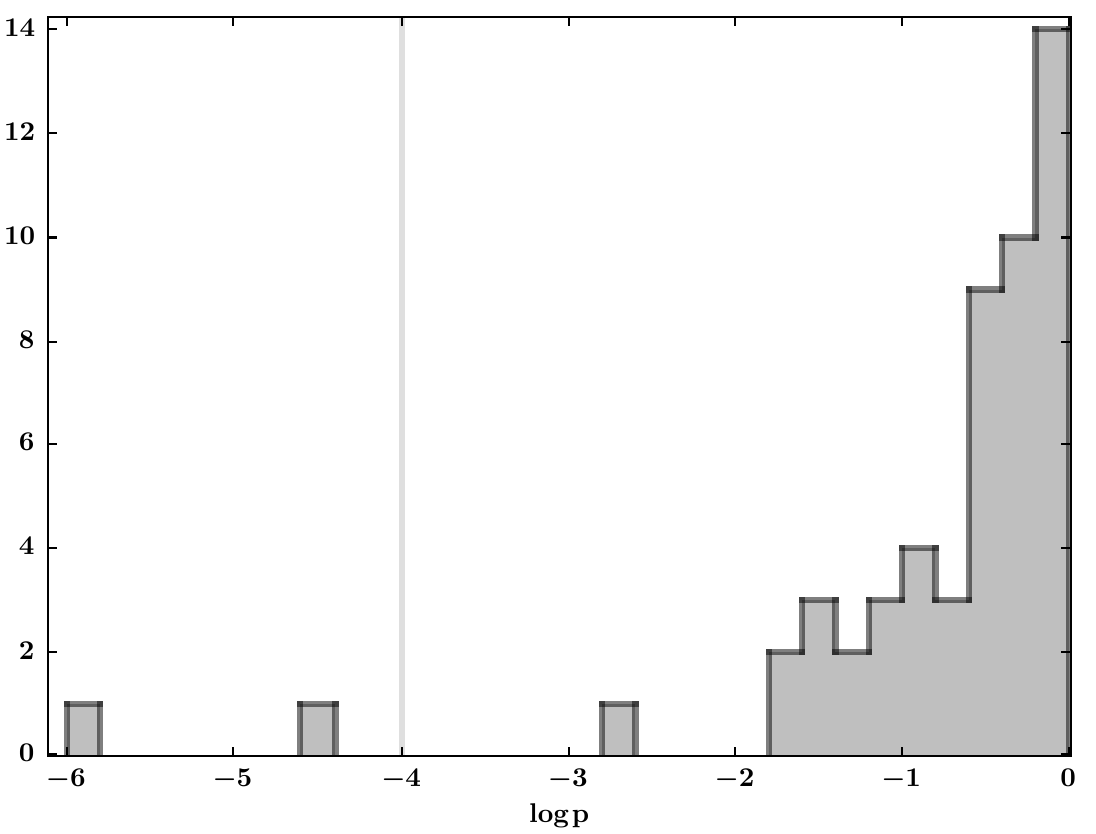}
\end{center}
\caption{$\log{p}$ distribution of the sample. The significance level $\log{p}< -4.0$ is marked by the vertical line.}
\label{fig:logp}
\end{figure}

The p-values listed in Table~\ref{tab:spec} are high for most stars with no indication of RV variability (see also Fig~\ref{fig:logp}). Only two stars in the sample show evidence for significant RV variations. These are the sdOB J180313.45+234000.1 and the sdB J102439.43+383917.9, both considered members of category 1. The former has $\log{p}=-4.43$, which is marginally above our detection threshold. The four SDSS spectra, however, show a higher $\log{p}$ indicative of constant RV and the deviation stems only from one follow-up spectrum taken with FORS1. We, therefore, treat this star as a candidate for RV variability only. The latter on the other hand shows both a lower $\log{p}=-5.87$ and a more consistent variation between the 9 individual spectra taken with SDSS and ISIS. We therefore classify J102439.43+383917.9 as a close binary.

\begin{figure*}
     \begin{subfigure}[b]{0.32\textwidth}
         \centering
          \includegraphics[width=\textwidth]{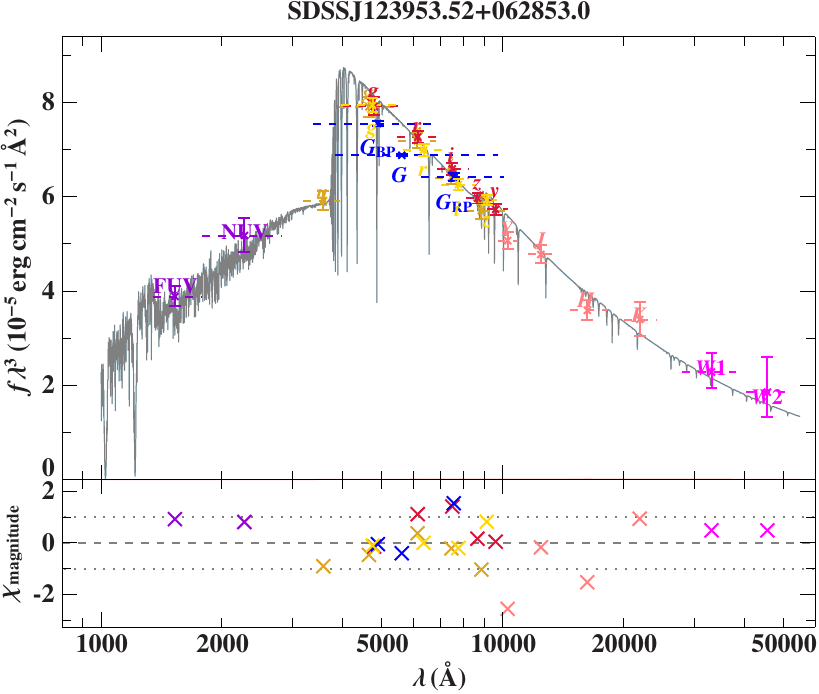} 
     \end{subfigure}
     \hfill
     \begin{subfigure}[b]{0.32\textwidth}
         \centering
         \includegraphics[width=\textwidth]{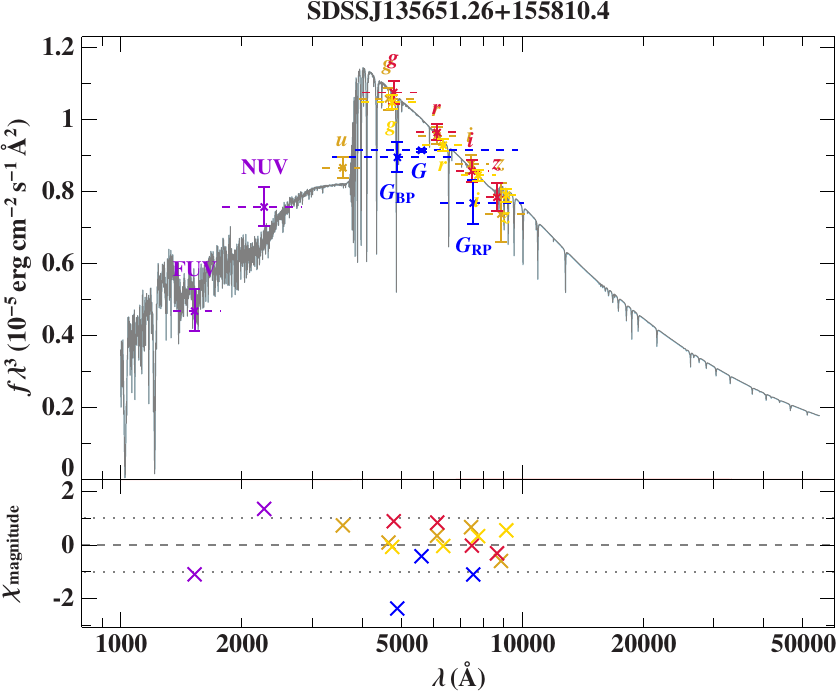} 
     \end{subfigure}
     \hfill
     \begin{subfigure}[b]{0.32\textwidth}
         \centering
         \includegraphics[width=\textwidth]{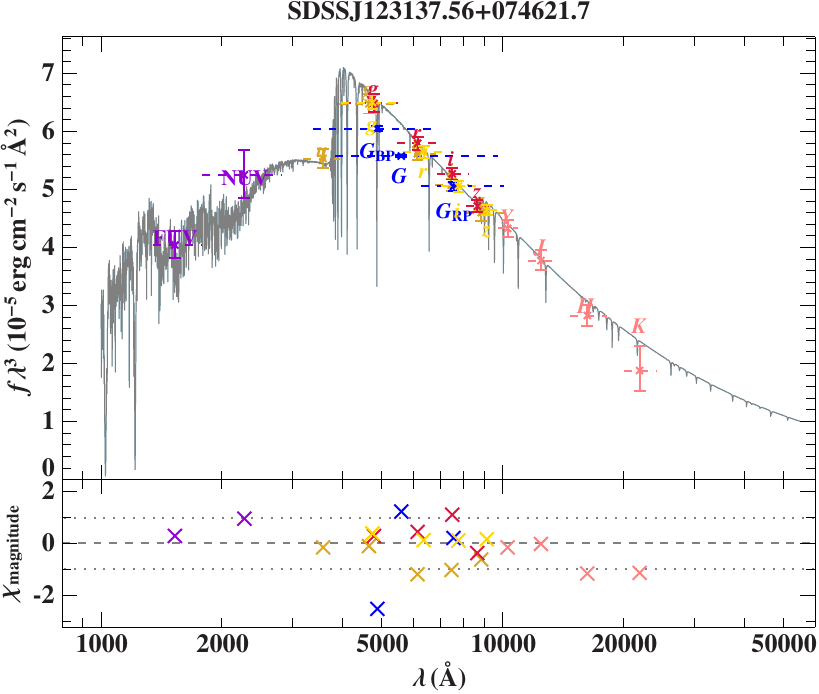}
         \end{subfigure}
        \begin{subfigure}[b]{0.32\textwidth}
         \centering
          \includegraphics[width=\textwidth]{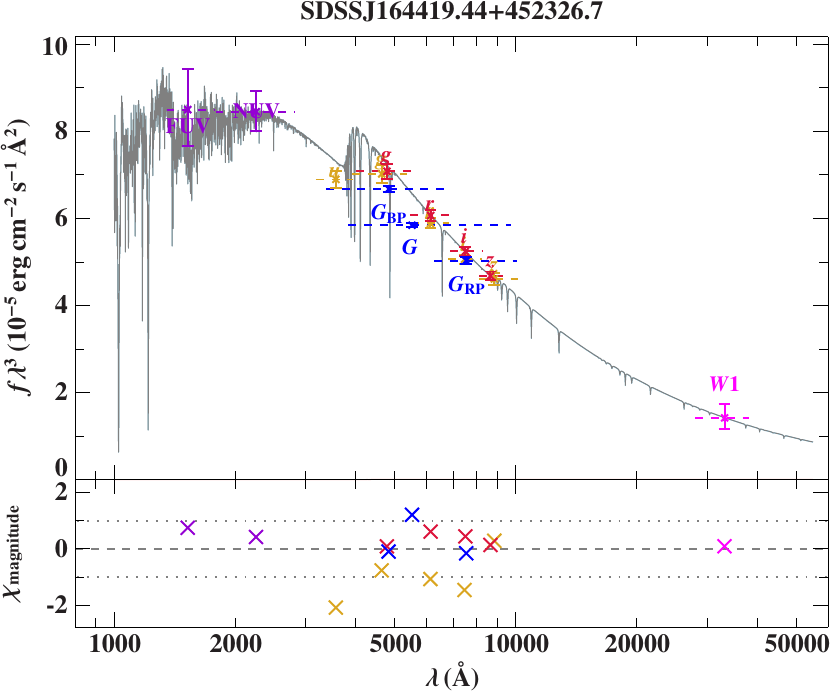} 
     \end{subfigure}
     \hfill
     \begin{subfigure}[b]{0.32\textwidth}
         \centering
         \includegraphics[width=\textwidth]{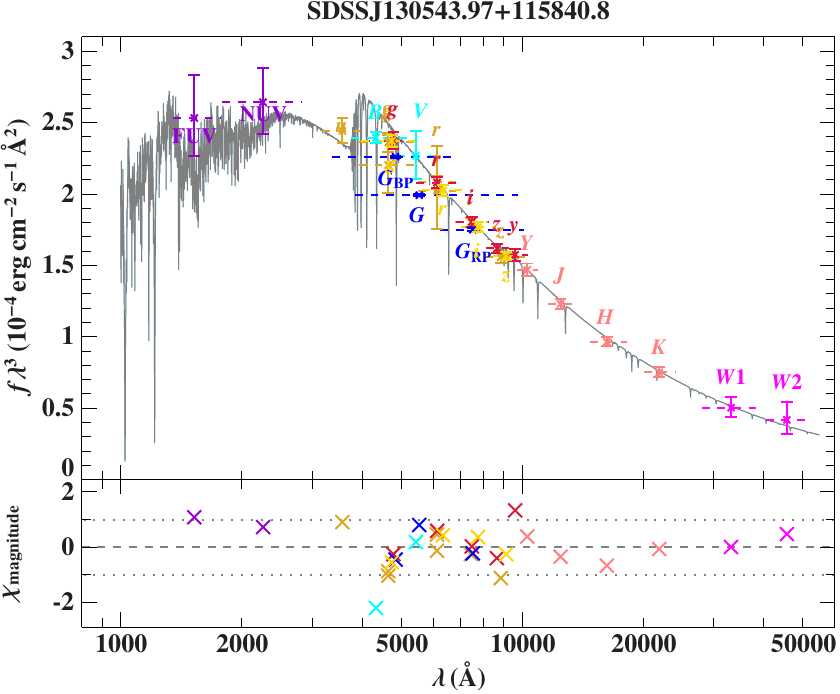}
     \end{subfigure}
     \hfill
     \begin{subfigure}[b]{0.32\textwidth}
         \centering
         \includegraphics[width=\textwidth]{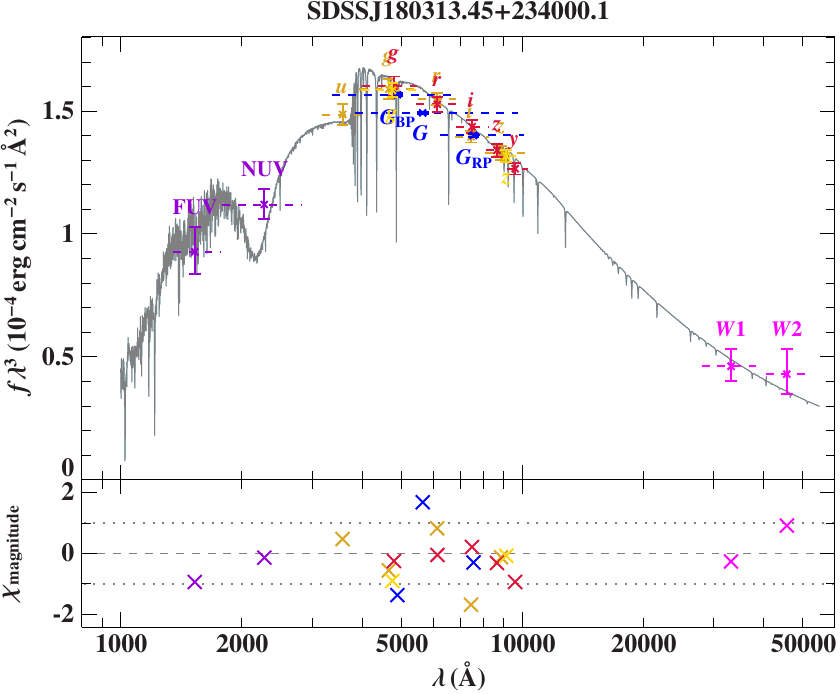}
         \end{subfigure}
     \begin{subfigure}[b]{0.32\textwidth}
         \centering
          \includegraphics[width=\textwidth]{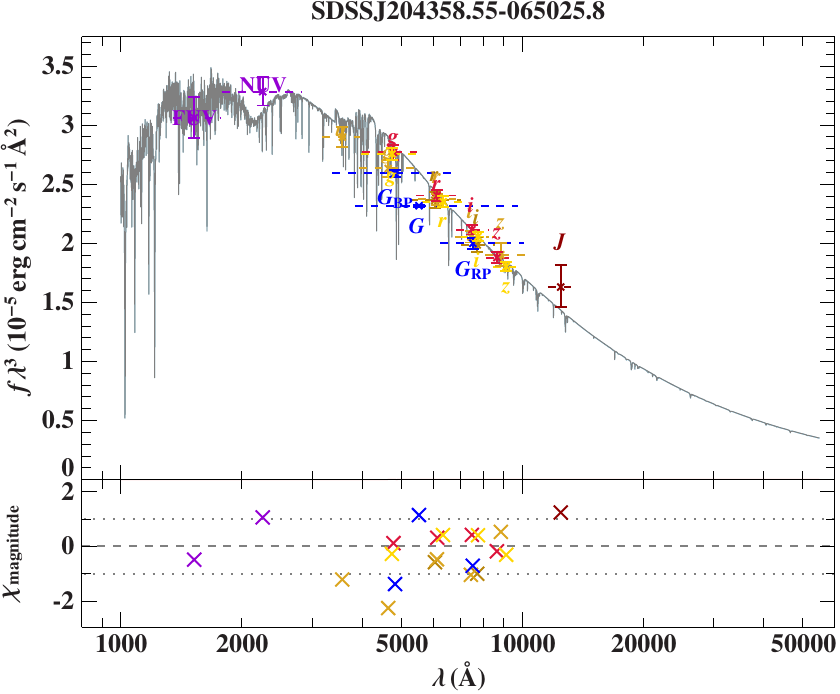} 
     \end{subfigure}
     \hfill
     \begin{subfigure}[b]{0.32\textwidth}
         \centering
         \includegraphics[width=\textwidth]{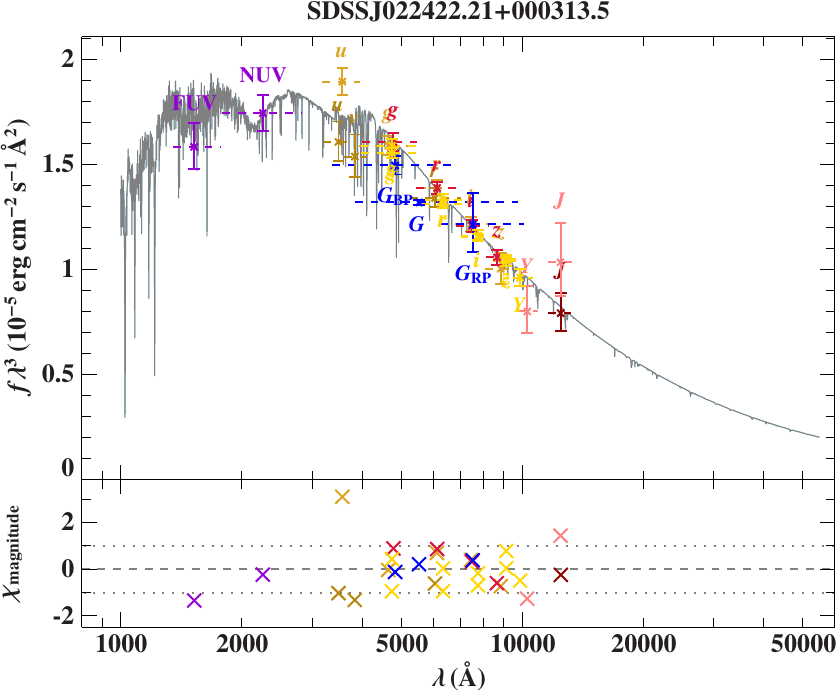} 
     \end{subfigure}
     \hfill
     \begin{subfigure}[b]{0.32\textwidth}
         \centering
         \includegraphics[width=\textwidth]{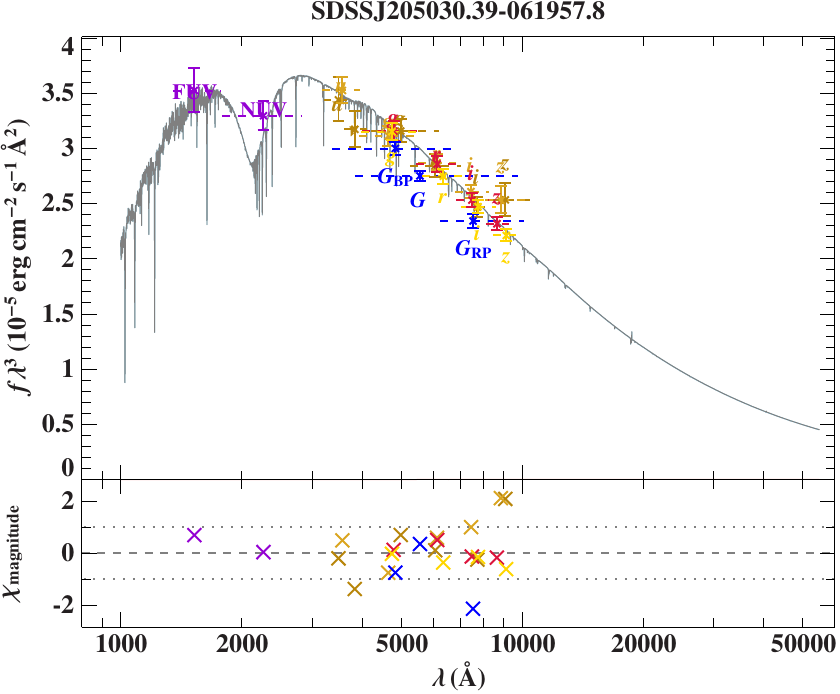}
         \end{subfigure}
     \caption{Fits of the SEDs of selected programme stars. Each plot consists of two panels; the upper one compares the observed fluxes to the synthetic SED. To ease the slope of the distribution the flux is multiplied by the wavelength to the power of three. 
     Photometric fluxes are displayed as coloured data points (for details see Table \ref{tab:photo_surveys}) with their respective uncertainties and filter widths (dashed lines). The best-fit models are drawn as grey full drawn lines. The lower panels give reduce $\chi$ to demonstrate the quality of the fit.
     Example fits are selected to demonstrate relevant issues. In each row the stars are arranged with increasing temperature.
    \textit{Top row}: He weak sdB stars: The cool sdB J123953.52+062853.0, which is the most distant programme star, J135651.26+155810.4 the sdB with the largest (positive) heliocentric RV, and J123137.56+074621.7. \textit{Middle row}: sdOB stars: J164419.44+452326.7, the low gravity J130543.97+115840.8, and the highly reddened J180313.45+234000.1. \textit{Bottom row}: He-sdO stars: The iHe-sdOB J204358.55-065025.8, the best observed iHe-sdOB star J022422.21+000313.5, and the eHe-sdO J205030.39-061957.8, which was considered a spectroscopic twin to US\,708 by Ziegerer et al. (\cite{ziegerer17}). 
     The strong colour excess of J180313.45+234000.1 and J205030.39-061957.8 is most obvious from the strength of the $2200\,{\rm \AA}$ UV feature effecting the {\em GALEX} NUV band.}
        \label{fig:sed}
\end{figure*}

\section{Spectral energy distribution and spectroscopic distances}\label{sec:sed}

Photometric measurements are obtained by querying several public data bases. The optical spectral range is well covered by observations in the Sloan filters and broad band \textit{Gaia} measurements. 
Most stars have UV observations from the {\em GALEX} mission, while infrared magnitudes are available for a few stars only. See Table \ref{tab:photo_surveys} for a list of the photometric surveys used.

We fit the observed magnitudes with synthetic SEDs computed with ATLAS12. The stellar parameters were set to the values determined from spectroscopy. The angular diameter $\Theta$, and the interstellar colour excess $E(44-55)$ remained as free parameters to match the observed SED (see Heber et al.  (\cite{heber18}) for more details). To model interstellar extinction, the function given by Fitzpatrick et al. (\cite{fitzpatrick19}) was used, assuming a standard extinction coefficient of $R(55)=3.02$. For illustration we show the fits for nine stars of different spectral types in Fig. \ref{fig:sed}.

Adopting the theoretically predicted degenerate core mass ($0.47\,M_{\rm \odot}$) for the helium flash in low-mass stars, with an uncertainty of $\pm0.05\,M_{\rm \odot}$ to allow for different metallicities, the radii $R$ of our sample stars are calculated using the surface gravity and $R=\sqrt{GM/g}$. The spectroscopic distances are determined using the angular diameter $d=2R/\Theta$ (see Table \ref{tab:sdistance}).

\begin{table}
\caption{Results of the SED analysis: Angular diameter $\Theta$, interstellar reddening parameter $E(44-55)$, and spectroscopic distance $d_{\rm spec}$.}\label{tab:sdistance}
\begin{center}
\renewcommand{\arraystretch}{1.2}
\setlength{\tabcolsep}{0.25em}
\begin{tabular}{llll}
\hline\hline

SDSS star & $\log(\Theta)$& $E(44-55)$& $d_{\rm spec}$\\
 & [rad] & [mag] & [kpc]\\
\hline
J022422.21+000313.5 & $-12.087 \pm 0.009$ & $0.054 \pm 0.005$ & $6.7^{+ 1.0}_{- 0.9}$ \\
J080833.76+180221.8 & $-11.687^{+ 0.008}_{- 0.007}$ & $0.046 \pm 0.008$ & $2.8 \pm 0.4$ \\
J082802.03+404008.9 & $-11.993^{+ 0.011}_{- 0.010}$ & $0.007^{+ 0.010}_{- 0.008}$ & $7.9^{+ 1.1}_{- 1}$ \\
J084556.85+135211.3 & $-11.637^{+ 0.010}_{- 0.009}$ & $0.017 \pm 0.006$ & $3.5^{+ 0.6}_{- 0.5}$ \\
J090252.99+073533.9 & $-11.744^{+ 0.010}_{- 0.012}$ & $0.074 \pm 0.007$ & $3.2^{+ 0.6}_{- 0.5}$ \\
J091512.06+191114.6 & $-12.116^{+ 0.010}_{- 0.011}$ & $0.044 \pm 0.006$ & $8.3^{+ 1.2}_{- 1.1}$ \\
J094850.47+551631.6 & $-12.015 \pm 0.010$ & $0.035 \pm 0.006$ & $5.3^{+ 0.8}_{- 0.7}$ \\
J102057.16+013751.2 & $-11.564^{+ 0.011}_{- 0.010}$ & $0.072 \pm 0.009$ & $3.4 \pm 0.5$ \\
J102439.43+383917.9 & $-11.581 \pm 0.012$ & $0.042 \pm 0.012$ & $2.6 \pm 0.4$ \\
J103810.94+253204.8 & $-11.500 \pm 0.008$ & $0.060^{+ 0.004}_{- 0.005}$ & $5.6^{+ 0.8}_{- 0.7}$ \\
J120352.24+235343.3 & $-11.744^{+ 0.012}_{- 0.010}$ & $0.028 \pm 0.004$ & $3.2^{+ 0.5}_{- 0.4}$ \\
J120521.48+224702.2 & $-12.009^{+ 0.018}_{- 0.015}$ & $0.011^{+ 0.023}_{- 0.012}$ & $9.6^{+ 1.5}_{- 1.3}$ \\
J121703.12+454539.3 & $-11.821^{+ 0.009}_{- 0.010}$ & $0.034 \pm 0.010$ & $2.8^{+ 0.5}_{- 0.4}$ \\
J123137.56+074621.7 & $-11.634 \pm 0.009$ & $0.043 \pm 0.005$ & $5.3^{+ 0.8}_{- 0.7}$ \\
J123428.30+262757.9 & $-12.016^{+ 0.011}_{- 0.010}$ & $0.038 \pm 0.008$ & $7.1^{+ 1.1}_{- 1}$ \\
J123953.52+062853.0 & $-11.528^{+ 0.008}_{- 0.007}$ & $0.008 \pm 0.005$ & $6.2^{+ 0.9}_{- 0.8}$ \\
J124248.89+133632.6 & $-11.819^{+ 0.010}_{- 0.009}$ & $0.038 \pm 0.005$ & $7.1^{+ 1.1}_{- 1}$ \\
J124310.58+343358.4 & $-11.873 \pm 0.010$ & $0.019 \pm 0.007$ & $4.9 \pm 0.7$ \\
J124819.08+035003.2 & $-11.805 \pm 0.007$ & $0.022 \pm 0.005$ & $3.3 \pm 0.5$ \\
J130543.97+115840.8 & $-11.439 \pm 0.010$ & $0.027^{+ 0.003}_{- 0.004}$ & $2.4^{+ 0.4}_{- 0.3}$ \\
J133135.41+020919.8 & $-11.670^{+ 0.010}_{- 0.009}$ & $0.003 \pm 0.004$ & $3.0^{+ 0.5}_{- 0.4}$ \\
J133417.10+173850.7 & $-12.000 \pm 0.011$ & $0.029 \pm 0.008$ & $8.3^{+ 1.3}_{- 1.1}$ \\
J135651.26+155810.4 & $-11.991^{+ 0.016}_{- 0.017}$ & $0.040 \pm 0.007$ & $15^{+ 5}_{- 4}$ \\
J140532.34+410626.1 & $-12.129 \pm 0.011$ & $0.054 \pm 0.008$ & $9.5^{+ 1.8}_{- 1.5}$ \\
J143127.88+014416.2 & $-12.001^{+ 0.011}_{- 0.010}$ & $0.056 \pm 0.008$ & $8.9^{+ 1.4}_{- 1.2}$ \\
J143257.99+011857.2 & $-11.783^{+ 0.010}_{- 0.009}$ & $0.037 \pm 0.006$ & $5.0 \pm 0.7$ \\
J144209.90+105733.9 & $-11.749 \pm 0.010$ & $0.058 \pm 0.007$ & $4.6^{+ 0.7}_{- 0.6}$ \\
J145141.40+090645.1 & $-11.556^{+ 0.007}_{- 0.006}$ & $0.034 \pm 0.005$ & $7.3^{+ 1.1}_{- 1}$ \\
J145930.70+175846.1 & $-11.955 \pm 0.010$ & $0.051 \pm 0.006$ & $4.8^{+ 0.8}_{- 0.7}$ \\
J150222.35+320220.9 & $-11.808^{+ 0.013}_{- 0.010}$ & $0.004^{+ 0.010}_{- 0.005}$ & $4.7^{+ 0.7}_{- 0.6}$ \\
J151248.61+042205.5 & $-11.993^{+ 0.010}_{- 0.009}$ & $0.057 \pm 0.007$ & $6.3^{+ 1}_{- 0.9}$ \\
J153419.42+372557.2 & $-12.119^{+ 0.010}_{- 0.009}$ & $0.046 \pm 0.009$ & $6.8^{+ 1.1}_{- 0.9}$ \\
J154958.29+043820.1 & $-11.961 \pm 0.011$ & $0.078 \pm 0.007$ & $5.5^{+ 0.8}_{- 0.9}$ \\
J161143.29+554044.3 & $-12.088 \pm 0.016$ & $0.054 \pm 0.021$ & $14.7^{+ 2.2}_{- 2.1}$ \\
J163213.05+205124.0 & $-11.703 \pm 0.010$ & $0.073 \pm 0.005$ & $3.8^{+ 0.6}_{- 0.5}$ \\
J164419.44+452326.7 & $-11.736 \pm 0.010$ & $0.016 \pm 0.006$ & $3.3 \pm 0.5$ \\
J164853.26+121703.0 & $-11.896^{+ 0.011}_{- 0.010}$ & $0.073^{+ 0.006}_{- 0.007}$ & $7.8^{+ 1.4}_{- 1.2}$ \\
J165924.75+273244.4 & $-12.294 \pm 0.016$ & $0.130 \pm 0.020$ & $13^{+ 4}_{- 2.5}$ \\
J170256.38+241757.9 & $-11.996 \pm 0.012$ & $0.093 \pm 0.010$ & $9.7^{+ 1.7}_{- 1.4}$ \\
J171533.84+365214.8 & $-11.569 \pm 0.007$ & $0.046 \pm 0.005$ & $6.2^{+ 0.9}_{- 0.8}$ \\
J172736.02+361706.3 & $-11.999^{+ 0.011}_{- 0.010}$ & $0.033 \pm 0.008$ & $6.7^{+ 1}_{- 0.9}$ \\
J174211.75+643009.9 & $-11.218^{+ 0.006}_{- 0.007}$ & $0.059 \pm 0.008$ & $7.7^{+ 1.1}_{- 1}$ \\
J180313.45+234000.1 & $-11.485 \pm 0.010$ & $0.155^{+ 0.003}_{- 0.004}$ & $3.8^{+ 0.6}_{- 0.5}$ \\
J184832.52+181540.0 & $-11.643 \pm 0.011$ & $0.354 \pm 0.008$ & $3.3^{+ 0.6}_{- 0.5}$ \\
J204358.55-065025.8 & $-11.967 \pm 0.009$ & $0.050 \pm 0.005$ & $4.2 \pm 0.6$ \\
J205030.39-061957.8 & $-11.986 \pm 0.006$ & $0.095 \pm 0.005$ & $5.6 \pm 0.8$ \\
J210907.28+103640.6 & $-11.978^{+ 0.016}_{- 0.014}$ & $0.182^{+ 0.007}_{- 0.006}$ & $5.0^{+ 1.6}_{- 1.1}$ \\
J212300.31+043453.0 & $-12.075^{+ 0.013}_{- 0.010}$ & $0.117 \pm 0.006$ & $7.5^{+ 1.4}_{- 1.5}$ \\
J212449.22+061956.4 & $-12.102 \pm 0.009$ & $0.087 \pm 0.006$ & $5.9^{+ 0.9}_{- 0.8}$ \\
J215648.71+003620.7 & $-11.774 \pm 0.010$ & $0.058 \pm 0.004$ & $4.5^{+ 0.7}_{- 0.6}$ \\
J220759.08+204505.9 & $-11.407 \pm 0.009$ & $0.132 \pm 0.009$ & $4.4^{+ 0.7}_{- 0.6}$ \\
J221728.35+121642.6 & $-11.535^{+ 0.009}_{- 0.008}$ & $0.082 \pm 0.006$ & $2.5^{+ 0.4}_{- 0.3}$ \\
J222515.34-011156.8 & $-11.995^{+ 0.009}_{- 0.010}$ & $0.086 \pm 0.005$ & $5.2 \pm 0.8$ \\
\hline
\end{tabular}
\end{center}
\end{table}

\section{Astrometry and kinematic analysis}
\label{sec:kinematic}

Due to the large distances of the stars in our sample, most \textit{Gaia} parallaxes are of insufficient accuracy to derive meaningful distances from astrometry only (see Table~\ref{tab:astro}). 
Instead we calculated the spectroscopic distances as described in Sect.~\ref{sec:sed} using the improved spectroscopic parameters derived from the follow-up analysis. Proper motions and astrometric correlation coefficients were taken from \textit{Gaia} Early Data Release 3 (EDR3; {\em Gaia} Collaboration et al. \cite{gaia21}, see Table~\ref{tab:astro}). 

The Galactic velocity components ($U$ points away from the Galactic centre, $V$ points in the direction of Galactic rotation, and $W$ points toward Galactic north) and the total Galactic restframe velocities have been updated accordingly.

Four stars of our sample have \textit{Gaia} parallaxes with uncertainties smaller than $20\%$, accurate enough to compare them directly to the spectroscopic distances (see Table~\ref{tab:spec_par}). The parallax distances are consistent with the spectroscopic distances within the uncertainties for all those stars confirming the validity of our approach (see Table \ref{tab:spec_par}).

\begin{table}
\centering
\caption{Comparison of spectroscopic vs. parallax $\varpi$ based distances. The inversion accounts for asymmetry and zero point offset (Lindegren et al. \cite{lindegren20}). }\label{tab:spec_par}
\renewcommand{\arraystretch}{1.25}
\begin{tabular}{lrr}
\hline\hline
\noalign{\smallskip}
Star & $d_\textrm{spec}$ (kpc) & $d_\varpi$ (kpc) \\
\hline
\noalign{\smallskip}     
J130543.96+115840.8     &  $2.4_{-0.3}^{+0.4}$ & $2.4_{-0.3}^{+0.5}$ \\
J133135.42+020919.8     &  $3.0_{-0.4}^{+0.5}$ & $2.2_{-0.4}^{+0.6}$ \\ 
J180313.45+234000.2     &  $3.8_{-0.5}^{+0.6}$ & $3.0_{-0.4}^{+0.5}$ \\
J221728.34+121642.6     &  $2.5_{-0.3}^{+0.4}$ & $2.6_{-0.4}^{+0.6}$ \\
\hline
\noalign{\smallskip}     
\end{tabular}
\end{table}

The Galactic orbits were calculated as described in Irrgang et al. (\cite{irrgang18a}) using Model I of Irrgang et al. (\cite{irrgang13}) for the Galactic potential. The orbits were traced back and the eccentricities of the Galactic orbits $e$ as well as the components of the angular momentum in $Z$-direction $L_{\rm Z}$ determined. The uncertainties in the input parameters were propagated with Monte Carlo simulations assuming Gaussian distributions while also accounting for asymmetric errors and the correlations between the proper motion components. 

The kinematic parameters of our sample are provided in Table~\ref{tab:kinematic}. Besides the Galactic velocity components $U$, $V$ and $W$ as well as the total Galactic rest frame velocity $v_{\rm grf}$ corrected for the motion of the local standard of rest, we provide $e$ and $L_{\rm Z}$ as well as the time of flight $t_{\rm d}$ since the last crossing of the Galactic disc within $100\,{\rm pc}$ of $Z=0$.

\begin{figure}[t!]
\begin{center}
\includegraphics[width=\columnwidth]{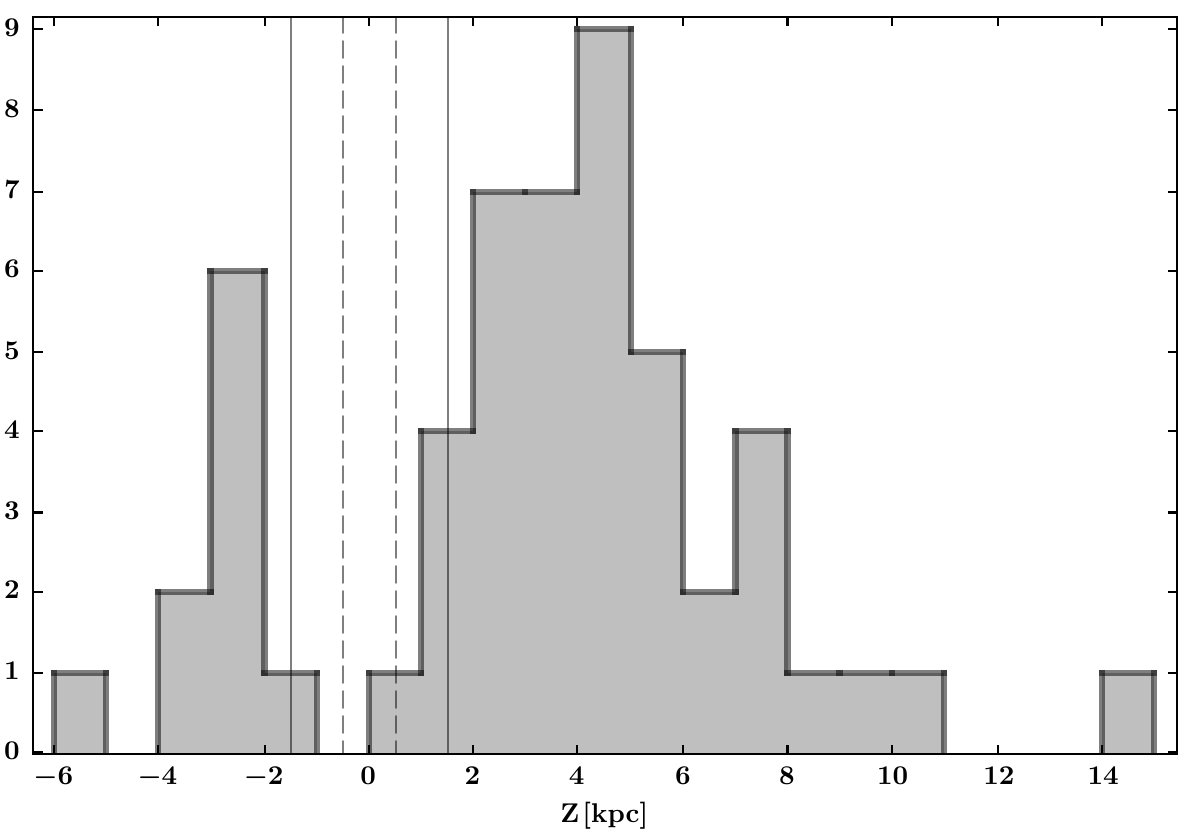}
\end{center} 
\caption{Distribution of the distances of our sample from the Galactic plane. The dashed vertical lines mark the scale height of the Galactic thin disc, the solid vertical lines mark the scale height of the thick disk.}
\label{fig:zdist}
\end{figure}

\begin{figure*}[t!]
\begin{center}
\includegraphics[width=0.49\textwidth]{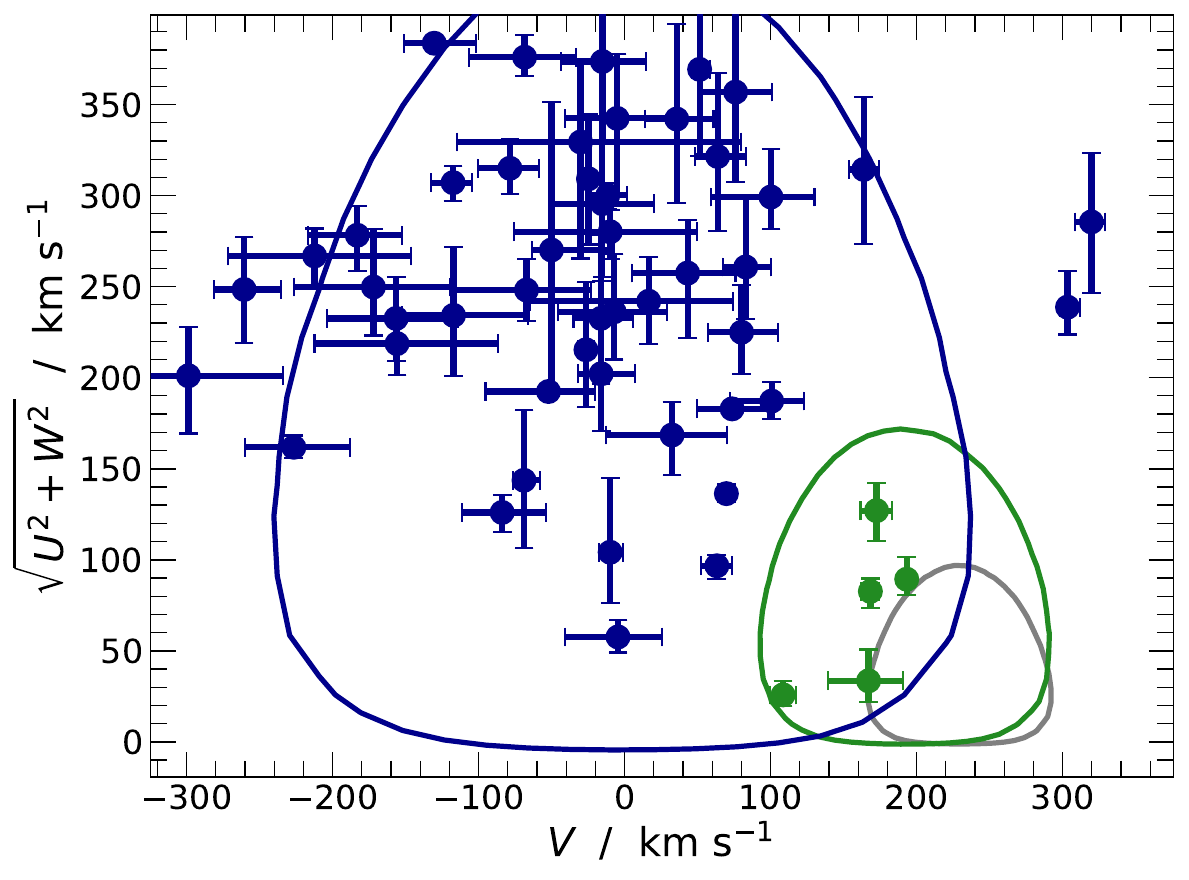}
\includegraphics[width=0.47\textwidth]{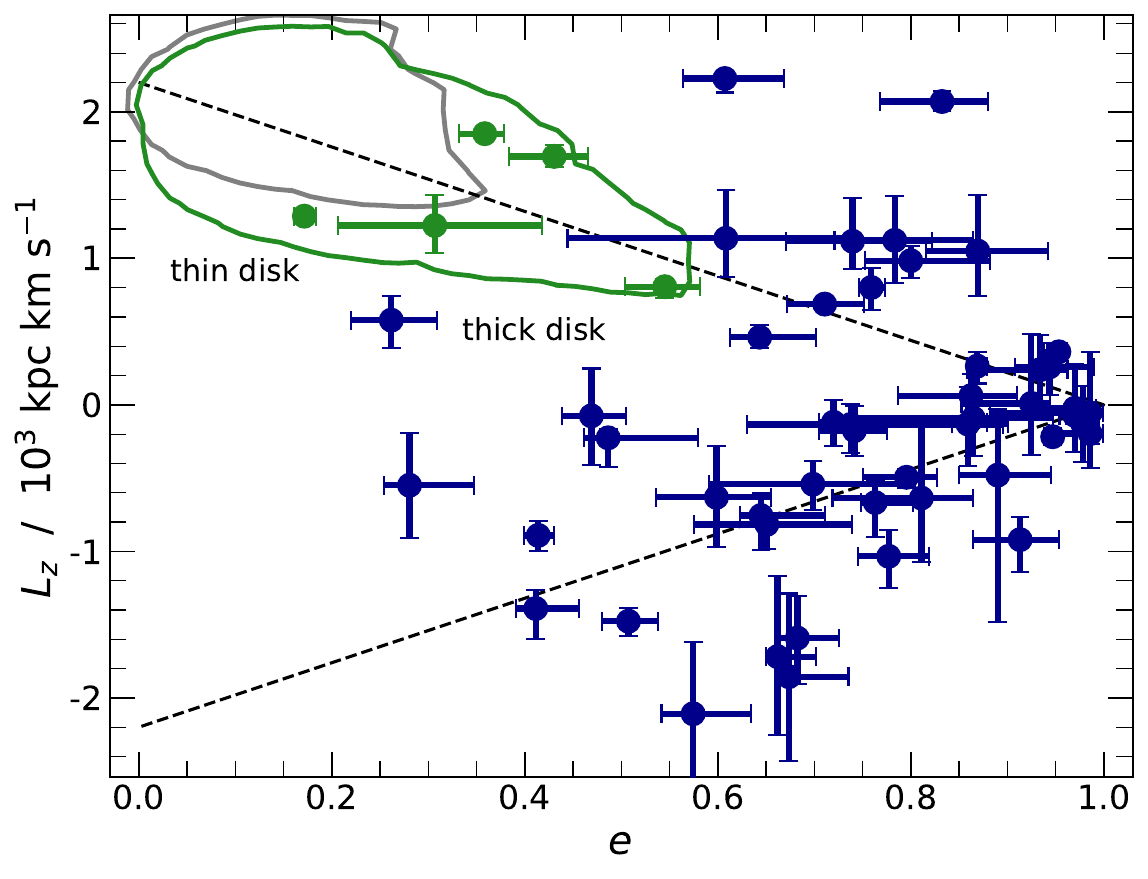}
\end{center} 
\caption{\textit{Left panel}: Toomre diagram of the sample. The $2\sigma$ contours of the thin disc (grey), the thick disc (green), and the halo (blue) are constructed from the $U$, $V$ and $W$ distributions of Anguiano et al.\ (\cite{Anguiano2020}). Objects with halo kinematics are marked in blue, the few objects also consistent with thick disc kinematics are marked in green. 
\textit{Right panel}: $e-L_{\rm Z}$ diagram of the sample using the same colour coding for the stars. The green and grey 2\,$\sigma$ contours in the upper left denote the locations of thin and thick disc stars, respectively, based on the GALAH + \textit{Gaia} EDR3 sample of Buder et al. (\cite{buder21}). The dashed lines mark prograde (positive $L_{\rm Z}$) and retrograde (negative $L_{\rm Z}$) directions of Galactic rotation, respectively.}
\label{fig:toomre}
\end{figure*}

Fig.~\ref{fig:zdist} shows the distribution of distances from the Galactic plane while the Toomre and the $L_{\rm Z}-e$ diagrams are shown in Fig.~\ref{fig:toomre}. Most stars are situated well above the Galactic disc in the halo, but all of them are bound to the Galaxy -- the prototype US\,708 remains the only unbound HVS sdO/B. The sample represents a kinematically extreme population in the Galactic halo and contains only very few stars, which could also belong to the disc population. 

\section{The sample of fast hot subdwarf stars - ejected or born in the halo?}

After a thorough analysis of our sample of fast hot subdwarfs, we can now evaluate whether the properties of the sample are consistent with a genuine halo population or whether there is evidence for a (sub-)population of stars ejected from the disc.

Although bound to the Galaxy, the extreme kinematics of our sample might still hint at the SN\,Ia scenario or other ejection scenarios from the disc. Since the core-collapse and the thermonuclear supernova scenarios require either the explosion of a massive star or the prior evolution of an intermediate mass close binary (e.g. Geier et al. \cite{geier13}), runaways from these scenarios must originate from young stellar populations, likely in the Galactic disc. Since the lifetime of sdO/B stars is of the order of $100\,{\rm Myr}$ at most, this defines a maximum possible time of flight since the ejection. We have calculated the flight times from the last disc crossing for our sample, the location of the crossings in Galactic coordinates, and the velocity at the moment of the crossing (or ``ejection velocity'', $v_{\rm ej}$). Out of 53 stars, 36 ($68\%$) could have been ejected from the disc within the uncertainties (see Table~\ref{tab:crossing}), assuming a disc radius of $15\,{\rm kpc}$. This means that the SN ejection scenario is likely excluded for the remaining 32\%. 

\begin{figure}[t!]
\begin{center}
\includegraphics[width=\columnwidth]{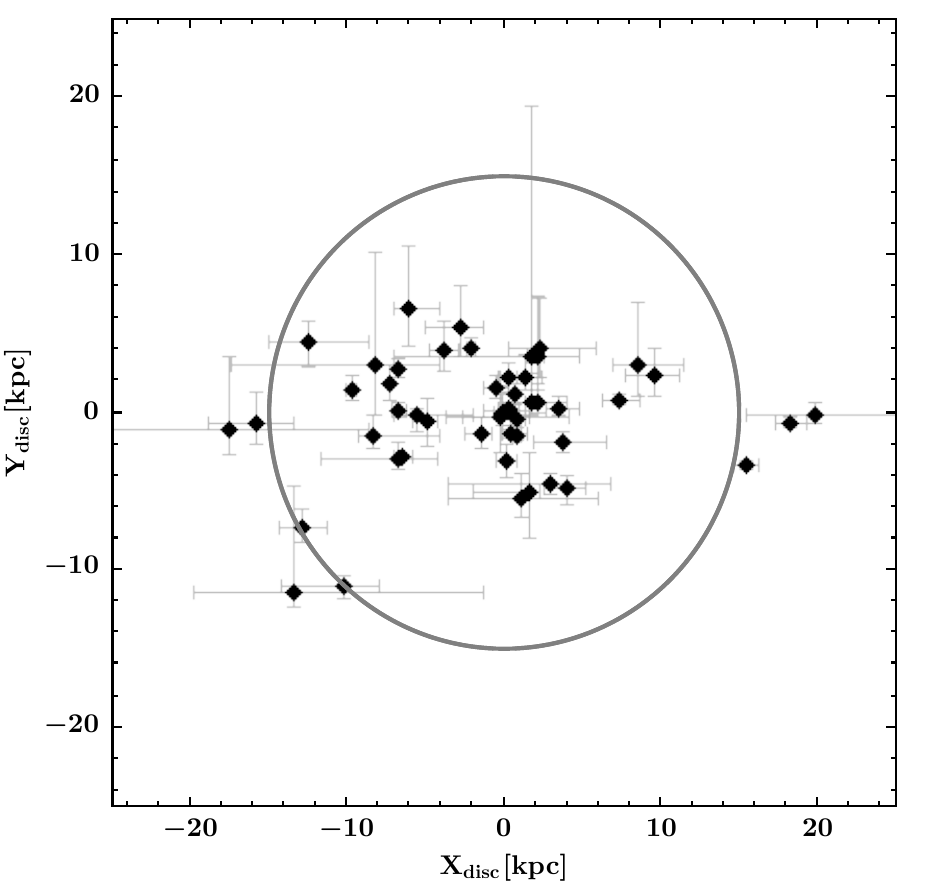}\\
\includegraphics[width=0.45\textwidth]{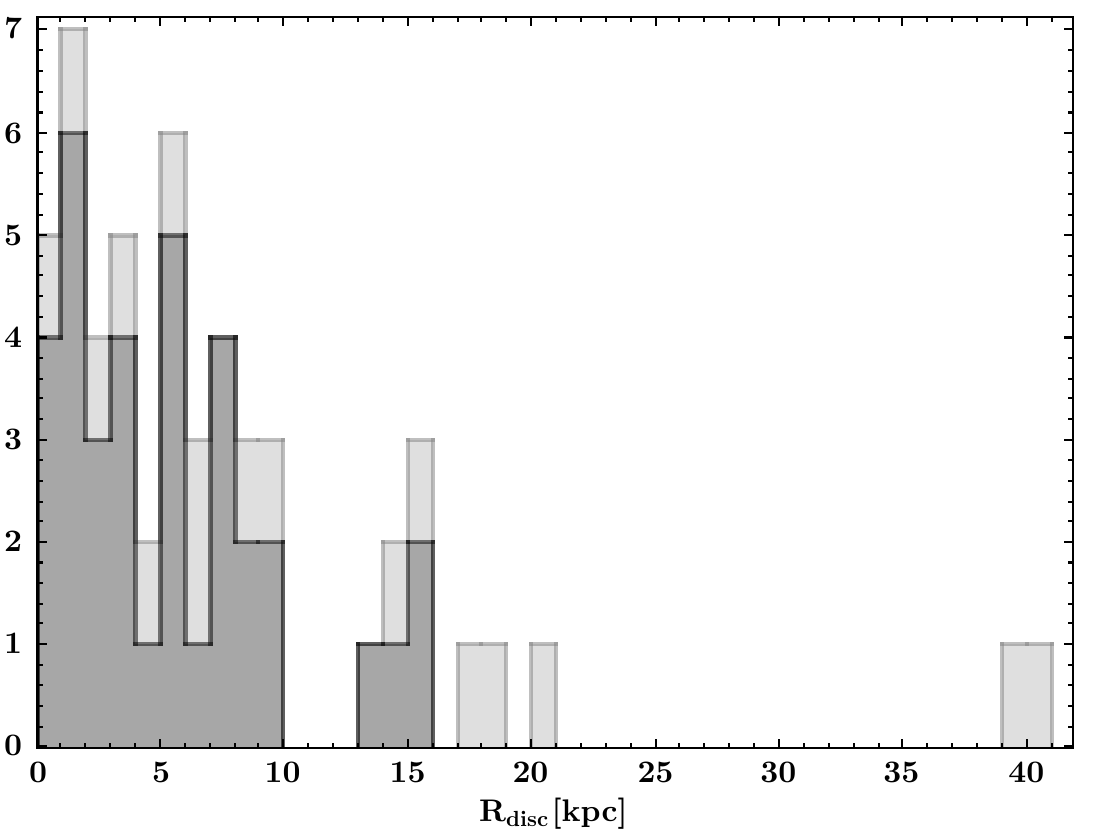}
\end{center} 
\caption{
\textit{Upper panel}: Points of last Galactic disc crossings of the stars in our sample in the $X$-$Y$ plane. The two stars with the largest distances have been omitted here for better visualisation. \textit{Lower panel}: Distribution of the Galactocentric radii at disc crossing for all stars of the sample. Stars which could have been ejected from the disc are marked in dark grey, the others in light grey.}
\label{fig:crossing1}
\end{figure}
    
\begin{figure}
\begin{center}
\includegraphics[width=0.50\textwidth]{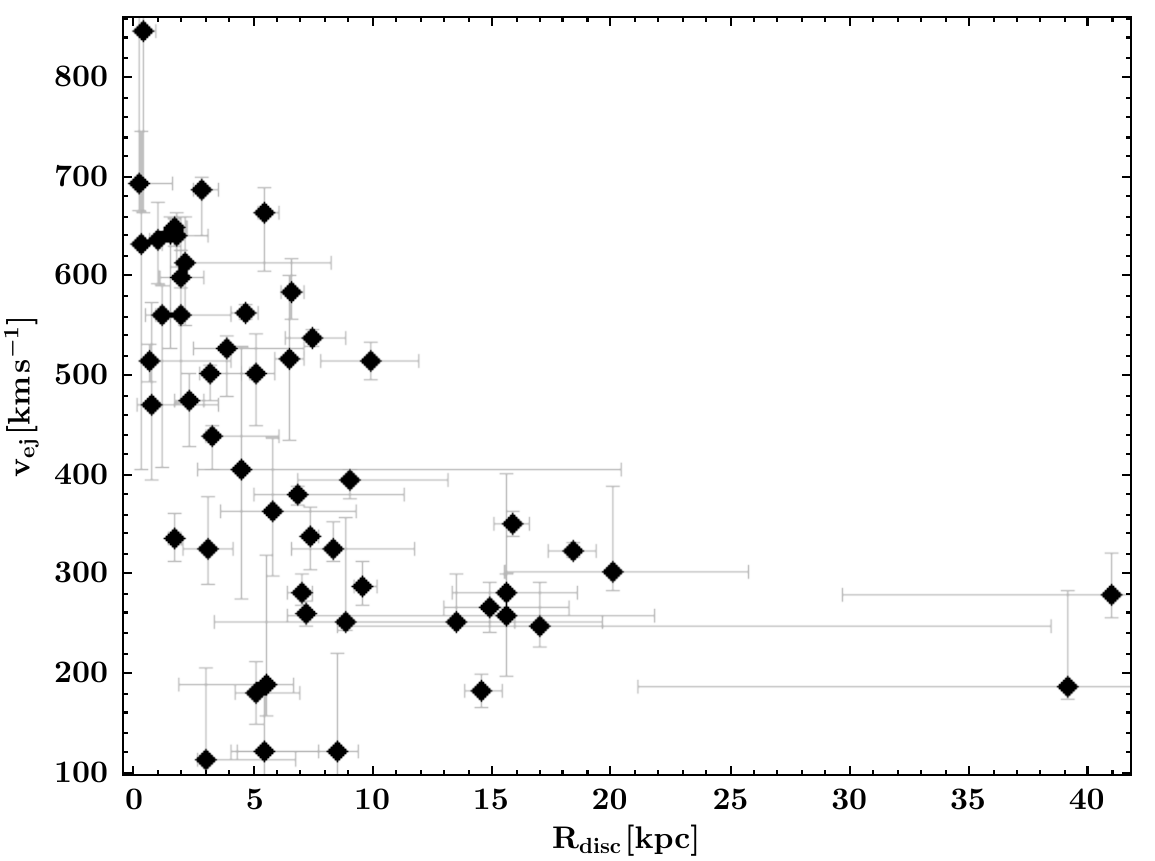}
\end{center} 
\caption{Ejection or disc crossing velocities plotted against Galactocentric distance.}
\label{fig:crossing2}
\end{figure}

As shown in Fig.~\ref{fig:crossing1}, most Galactic plane crossing locations are at $R_{\rm disc}<15\,{\rm kpc}$; only two stars cross further out. The disc crossings are quite concentrated towards the central region of the Galaxy and fall off sharply at larger Galactocentric distances. There is also no difference in the crossing locations between stars, which fulfill the criteria for an ejection and the rest of the sample. This is expected in a halo population, because the density of stars in the halo is known to be highest in the central bulge region and to drop off in radial direction (e.g. Bland-Hawthorn \& Gerhard \cite{bland16} and references therein). In addition, we know that the $32\%$ stars of our sample are very likely halo stars. 

The distribution of disc crossing velocities is shown in Fig.~\ref{fig:crossing2}.  
Unlike the (hyper-)runaway A/B-type stars studied by Irrgang et al. (\cite{irrgang21}), the present sample does not contain stars with high ejection velocities at large $R_{\rm disc}$.  On the contrary, the ejection velocities are in fact highest towards the Galactic center and decrease with Galactocentric distance. This is in part a bias introduced by our selection of stars by their high current space velocity. Since the gravitational attraction of the Galaxy decreases with Galactocentric distance, the stars coming from the central part of the Galaxy require higher disc crossing velocities than those from further out in the disc to remain fast at their present positions. 

The halo scenario is further corroborated by the diversity of spectral types in our sample. Although we did not derive spectroscopic distances of the composite binaries we identified in our survey (except the well-studied system PB\,3877, Nemeth et al. \cite{nemeth16}), their locations at high Galactic latitudes and their faintness point to a halo membership as well. Our extended sample, therefore, consists of almost all known types of hot subdwarfs including composite binaries, which is another indication that we see a halo population rather than a sample of ejected stars, which would have to be less diverse and should consist exclusively of single stars. In addition, other members of this halo population with similarily extreme kinematics as our fast sdO/Bs, but in earlier evolutionary stages on the main sequence and the red giant branch have been discovered (Scholz et al. \cite{scholz15}; Hattori et al. \cite{hattori18}; Li et al. \cite{li21}). 

On the other hand, the distribution of the stars in the $T_{\rm eff}-\log{g}$ and the $T_{\rm eff}-\log{n({\rm He})/n({\rm H})}$ diagrams (see Figs.~\ref{spectro_kiel} and \ref{spectro_he}) resembles the typical field population of sdO/Bs (e.g. Heber \cite{heber16}; Luo et al. \cite{luo21}; Geier et al. \cite{geier22}) rather than the markedly different distribution seen in the old halo globular cluster $\omega$\,Cen (Latour et al. \cite{latour18,latour23}).  It is, however, not clear whether $\omega$\,Cen is a good proxy for the halo population of sdO/Bs as such, because other old GCs (e.g. NGC\,6752) contain a different mixture of hot subdwarf types (Latour et al. \cite{latour23}). 

The extremely small fraction of RV-variable close binaries in our sample is especially for the hydrogen-rich subtypes completely different than what is observed in the disc population (see Geier et al. \cite{geier22} and references therein). We only found one star in our sample to be significantly variable (J102439.43+383917.9), next to a single candidate for RV variability (J180313.45+234000.1). Both stars are kinematically more likely to be members of the thick disc rather than the halo population -- we do not find any RV variable close binaries in our sample of hot subdwarf stars with halo kinematics. This is  consistent with the lack of such binaries in $\omega$\,Cen (Latour et al. \cite{latour18}). A possible explanation might be the age dependency of the contribution of the different binary formation channels. For older populations, the relative contribution of mergers is predicted to rise, leading to a higher fraction of single stars (Han \cite{han08}; Lisker \& Han \cite{lisker08}). This strongly indicates that the lack of hot subdwarfs in close binary systems in GCs is an age rather than an environmental effect.

We therefore conclude that we neither found another unbound HVS sdO/B similar to the prototype US\,708 nor strong candidates for bound ejected companions of SN\,Ia in our spectroscopic survey. Of the 53 stars in our sample, 48 are very likely halo sdO/Bs with extreme kinematics and 5 are more likely members of the thick disc population. 

\begin{table*}
%\begin{table*}
\centering
\caption{\label{tab:kinematic} Kinematic parameters of the fast star candidates.}
\renewcommand{\arraystretch}{1.25}
\begin{tabular}{lrrrrrrr}
\hline\hline
\noalign{\smallskip}
Name                &         $X$                       &         $Y$                       &         $Z$                       &   $U$                       &  $V$                         &  $W$                         &  $v_{\rm grf}$           \\
                    &         ${\rm [kpc]}$             &         ${\rm [kpc]}$             &         ${\rm [kpc]}$             &   ${\rm [km\,s^{-1}]}$      &  ${\rm [km\,s^{-1}]}$        &  ${\rm [km\,s^{-1}]}$        &  ${\rm [km\,s^{-1}]}$    \\
\noalign{\smallskip}                                                                                                                                                                                                                                  
\hline                                                                                                                                                                                                                                                
\noalign{\smallskip}
J022422.21+000313.5 & -12.20$^{+0.50 }_{-0.48 }$        & 0.93  $^{+0.13 }_{-0.11 }$        & -5.51 $^{+0.69 }_{-0.73 }$        & -131$^{+8  }_{-8  }$        &  -52 $^{+32 }_{-43 }$        &  141 $^{+8  }_{-8  }$        & 198 $^{+10 }_{-4  }$     \\
J080833.76+180221.8 & -10.74$^{+0.30 }_{-0.31 }$        & -1.06 $^{+0.13 }_{-0.15 }$        & 1.17  $^{+0.17 }_{-0.13 }$        & -187$^{+10 }_{-11 }$        &  101 $^{+22 }_{-29 }$        &  -13 $^{+3  }_{-3  }$        & 211 $^{+4  }_{-3  }$     \\
J082802.03+404008.9 & -14.86$^{+0.81 }_{-0.88 }$        & -0.06 $^{+0.01 }_{-0.01 }$        & 4.52  $^{+0.61 }_{-0.56 }$        & -160$^{+4  }_{-5  }$        &  74  $^{+25 }_{-24 }$        &  -87 $^{+5  }_{-5  }$        & 196 $^{+9  }_{-9  }$     \\
J084556.85+135211.3 & -10.90$^{+0.35 }_{-0.37 }$        & -1.63 $^{+0.24 }_{-0.22 }$        & 1.83  $^{+0.27 }_{-0.25 }$        & 77  $^{+7  }_{-6  }$        &  168 $^{+6  }_{-6  }$        &  -26 $^{+13 }_{-14 }$        & 188 $^{+3  }_{-3  }$     \\
J090252.99+073533.9 & -10.41$^{+0.25 }_{-0.33 }$        & -1.80 $^{+0.23 }_{-0.28 }$        & 1.73  $^{+0.27 }_{-0.22 }$        & -294$^{+16 }_{-29 }$        &  100 $^{+30 }_{-41 }$        &  -44 $^{+4  }_{-4  }$        & 314 $^{+13 }_{-8  }$     \\
J091512.06+191114.6 & -13.90$^{+0.71 }_{-0.77 }$        & -3.19 $^{+0.43 }_{-0.42 }$        & 5.37  $^{+0.71 }_{-0.73 }$        & 106 $^{+30 }_{-29 }$        &  76  $^{+25 }_{-25 }$        &  -337$^{+39 }_{-47 }$        & 350 $^{+56 }_{-31 }$     \\
J094850.47+551631.6 & -11.77$^{+0.44 }_{-0.50 }$        & 1.27  $^{+0.18 }_{-0.17 }$        & 3.87  $^{+0.53 }_{-0.52 }$        & -32 $^{+7  }_{-7  }$        &  -5  $^{+30 }_{-36 }$        &  -48 $^{+7  }_{-7  }$        & 62  $^{+11 }_{-9  }$     \\
J102057.16+013751.2 & -9.55 $^{+0.16 }_{-0.15 }$        & -2.09 $^{+0.24 }_{-0.31 }$        & 2.45  $^{+0.36 }_{-0.28 }$        & 88  $^{+6  }_{-5  }$        &  70  $^{+6  }_{-5  }$        &  103 $^{+10 }_{-9  }$        & 154 $^{+6  }_{-7  }$     \\
J102439.43+383917.9 & -9.83 $^{+0.19 }_{-0.19 }$        & -0.07 $^{+0.01 }_{-0.01 }$        & 2.23  $^{+0.29 }_{-0.29 }$        & 124 $^{+14 }_{-15 }$        &  173 $^{+10 }_{-11 }$        &  -27 $^{+10 }_{-10 }$        & 214 $^{+3  }_{-2  }$     \\
J103810.94+253204.8 & -10.85$^{+0.29 }_{-0.35 }$        & -1.33 $^{+0.17 }_{-0.17 }$        & 4.90  $^{+0.63 }_{-0.61 }$        & 235 $^{+23 }_{-27 }$        &  -156$^{+42 }_{-47 }$        &  14  $^{+23 }_{-22 }$        & 286 $^{+39 }_{-49 }$     \\
J120352.24+235343.3 & -8.84 $^{+0.08 }_{-0.07 }$        & -0.46 $^{+0.06 }_{-0.06 }$        & 3.21  $^{+0.40 }_{-0.42 }$        & 362 $^{+43 }_{-50 }$        &  -15 $^{+30 }_{-29 }$        &  97  $^{+11 }_{-12 }$        & 364 $^{+48 }_{-37 }$     \\
J120521.48+224702.2 & -9.53 $^{+0.15 }_{-0.18 }$        & -1.50 $^{+0.20 }_{-0.22 }$        & 9.42  $^{+1.40 }_{-1.24 }$        & 249 $^{+40 }_{-37 }$        &  83  $^{+17 }_{-16 }$        &  90  $^{+9  }_{-10 }$        & 275 $^{+28 }_{-25 }$     \\
J121703.12+454539.3 & -9.15 $^{+0.11 }_{-0.12 }$        & 0.61  $^{+0.09 }_{-0.08 }$        & 2.69  $^{+0.39 }_{-0.37 }$        & 117 $^{+30 }_{-30 }$        &  -69 $^{+35 }_{-38 }$        &  -359$^{+4  }_{-5  }$        & 379 $^{+20 }_{-14 }$     \\
J123137.56+074621.7 & -7.82 $^{+0.09 }_{-0.09 }$        & -1.73 $^{+0.22 }_{-0.22 }$        & 5.02  $^{+0.65 }_{-0.63 }$        & 90  $^{+26 }_{-26 }$        &  -130$^{+29 }_{-21 }$        &  372 $^{+9  }_{-9  }$        & 401 $^{+7  }_{-3  }$     \\
J123428.30+262757.9 & -8.76 $^{+0.07 }_{-0.07 }$        & -0.32 $^{+0.04 }_{-0.04 }$        & 7.12  $^{+0.99 }_{-0.97 }$        & -226$^{+28 }_{-33 }$        &  -7  $^{+36 }_{-38 }$        &  70  $^{+9  }_{-9  }$        & 232 $^{+33 }_{-26 }$     \\
J123953.52+062853.0 & -7.47 $^{+0.13 }_{-0.13 }$        & -1.98 $^{+0.24 }_{-0.29 }$        & 5.75  $^{+0.83 }_{-0.68 }$        & 233 $^{+46 }_{-39 }$        &  -16 $^{+22 }_{-19 }$        &  30  $^{+9  }_{-9  }$        & 233 $^{+45 }_{-38 }$     \\
J124248.89+133632.6 & -7.71 $^{+0.11 }_{-0.10 }$        & -1.55 $^{+0.20 }_{-0.22 }$        & 6.85  $^{+1.12 }_{-0.78 }$        & 77  $^{+24 }_{-17 }$        &  -67 $^{+44 }_{-48 }$        &  -234$^{+10 }_{-12 }$        & 249 $^{+31 }_{-22 }$     \\
J124310.58+343358.4 & -8.88 $^{+0.08 }_{-0.08 }$        & 0.46  $^{+0.06 }_{-0.06 }$        & 4.93  $^{+0.63 }_{-0.63 }$        & 178 $^{+19 }_{-19 }$        &  -212$^{+66 }_{-59 }$        &  198 $^{+5  }_{-4  }$        & 331 $^{+55 }_{-45 }$     \\
J124819.08+035003.2 & -7.72 $^{+0.10 }_{-0.10 }$        & -1.13 $^{+0.14 }_{-0.15 }$        & 3.06  $^{+0.39 }_{-0.39 }$        & -317$^{+37 }_{-31 }$        &  -5  $^{+41 }_{-36 }$        &  -128$^{+19 }_{-18 }$        & 336 $^{+40 }_{-37 }$     \\
J130543.97+115840.8 & -7.93 $^{+0.08 }_{-0.08 }$        & -0.44 $^{+0.05 }_{-0.06 }$        & 2.32  $^{+0.30 }_{-0.30 }$        & 288 $^{+35 }_{-41 }$        &  -15 $^{+35 }_{-33 }$        &  -67 $^{+2  }_{-2  }$        & 293 $^{+36 }_{-40 }$     \\
J133135.41+020919.8 & -7.26 $^{+0.16 }_{-0.15 }$        & -0.78 $^{+0.10 }_{-0.10 }$        & 2.73  $^{+0.37 }_{-0.35 }$        & 339 $^{+50 }_{-47 }$        &  36  $^{+25 }_{-22 }$        &  56  $^{+10 }_{-9  }$        & 343 $^{+48 }_{-43 }$     \\
J133417.10+173850.7 & -6.49 $^{+0.31 }_{-0.24 }$        & -0.29 $^{+0.04 }_{-0.05 }$        & 7.97  $^{+1.25 }_{-0.99 }$        & -253$^{+41 }_{-37 }$        &  320 $^{+10 }_{-11 }$        &  -135$^{+14 }_{-16 }$        & 427 $^{+32 }_{-31 }$     \\
J135651.26+155810.4 & -3.42 $^{+1.35 }_{-1.20 }$        & -0.01 $^{+0.00 }_{0.00  }$        & 14.16 $^{+4.39 }_{-3.10 }$        & 26  $^{+43 }_{-38 }$        &  17  $^{+58 }_{-81 }$        &  239 $^{+21 }_{-20 }$        & 247 $^{+25 }_{-23 }$     \\
J140532.34+410626.1 & -7.82 $^{+0.11 }_{-0.10 }$        & 3.27  $^{+0.63 }_{-0.46 }$        & 8.79  $^{+1.71 }_{-1.24 }$        & -200$^{+42 }_{-53 }$        &  -16 $^{+23 }_{-16 }$        &  -55 $^{+19 }_{-18 }$        & 201 $^{+51 }_{-32 }$     \\
J143127.88+014416.2 & -3.43 $^{+0.79 }_{-0.60 }$        & -0.83 $^{+0.10 }_{-0.13 }$        & 7.20  $^{+1.17 }_{-0.83 }$        & -102$^{+15 }_{-15 }$        &  32  $^{+38 }_{-45 }$        &  -131$^{+17 }_{-17 }$        & 169 $^{+17 }_{-15 }$     \\
J143258.05+011857.9 & -5.52 $^{+0.36 }_{-0.39 }$        & -0.48 $^{+0.06 }_{-0.06 }$        & 4.10  $^{+0.52 }_{-0.54 }$        & 189 $^{+39 }_{-29 }$        &  -117$^{+51 }_{-36 }$        &  136 $^{+15 }_{-14 }$        & 256 $^{+53 }_{-47 }$     \\
J144209.90+105733.9 & -6.08 $^{+0.33 }_{-0.29 }$        & 0.26  $^{+0.04 }_{-0.03 }$        & 3.94  $^{+0.56 }_{-0.49 }$        & -292$^{+39 }_{-32 }$        &  164 $^{+11 }_{-10 }$        &  -124$^{+20 }_{-21 }$        & 353 $^{+31 }_{-32 }$     \\
J145141.40+090645.1 & -4.37 $^{+0.57 }_{-0.50 }$        & 0.44  $^{+0.06 }_{-0.06 }$        & 6.12  $^{+0.83 }_{-0.78 }$        & -262$^{+38 }_{-37 }$        &  64  $^{+19 }_{-15 }$        &  -188$^{+20 }_{-25 }$        & 323 $^{+45 }_{-33 }$     \\
J145930.70+175846.1 & -6.11 $^{+0.34 }_{-0.33 }$        & 0.95  $^{+0.13 }_{-0.14 }$        & 4.16  $^{+0.59 }_{-0.60 }$        & 243 $^{+27 }_{-28 }$        &  43  $^{+33 }_{-38 }$        &  77  $^{+25 }_{-22 }$        & 255 $^{+26 }_{-27 }$     \\
J150222.35+320220.9 & -6.97 $^{+0.22 }_{-0.17 }$        & 1.77  $^{+0.24 }_{-0.22 }$        & 4.21  $^{+0.54 }_{-0.56 }$        & -357$^{+49 }_{-61 }$        &  51  $^{+7  }_{-5  }$        &  -99 $^{+5  }_{-5  }$        & 375 $^{+56 }_{-48 }$     \\
J151248.61+042205.5 & -4.33 $^{+0.63 }_{-0.50 }$        & 0.37  $^{+0.05 }_{-0.05 }$        & 4.79  $^{+0.70 }_{-0.61 }$        & -89 $^{+30 }_{-39 }$        &  -172$^{+53 }_{-54 }$        &  -234$^{+17 }_{-20 }$        & 305 $^{+52 }_{-53 }$     \\
J153419.42+372557.2 & -6.43 $^{+0.30 }_{-0.27 }$        & 3.43  $^{+0.52 }_{-0.46 }$        & 5.52  $^{+0.83 }_{-0.74 }$        & -52 $^{+14 }_{-23 }$        &  80  $^{+25 }_{-23 }$        &  217 $^{+23 }_{-19 }$        & 238 $^{+17 }_{-12 }$     \\
J154958.29+043820.1 & -4.41 $^{+0.52 }_{-0.65 }$        & 0.93  $^{+0.13 }_{-0.15 }$        & 3.67  $^{+0.50 }_{-0.58 }$        & 176 $^{+6  }_{-6  }$        &  -156$^{+69 }_{-57 }$        &  119 $^{+33 }_{-30 }$        & 262 $^{+45 }_{-51 }$     \\
J161143.29+554044.3 & -7.62 $^{+0.13 }_{-0.12 }$        & 10.52 $^{+1.54 }_{-1.46 }$        & 10.30 $^{+1.51 }_{-1.43 }$        & -91 $^{+11 }_{-12 }$        &  -12 $^{+13 }_{-15 }$        &  -285$^{+7  }_{-9  }$        & 300 $^{+7  }_{-7  }$     \\
J163213.05+205124.0 & -6.10 $^{+0.29 }_{-0.30 }$        & 1.81  $^{+0.25 }_{-0.22 }$        & 2.40  $^{+0.33 }_{-0.29 }$        & 128 $^{+8  }_{-11 }$        &  -84 $^{+30 }_{-27 }$        &  -4  $^{+19 }_{-18 }$        & 147 $^{+8  }_{-3  }$     \\
J164419.44+452326.7 & -7.57 $^{+0.12 }_{-0.12 }$        & 2.38  $^{+0.31 }_{-0.31 }$        & 2.18  $^{+0.29 }_{-0.29 }$        & -261$^{+45 }_{-45 }$        &  -25 $^{+9  }_{-5  }$        &  -168$^{+5  }_{-5  }$        & 311 $^{+34 }_{-36 }$     \\
J164853.26+121703.0 & -2.75 $^{+0.94 }_{-0.83 }$        & 3.34  $^{+0.52 }_{-0.52 }$        & 4.24  $^{+0.66 }_{-0.66 }$        & -255$^{+72 }_{-82 }$        &  -50 $^{+43 }_{-13 }$        &  79  $^{+19 }_{-17 }$        & 265 $^{+83 }_{-69 }$     \\
J165924.75+273244.4 & -1.55 $^{+1.68 }_{-1.31 }$        & 7.79  $^{+1.93 }_{-1.47 }$        & 7.41  $^{+1.83 }_{-1.40 }$        & -302$^{+83 }_{-56 }$        &  -30 $^{+110}_{-84 }$        &  -149$^{+25 }_{-22 }$        & 296 $^{+65 }_{-40 }$     \\
J170256.38+241757.9 & -2.74 $^{+1.02 }_{-0.77 }$        & 5.76  $^{+0.97 }_{-0.83 }$        & 5.45  $^{+0.92 }_{-0.79 }$        & 256 $^{+10 }_{-13 }$        &  -10 $^{+59 }_{-66 }$        &  106 $^{+19 }_{-19 }$        & 276 $^{+19 }_{-13 }$     \\
J171533.84+365214.8 & -5.91 $^{+0.35 }_{-0.30 }$        & 4.48  $^{+0.58 }_{-0.58 }$        & 3.50  $^{+0.45 }_{-0.45 }$        & -125$^{+10 }_{-11 }$        &  303 $^{+8  }_{-5  }$        &  206 $^{+14 }_{-14 }$        & 385 $^{+18 }_{-13 }$     \\
J172736.02+361706.3 & -5.61 $^{+0.43 }_{-0.34 }$        & 5.00  $^{+0.71 }_{-0.65 }$        & 3.55  $^{+0.51 }_{-0.46 }$        & 17  $^{+8  }_{-10 }$        &  109 $^{+9  }_{-9  }$        &  18  $^{+11 }_{-10 }$        & 112 $^{+10 }_{-9  }$     \\
J174211.75+643009.9 & -8.87 $^{+0.08 }_{-0.08 }$        & 6.56  $^{+0.87 }_{-0.84 }$        & 4.06  $^{+0.54 }_{-0.52 }$        & -286$^{+27 }_{-21 }$        &  -79 $^{+20 }_{-22 }$        &  -144$^{+17 }_{-13 }$        & 324 $^{+10 }_{-9  }$     \\
J180313.45+234000.1 & -6.08 $^{+0.33 }_{-0.31 }$        & 2.70  $^{+0.43 }_{-0.32 }$        & 1.34  $^{+0.21 }_{-0.16 }$        & 47  $^{+7  }_{-4  }$        &  193 $^{+4  }_{-4  }$        &  -76 $^{+7  }_{-10 }$        & 214 $^{+4  }_{-4  }$     \\
J184832.52+181540.0 & -6.23 $^{+0.32 }_{-0.29 }$        & 2.47  $^{+0.38 }_{-0.31 }$        & 0.51  $^{+0.08 }_{-0.06 }$        & -93 $^{+27 }_{-41 }$        &  -10 $^{+9  }_{-8  }$        &  47  $^{+10 }_{-8  }$        & 103 $^{+42 }_{-27 }$     \\
J204358.55-065025.8 & -5.55 $^{+0.41 }_{-0.34 }$        & 2.42  $^{+0.31 }_{-0.32 }$        & -2.01 $^{+0.27 }_{-0.25 }$        & 125 $^{+21 }_{-37 }$        &  -227$^{+38 }_{-33 }$        &  -115$^{+38 }_{-33 }$        & 263 $^{+40 }_{-22 }$     \\
J205030.39-061957.8 & -4.75 $^{+0.51 }_{-0.48 }$        & 3.20  $^{+0.45 }_{-0.42 }$        & -2.72 $^{+0.35 }_{-0.38 }$        & 151 $^{+37 }_{-46 }$        &  -261$^{+25 }_{-21 }$        &  204 $^{+7  }_{-8  }$        & 358 $^{+6  }_{-7  }$     \\
J210907.28+103640.6 & -6.13 $^{+0.70 }_{-0.45 }$        & 3.95  $^{+1.22 }_{-0.79 }$        & -2.06 $^{+0.42 }_{-0.62 }$        & -33 $^{+13 }_{-17 }$        &  167 $^{+24 }_{-28 }$        &  2   $^{+13 }_{-16 }$        & 169 $^{+26 }_{-25 }$     \\
J212300.31+043453.0 & -4.87 $^{+0.60 }_{-0.70 }$        & 5.38  $^{+0.97 }_{-1.02 }$        & -3.79 $^{+0.72 }_{-0.68 }$        & 203 $^{+24 }_{-28 }$        &  -299$^{+65 }_{-52 }$        &  3   $^{+43 }_{-32 }$        & 347 $^{+34 }_{-33 }$     \\
J212449.22+061956.4 & -5.77 $^{+0.43 }_{-0.31 }$        & 4.43  $^{+0.62 }_{-0.58 }$        & -2.97 $^{+0.39 }_{-0.42 }$        & -147$^{+44 }_{-40 }$        &  -69 $^{+11 }_{-8  }$        &  25  $^{+17 }_{-16 }$        & 160 $^{+30 }_{-32 }$     \\
J215648.71+003620.7 & -6.59 $^{+0.25 }_{-0.24 }$        & 3.00  $^{+0.42 }_{-0.37 }$        & -2.92 $^{+0.38 }_{-0.39 }$        & 65  $^{+4  }_{-4  }$        &  63  $^{+10 }_{-11 }$        &  70  $^{+7  }_{-7  }$        & 115 $^{+10 }_{-11 }$     \\
J220759.08+204505.9 & -7.66 $^{+0.11 }_{-0.10 }$        & 3.91  $^{+0.48 }_{-0.51 }$        & -2.10 $^{+0.27 }_{-0.27 }$        & 272 $^{+20 }_{-23 }$        &  -183$^{+30 }_{-34 }$        &  60  $^{+14 }_{-14 }$        & 331 $^{+34 }_{-31 }$     \\
J221728.38+121643.8 & -7.86 $^{+0.09 }_{-0.08 }$        & 1.93  $^{+0.27 }_{-0.23 }$        & -1.44 $^{+0.17 }_{-0.21 }$        & -215$^{+34 }_{-38 }$        &  -26 $^{+6  }_{-5  }$        &  25  $^{+13 }_{-18 }$        & 218 $^{+36 }_{-33 }$     \\
J222515.34-011156.8 & -6.78 $^{+0.24 }_{-0.22 }$        & 3.20  $^{+0.47 }_{-0.43 }$        & -3.77 $^{+0.50 }_{-0.56 }$        & 103 $^{+10 }_{-11 }$        &  -117$^{+13 }_{-15 }$        &  289 $^{+7  }_{-7  }$        & 329 $^{+5  }_{-5  }$     \\
\noalign{\smallskip}                                                                                                                                                                                                                 
\hline\hline                        
\end{tabular}                       
\end{table*}

\begin{table*}
%\begin{table*}
\centering
\caption{\label{tab:crossing} Parameters of the fast star candidates at the last Galactic disc passage.}
\renewcommand{\arraystretch}{1.25}
\begin{tabular}{lrrrrrrrrrrrrrr}
\hline\hline
\noalign{\smallskip}
Name                &         $X_{\rm disc}$            &         $Y_{\rm disc}$            &         $R_{\rm disc}$             &         $v_{\rm ej}$            & $t_{\rm disc}$                 &  $e$                       &  $L_{\rm Z}$                       \\
                    &         ${\rm [kpc]}$             &         ${\rm [kpc]}$             &         ${\rm [kpc]}$              &         ${\rm [km\,s^{-1}]}$    &  [Myr]                         &                            &  ${\rm [10^{3}\,kpc\,km\,s^{-1}]}$ \\
\noalign{\smallskip}                                                                                                                                                 
\hline                                                                                                                                                               
\noalign{\smallskip}
J022422.21+000313.5 & 1.73  $^{+0.49 }_{-0.43 }$        & 0.64  $^{+1.88 }_{-0.75 }$        & 1.82  $^{+1.34 }_{-0.52 }$         &  641  $^{+24   }_{-32   }$      &  171.9 $^{+21.1 }_{-14.1 }$    &  0.811$^{+0.054}_{-0.092}$  & 0.65 $^{+0.55}_{-0.44}$           \\
J080833.76+180221.8 & -12.88$^{+1.61 }_{-1.40 }$        & -7.40 $^{+1.21 }_{-0.94 }$        & 14.58 $^{+0.87 }_{-0.72 }$         &  182  $^{+17   }_{-17   }$      &  78.9  $^{+4.4  }_{-4.6  }$    &  0.740$^{+0.083}_{-0.069}$  & -1.14$^{+0.30}_{-0.19}$           \\
J082802.03+404008.9 & 0.19  $^{+0.64 }_{-0.64 }$        & -3.09 $^{+1.07 }_{-1.08 }$        & 3.16  $^{+1.04 }_{-1.08 }$         &  326  $^{+52   }_{-36   }$      &  200.9 $^{+9.0  }_{-7.9  }$    &  0.783$^{+0.081}_{-0.062}$  & -1.15$^{+0.31}_{-0.30}$           \\
J084556.85+135211.3 & 1.09  $^{+4.89 }_{-4.63 }$        & -5.53 $^{+1.59 }_{-1.21 }$        & 5.48  $^{+2.30 }_{-1.39 }$         &  123  $^{+60   }_{-28   }$      &  39.2  $^{+36.8 }_{-20.3 }$    &  0.359$^{+0.020}_{-0.026}$  & -1.89$^{+0.04}_{-0.04}$           \\
J090252.99+073533.9 & -13.32$^{+12.06}_{-6.36 }$        & -11.49$^{+6.80 }_{-0.94 }$        & 15.61 $^{+6.21 }_{-9.21 }$         &  260  $^{+143  }_{-61   }$      &  32.7  $^{+370.1}_{-19.6 }$    &  0.869$^{+0.072}_{-0.054}$  & -1.07$^{+0.39}_{-0.31}$           \\
J091512.06+191114.6 & 40.21 $^{+1.81 }_{-10.70}$        & 4.31  $^{+3.13 }_{-2.15 }$        & 40.94 $^{+1.88 }_{-11.27}$         &  280  $^{+43   }_{-23   }$      &  478.8 $^{+384.2}_{-192.4}$    &  0.608$^{+0.129}_{-0.164}$  & -1.16$^{+0.34}_{-0.27}$           \\
J094850.47+551631.6 & -0.05 $^{+0.27 }_{-0.49 }$        & -0.01 $^{+1.23 }_{-0.78 }$        & 0.31  $^{+0.89 }_{-0.19 }$         &  633  $^{+113  }_{-226  }$      &  77.8  $^{+4.8  }_{-3.1  }$    &  0.986$^{+0.013}_{-0.103}$  & 0.06 $^{+0.42}_{-0.39}$           \\
J102057.16+013751.2 & -6.49 $^{+0.71 }_{-0.59 }$        & -2.82 $^{+0.16 }_{-0.11 }$        & 7.08  $^{+0.47 }_{-0.61 }$         &  282  $^{+20   }_{-13   }$      &  19.3  $^{+3.5  }_{-3.2  }$    &  0.711$^{+0.041}_{-0.039}$  & -0.70$^{+0.05}_{-0.04}$           \\
J102439.43+383917.9 & 2.94  $^{+3.82 }_{-6.49 }$        & -4.55 $^{+0.63 }_{-0.71 }$        & 5.16  $^{+1.85 }_{-0.88 }$         &  181  $^{+32   }_{-30   }$      &  37.6  $^{+26.2 }_{-22.3 }$    &  0.431$^{+0.035}_{-0.047}$  & -1.73$^{+0.08}_{-0.07}$           \\
J103810.94+253204.8 & 2.26  $^{+3.64 }_{-2.06 }$        & 4.10  $^{+3.14 }_{-1.80 }$        & 2.13  $^{+6.14 }_{-0.20 }$         &  614  $^{+45   }_{-64   }$      &  41.5  $^{+11.7 }_{-6.4  }$    &  0.662$^{+0.040}_{-0.012}$  & 1.76 $^{+0.57}_{-0.55}$           \\
J120352.24+235343.3 & 0.37  $^{+1.35 }_{-1.65 }$        & 0.12  $^{+0.52 }_{-0.45 }$        & 0.26  $^{+1.36 }_{-0.17 }$         &  694  $^{+236  }_{-26   }$      &  20.0  $^{+1.6  }_{-1.4  }$    &  0.978$^{+0.017}_{-0.008}$  & 0.13 $^{+0.26}_{-0.27}$           \\
J120521.48+224702.2 & 3.76  $^{+2.82 }_{-1.84 }$        & -1.89 $^{+0.69 }_{-0.64 }$        & 3.27  $^{+2.76 }_{-0.27 }$         &  439  $^{+10   }_{-33   }$      &  43.5  $^{+6.9  }_{-4.4  }$    &  0.759$^{+0.014}_{-0.013}$  & -0.82$^{+0.14}_{-0.16}$           \\
J121703.12+454539.3 & 19.89 $^{+5.84 }_{-4.41 }$        & -0.14 $^{+0.74 }_{-0.50 }$        & 20.05 $^{+5.72 }_{-4.53 }$         &  303  $^{+86   }_{-18   }$      &  154.4 $^{+457.7}_{-97.5 }$    &  0.599$^{+0.057}_{-0.062}$  & 0.65 $^{+0.36}_{-0.34}$           \\
J123137.56+074621.7 & -6.67 $^{+0.55 }_{-0.44 }$        & 0.02  $^{+0.50 }_{-0.36 }$        & 6.57  $^{+0.51 }_{-0.45 }$         &  584  $^{+35   }_{-28   }$      &  12.3  $^{+1.9  }_{-1.6  }$    &  0.777$^{+0.042}_{-0.032}$  & 1.06 $^{+0.18}_{-0.22}$           \\
J123428.30+262757.9 & -15.74$^{+2.39 }_{-3.00 }$        & -0.68 $^{+2.04 }_{-1.36 }$        & 15.63 $^{+3.00 }_{-2.30 }$         &  282  $^{+19   }_{-19   }$      &  55.6  $^{+7.7  }_{-6.9  }$    &  0.469$^{+0.036}_{-0.031}$  & 0.08 $^{+0.33}_{-0.34}$           \\
J123953.52+062853.0 & 2.19  $^{+1.92 }_{-1.41 }$        & 0.58  $^{+1.25 }_{-0.70 }$        & 0.66  $^{+3.41 }_{-0.11 }$         &  516  $^{+17   }_{-21   }$      &  33.6  $^{+4.4  }_{-3.0  }$    &  0.720$^{+0.159}_{-0.015}$  & 0.12 $^{+0.16}_{-0.17}$           \\
J124248.89+133632.6 & 8.56  $^{+2.97 }_{-1.59 }$        & 2.98  $^{+4.00 }_{-1.93 }$        & 9.03  $^{+4.13 }_{-2.17 }$         &  394  $^{+10   }_{-18   }$      &  152.1 $^{+52.3 }_{-25.8 }$    &  0.281$^{+0.067}_{-0.027}$  & 0.56 $^{+0.36}_{-0.37}$           \\
J124310.58+343358.4 & -3.77 $^{+0.97 }_{-0.88 }$        & 3.93  $^{+1.80 }_{-1.28 }$        & 5.51  $^{+0.59 }_{-0.22 }$         &  663  $^{+26   }_{-59   }$      &  20.6  $^{+2.0  }_{-2.0  }$    &  0.674$^{+0.062}_{-0.013}$  & 1.90 $^{+0.58}_{-0.59}$           \\
J124819.08+035003.2 & -4.92 $^{+0.68 }_{-0.89 }$        & -0.56 $^{+1.49 }_{-1.55 }$        & 5.11  $^{+0.75 }_{-0.57 }$         &  503  $^{+40   }_{-52   }$      &  215.1 $^{+111.3}_{-57.2 }$    &  0.970$^{+0.022}_{-0.034}$  & 0.02 $^{+0.31}_{-0.31}$           \\
J130543.97+115840.8 & 3.58  $^{+1.34 }_{-0.80 }$        & 0.23  $^{+0.73 }_{-0.45 }$        & 3.23  $^{+1.69 }_{-0.47 }$         &  503  $^{+9    }_{-27   }$      &  31.8  $^{+1.2  }_{-0.8  }$    &  0.859$^{+0.042}_{-0.016}$  & 0.13 $^{+0.26}_{-0.29}$           \\
J133135.41+020919.8 & 0.71  $^{+3.39 }_{-3.39 }$        & -0.27 $^{+0.41 }_{-0.49 }$        & 0.80  $^{+2.77 }_{-0.63 }$         &  472  $^{+103  }_{-77   }$      &  17.6  $^{+7.1  }_{-5.0  }$    &  0.944$^{+0.046}_{-0.036}$  & -0.26$^{+0.17}_{-0.19}$           \\
J133417.10+173850.7 & -19.90$^{+21.64}_{-13.28}$        & -36.94$^{+18.32}_{-22.29}$        & 39.06 $^{+26.93}_{-17.96}$         &  188  $^{+96   }_{-13   }$      &  107.6 $^{+616.3}_{-65.7 }$    &  0.832$^{+0.047}_{-0.064}$  & -2.12$^{+0.07}_{-0.07}$           \\
J135651.26+155810.4 & -0.18 $^{+1.91 }_{-3.52 }$        & -0.29 $^{+2.00 }_{-2.28 }$        & 1.98  $^{+2.07 }_{-1.47 }$         &  561  $^{+99   }_{-88   }$      &  38.5  $^{+14.7 }_{-10.0 }$    &  0.865$^{+0.083}_{-0.121}$  & 0.09 $^{+0.11}_{-0.26}$           \\
J140532.34+410626.1 & -8.24 $^{+4.16 }_{-9.21 }$        & 2.98  $^{+7.15 }_{-3.20 }$        & 8.86  $^{+10.76}_{-5.53 }$         &  252  $^{+106  }_{-8    }$      &  134.6 $^{+22.7 }_{-18.9 }$    &  0.738$^{+0.158}_{-0.108}$  & 0.14 $^{+0.14}_{-0.20}$           \\
J143127.88+014416.2 & -1.46 $^{+0.70 }_{-1.02 }$        & -1.37 $^{+1.00 }_{-0.91 }$        & 2.34  $^{+0.64 }_{-0.57 }$         &  475  $^{+27   }_{-47   }$      &  83.6  $^{+16.8 }_{-10.8 }$    &  0.862$^{+0.048}_{-0.076}$  & -0.06$^{+0.15}_{-0.17}$           \\
J143258.05+011857.9 & -0.51 $^{+0.86 }_{-0.82 }$        & 1.53  $^{+0.84 }_{-0.65 }$        & 1.68  $^{+0.50 }_{-0.22 }$         &  650  $^{+8    }_{-19   }$      &  19.0  $^{+0.6  }_{-0.6  }$    &  0.763$^{+0.040}_{-0.015}$  & 0.68 $^{+0.16}_{-0.24}$           \\
J144209.90+105733.9 & -10.19$^{+2.28 }_{-4.04 }$        & -11.06$^{+0.62 }_{-0.74 }$        & 14.94 $^{+3.33 }_{-1.94 }$         &  266  $^{+26   }_{-25   }$      &  184.9 $^{+87.1 }_{-47.1 }$    &  0.800$^{+0.082}_{-0.048}$  & -1.01$^{+0.11}_{-0.12}$           \\
J145141.40+090645.1 & -6.67 $^{+2.58 }_{-4.93 }$        & -2.97 $^{+1.04 }_{-0.61 }$        & 6.86  $^{+4.43 }_{-1.82 }$         &  380  $^{+9    }_{-11   }$      &  182.1 $^{+96.1 }_{-52.7 }$    &  0.869$^{+0.010}_{-0.011}$  & -0.27$^{+0.10}_{-0.12}$           \\
J145930.70+175846.1 & 0.81  $^{+0.54 }_{-0.38 }$        & -0.50 $^{+0.32 }_{-0.54 }$        & 1.17  $^{+0.36 }_{-0.25 }$         &  560  $^{+29   }_{-154  }$      &  22.2  $^{+1.0  }_{-0.8  }$    &  0.934$^{+0.028}_{-0.076}$  & -0.25$^{+0.24}_{-0.23}$           \\
J150222.35+320220.9 & -17.45$^{+8.91 }_{-21.13}$        & -1.14 $^{+4.57 }_{-1.54 }$        & 17.05 $^{+21.38}_{-8.47 }$         &  247  $^{+44   }_{-21   }$      &  240.6 $^{+156.8}_{-82.4 }$    &  0.953$^{+0.008}_{-0.015}$  & -0.37$^{+0.04}_{-0.06}$           \\
J151248.61+042205.5 & 2.17  $^{+0.36 }_{-0.68 }$        & 3.50  $^{+3.84 }_{-2.06 }$        & 3.88  $^{+3.24 }_{-1.40 }$         &  528  $^{+13   }_{-48   }$      &  123.1 $^{+78.3 }_{-36.4 }$    &  0.651$^{+0.088}_{-0.075}$  & 0.83 $^{+0.06}_{-0.17}$           \\
J153419.42+372557.2 & -7.26 $^{+0.16 }_{-0.16 }$        & 1.84  $^{+1.18 }_{-1.00 }$        & 7.41  $^{+0.38 }_{-0.22 }$         &  338  $^{+29   }_{-33   }$      &  21.2  $^{+1.4  }_{-1.3  }$    &  0.262$^{+0.047}_{-0.041}$  & -0.59$^{+0.17}_{-0.19}$           \\
J154958.29+043820.1 & 0.27  $^{+0.52 }_{-0.48 }$        & 2.13  $^{+0.88 }_{-0.95 }$        & 2.01  $^{+0.97 }_{-0.92 }$         &  599  $^{+28   }_{-10   }$      &  17.6  $^{+0.5  }_{-0.5  }$    &  0.645$^{+0.066}_{-0.022}$  & 0.77 $^{+0.16}_{-0.24}$           \\
J161143.29+554044.3 & 4.10  $^{+1.21 }_{-1.41 }$        & -4.77 $^{+0.76 }_{-1.01 }$        & 6.56  $^{+0.65 }_{-0.58 }$         &  517  $^{+85   }_{-82   }$      &  274.0 $^{+360.7}_{-116.6}$    &  0.742$^{+0.033}_{-0.036}$  & 0.18 $^{+0.17}_{-0.18}$           \\
J163213.05+205124.0 & 0.67  $^{+0.14 }_{-0.19 }$        & 1.18  $^{+0.71 }_{-0.73 }$        & 1.07  $^{+0.72 }_{-0.31 }$         &  636  $^{+39   }_{-44   }$      &  27.0  $^{+0.6  }_{-0.7  }$    &  0.699$^{+0.106}_{-0.108}$  & 0.55 $^{+0.16}_{-0.18}$           \\
J164419.44+452326.7 & 0.37  $^{+0.69 }_{-0.71 }$        & 0.15  $^{+0.28 }_{-0.13 }$        & 0.41  $^{+0.49 }_{-0.13 }$         &  848  $^{+41   }_{-183  }$      &  186.0 $^{+65.1 }_{-44.6 }$    &  0.986$^{+0.014}_{-0.043}$  & 0.20 $^{+0.05}_{-0.07}$           \\
J164853.26+121703.0 & -6.04 $^{+1.96 }_{-0.97 }$        & 6.48  $^{+3.94 }_{-2.41 }$        & 8.34  $^{+3.41 }_{-1.72 }$         &  326  $^{+27   }_{-12   }$      &  25.7  $^{+9.2  }_{-6.0  }$    &  0.486$^{+0.093}_{-0.025}$  & 0.23 $^{+0.06}_{-0.20}$           \\
J165924.75+273244.4 & 1.79  $^{+3.09 }_{-8.77 }$        & 3.54  $^{+15.86}_{-3.08 }$        & 4.54  $^{+15.85}_{-1.81 }$         &  405  $^{+124  }_{-131  }$      &  207.6 $^{+212.5}_{-88.1 }$    &  0.890$^{+0.055}_{-0.040}$  & 0.49 $^{+0.46}_{-1.02}$           \\
J170256.38+241757.9 & 0.49  $^{+1.01 }_{-0.54 }$        & -1.37 $^{+0.26 }_{-0.25 }$        & 1.53  $^{+0.49 }_{-0.30 }$         &  644  $^{+18   }_{-115  }$      &  23.4  $^{+2.8  }_{-0.9  }$    &  0.925$^{+0.019}_{-0.072}$  & -0.01$^{+0.48}_{-0.36}$           \\
J171533.84+365214.8 & -9.59 $^{+0.32 }_{-0.39 }$        & 1.41  $^{+0.91 }_{-0.60 }$        & 9.61  $^{+0.59 }_{-0.34 }$         &  288  $^{+25   }_{-19   }$      &  15.0  $^{+1.1  }_{-0.9  }$    &  0.607$^{+0.062}_{-0.044}$  & -2.28$^{+0.04}_{-0.10}$           \\
J172736.02+361706.3 & -5.51 $^{+3.61 }_{-1.40 }$        & -0.15 $^{+0.38 }_{-1.01 }$        & 5.61  $^{+1.17 }_{-3.68 }$         &  191  $^{+131  }_{-31   }$      &  38.0  $^{+7.2  }_{-10.5 }$    &  0.545$^{+0.037}_{-0.041}$  & -0.82$^{+0.05}_{-0.08}$           \\
J174211.75+643009.9 & 1.38  $^{+1.12 }_{-1.68 }$        & 2.20  $^{+1.45 }_{-0.73 }$        & 2.84  $^{+0.73 }_{-0.31 }$         &  687  $^{+12   }_{-46   }$      &  344.9 $^{+64.9 }_{-40.8 }$    &  0.913$^{+0.040}_{-0.048}$  & 0.94 $^{+0.16}_{-0.23}$           \\
J180313.45+234000.1 & 1.72  $^{+0.71 }_{-3.55 }$        & -5.13 $^{+2.57 }_{-2.86 }$        & 3.05  $^{+3.80 }_{-0.36 }$         &  113  $^{+93   }_{-25   }$      &  44.5  $^{+2.7  }_{-2.7  }$    &  0.172$^{+0.012}_{-0.011}$  & -1.32$^{+0.05}_{-0.05}$           \\
J184832.52+181540.0 & -6.71 $^{+0.10 }_{-0.09 }$        & 2.73  $^{+0.62 }_{-0.46 }$        & 7.22  $^{+0.20 }_{-0.13 }$         &  262  $^{+12   }_{-12   }$      &  8.8   $^{+1.7  }_{-1.5  }$    &  0.970$^{+0.018}_{-0.015}$  & 0.07 $^{+0.05}_{-0.06}$           \\
J204358.55-065025.8 & -2.11 $^{+0.23 }_{-0.21 }$        & 4.09  $^{+0.62 }_{-0.57 }$        & 4.70  $^{+0.54 }_{-0.59 }$         &  563  $^{+8    }_{-6    }$      &  13.2  $^{+1.5  }_{-1.3  }$    &  0.411$^{+0.044}_{-0.021}$  & 1.42 $^{+0.13}_{-0.21}$           \\
J205030.39-061957.8 & 18.33 $^{+1.04 }_{-0.97 }$        & -0.69 $^{+0.43 }_{-0.30 }$        & 18.38 $^{+1.00 }_{-1.02 }$         &  324  $^{+8    }_{-7    }$      &  121.8 $^{+14.3 }_{-10.4 }$    &  0.508$^{+0.031}_{-0.027}$  & 1.51 $^{+0.09}_{-0.10}$           \\
J210907.28+103640.6 & -8.28 $^{+4.27 }_{-0.96 }$        & -1.52 $^{+1.32 }_{-0.85 }$        & 8.54  $^{+0.83 }_{-4.21 }$         &  123  $^{+100  }_{-30   }$      &  40.7  $^{+5.9  }_{-14.0 }$    &  0.307$^{+0.111}_{-0.100}$  & -1.25$^{+0.21}_{-0.19}$           \\
J212300.31+043453.0 & 9.69  $^{+1.62 }_{-1.88 }$        & 2.27  $^{+1.69 }_{-1.37 }$        & 9.92  $^{+2.05 }_{-2.07 }$         &  515  $^{+18   }_{-19   }$      &  41.8  $^{+2.8  }_{-2.7  }$    &  0.575$^{+0.060}_{-0.033}$  & 2.16 $^{+0.50}_{-0.54}$           \\
J212449.22+061956.4 & -2.77 $^{+1.45 }_{-2.30 }$        & 5.39  $^{+2.59 }_{-1.90 }$        & 5.87  $^{+3.48 }_{-2.17 }$         &  364  $^{+74   }_{-65   }$      &  56.7  $^{+2.6  }_{-2.9  }$    &  0.796$^{+0.032}_{-0.045}$  & 0.51 $^{+0.06}_{-0.07}$           \\
J215648.71+003620.7 & 0.85  $^{+0.25 }_{-0.22 }$        & -1.54 $^{+0.19 }_{-0.21 }$        & 1.76  $^{+0.29 }_{-0.25 }$         &  337  $^{+25   }_{-24   }$      &  43.3  $^{+0.8  }_{-0.7  }$    &  0.643$^{+0.058}_{-0.031}$  & -0.47$^{+0.09}_{-0.08}$           \\
J220759.08+204505.9 & 7.40  $^{+1.40 }_{-1.09 }$        & 0.79  $^{+0.43 }_{-0.30 }$        & 7.50  $^{+1.38 }_{-1.16 }$         &  538  $^{+9    }_{-8    }$      &  40.8  $^{+1.1  }_{-0.9  }$    &  0.683$^{+0.043}_{-0.021}$  & 1.63 $^{+0.29}_{-0.32}$           \\
J221728.38+121643.8 & -12.45$^{+3.86 }_{-2.54 }$        & 4.39  $^{+1.27 }_{-1.65 }$        & 13.53 $^{+2.47 }_{-4.55 }$         &  252  $^{+48   }_{-3    }$      &  68.7  $^{+3.1  }_{-9.3  }$    &  0.947$^{+0.009}_{-0.009}$  & 0.22 $^{+0.04}_{-0.05}$           \\
J222515.34-011156.8 & 15.52 $^{+0.75 }_{-0.80 }$        & -3.34 $^{+0.28 }_{-0.29 }$        & 15.90 $^{+0.73 }_{-0.79 }$         &  351  $^{+14   }_{-12   }$      &  207.7 $^{+10.9 }_{-9.7  }$    &  0.414$^{+0.016}_{-0.014}$  & 0.91 $^{+0.10}_{-0.11}$           \\
\noalign{\smallskip}                                                                                                                                                                                                                 
\hline\hline                        
\end{tabular}                       
\end{table*}

\section{Discussion and outlook}                  
                          
The lack of other good sdO/B-type candidates for the single-degenerate SN\,Ia ejection scenario and the unclear relation to other fast (pre-)WDs (e.g. El-Badry et al. \cite{elbadry23}) makes the prototype US\,708 even more special. 

However, it might still be premature to draw strong conclusions on the birth rate of such objects and the number of corresponding SN\,Ia or other subtypes of stellar explosions. The observed lack of other candidates might be related to our search strategy. Neunteufel et al. (\cite{neunteufel21,neunteufel22}) simulated a model population of He-stars ejected by SNe\,Ia and determined the observable properties of this sample such as their positions and velocity components. The authors concluded that most of the ejected stars observed should be located in and around the Galactic bulge, where the stellar density is the highest. 
In addition, their simulation predicted that many objects with high absolute space velocities would have rather small proper motions, and would thus only be detectable through their large RVs, which requires spectral observations. 
Furthermore, most of those objects should orbit on bound trajectories with velocities around $500\,{\rm km\,s^{-1}}$ and should have masses around $0.8\,M_{\rm \odot}$ -- more massive than typical sdO/Bs. In these simulations, US\,708 represents an extreme case of a low-mass He-star ejected from the tightest possible binary progenitor, which has already been suggested by Geier et al. (\cite{geier15a}). 

Comparing these predictions with our selection criteria reveals some potentially quite significant biases. By selecting objects from the northern SDSS survey, we missed the Galactic bulge region, which is best accessible from the south. Focusing on objects with high tangential velocities and therefore relatively high proper motions instead of RVs, we did not impose a particularly strong selection criterion to single out the ejected sdO/B population. But probably most importantly, we selected sdO/Bs of all types, whereas the predicted higher mass He-stars should be He-sdOs with quite high luminosities. Since those stars have much shorter lifetimes on the EHB ($\sim10^{7}\,{\rm yr}$) than the bulk of the low-mass sdO/Bs ($\sim10^{8}\,{\rm yr}$, e.g. Han et al. \cite{han02}), they are rarer. If ejected from the disc at moderate velocities, they will not fly very deep into the halo and would be unable to reach our search area at high Galactic latitudes before turning to WDs. We therefore suspect that our selection criteria were likely not specific enough to find the best candidates for ejected companions in the thermonuclear SN scenario. 

We expect the situation to change for the better in the years to come. Using \textit{Gaia} astrometry and photometry, we are now able to compile extensive, all-sky  catalogues of hot subluminous stars (e.g. Culpan et al.  \cite{culpan22}), which will be used as input target lists for the next generation of spectroscopic surveys, in particular those covering the yet neglected southern sky (4MOST, de Jong et al. \cite{dejong19}; SDSS V, Kollmeier et al. \cite{kollmeier17}). In addition we will use simulations such as the one of Neunteufel et al. (\cite{neunteufel21,neunteufel22}) to define better selection criteria for targeted spectroscopic follow-up campaigns. Based on those spectra, which will reveal both the nature and the RV of tens of thousands of hot subluminous stars, we will be able to continue our search for the ejected companions of SN\,Ia and are confident that we will find the hidden siblings of US\,708.

\begin{acknowledgements}  

S.G. was supported by the Deutsche Forschungsgemeinschaft (DFG) through a Heisenberg fellowship (GE2506/8-1). V.S. was supported by the Deutsche Forschungsgemeinschaft (DFG) through grant GE2506/9-1. I.P.  acknowledges support by the Deutsche Forschungsgemeinschaft (DFG) through grant GE2506/12-1 and from a Royal Society University Research Fellowship (URF\textbackslash R1\textbackslash 231496). M.D. is supported by the German Aerospace Center (DLR) through grant 50OR2304.
P.N. acknowledges support from the Grant Agency of the Czech Republic (GA\v{C}R 18-20083S). N.R. is supported by the Deutsche Forschungsgemeinschaft (DFG) through grant RE3915/2-1.
R.R. acknowledges support from Grant RYC2021-030837-I funded by MCIN/AEI/10.13039/501100011033 and by “European Union NextGeneration EU/PRTR”. 
This work was partially supported by the AGAUR/Generalitat de Catalunya grant SGR-386/2021 and by the Spanish MINECO grant PID2020-117252GB-I00. 
We thank M. Pritzkuleit for his comments and suggestions. 

Based on observations at the La Silla-Paranal Observatory of the European Southern Observatory for programmes number 095.D-0055(A), 095.D-0588(A), 297.D-5004, 099.D-0208(A), 0102.D-0092(A). 

Based on observations with the William Herschel Telescope operated by the Isaac Newton Group at the Observatorio del Roque de los Muchachos of the Instituto de Astrofisica de Canarias on the island of La Palma, Spain.

Based on observations collected at the Centro Astron\'omico Hispano Alem\'an (CAHA) at Calar Alto, operated jointly by the Max-Planck Institut f\"ur Astronomie and the Instituto de Astrof\'isica de Andaluc\'ia (CSIC). 

Based on observations made with the Gran Telescopio Canarias (GTC), installed at the Spanish Observatorio del Roque de los Muchachos of the Instituto de Astrofisica de Canarias, in the island of La Palma.

Some of the data presented herein were obtained at the W. M. Keck Observatory, which is operated as a scientific partnership among the California Institute of Technology, the University of California and the National Aeronautics and Space Administration. The Observatory was made possible by the generous financial support of the W. M. Keck Foundation. The authors wish to recognize and acknowledge the very significant cultural role and reverence that the summit of Maunakea has always had within the indigenous Hawaiian community.  We are most fortunate to have the opportunity to conduct observations from this mountain. 

Funding for the SDSS and SDSS-II has been provided by the Alfred P. Sloan Foundation, the Participating Institutions, the National Science Foundation, the U.S. Department of Energy, the National Aeronautics and Space Administration, the Japanese Monbukagakusho, the Max Planck Society, and the Higher Education Funding Council for England. The SDSS Web Site is http://www.sdss.org/. The SDSS is managed by the Astrophysical Research Consortium for the Participating Institutions. The Participating Institutions are the American Museum of Natural History, Astrophysical Institute Potsdam, University of Basel, University of Cambridge, Case Western Reserve University, University of Chicago, Drexel University, Fermilab, the Institute for Advanced Study, the Japan Participation Group, Johns Hopkins University, the Joint Institute for Nuclear Astrophysics, the Kavli Institute for Particle Astrophysics and Cosmology, the Korean Scientist Group, the Chinese Academy of Sciences (LAMOST), Los Alamos National Laboratory, the Max-Planck-Institute for Astronomy (MPIA), the Max-Planck-Institute for Astrophysics (MPA), New Mexico State University, Ohio State University, University of Pittsburgh, University of Portsmouth, Princeton University, the United States Naval Observatory, and the University of Washington.   

Funding for the Sloan Digital Sky Survey IV has been provided by the Alfred P. Sloan Foundation, the U.S. Department of Energy Office of Science, and the Participating 
Institutions. SDSS-IV acknowledges support and resources from the Center for High Performance Computing  at the University of Utah. The SDSS website is www.sdss4.org. SDSS-IV is managed by the Astrophysical Research Consortium for the Participating Institutions of the SDSS Collaboration including the Brazilian Participation Group, 
the Carnegie Institution for Science, Carnegie Mellon University, Center for 
Astrophysics | Harvard \& Smithsonian, the Chilean Participation Group, the French Participation Group, Instituto de Astrof\'isica de Canarias, The Johns Hopkins 
University, Kavli Institute for the Physics and Mathematics of the Universe (IPMU) / University of Tokyo, the Korean Participation Group, Lawrence Berkeley National Laboratory, Leibniz Institut f\"ur Astrophysik Potsdam (AIP),  Max-Planck-Institut 
f\"ur Astronomie (MPIA Heidelberg), Max-Planck-Institut f\"ur Astrophysik (MPA Garching), Max-Planck-Institut f\"ur Extraterrestrische Physik (MPE), National Astronomical Observatories of China, New Mexico State University, New York University, University of Notre Dame, Observat\'ario Nacional / MCTI, The Ohio State University, Pennsylvania State University, Shanghai Astronomical Observatory, United Kingdom Participation Group, Universidad Nacional Aut\'onoma de M\'exico, University of Arizona, 
University of Colorado Boulder, University of Oxford, University of Portsmouth, University of Utah, University of Virginia, University of Washington, University of 
Wisconsin, Vanderbilt University, and Yale University.

\end{acknowledgements}

\onecolumn

\begin{appendix}

\begin{onecolumn}

\section{Additional material}

Table \ref{tab:astro}, \ref{tab:spec_overview}, and \ref{tab:photo_surveys} list the astrometric data from \textit{Gaia} EDR3, give an overview of the available spectra, and list the photometric surveys.

\begin{table*}
\centering
\caption{\label{tab:astro} Astrometric parameters.}
\begin{tabular}{llcrr}
\hline\hline
\noalign{\smallskip}
Name                 &  \textit{Gaia}\,DR3\,ID       & $\varpi$             & $\mu_{\rm \alpha}\cos{\delta}$ & $\mu_{\rm \delta}$ \\
                     &                      & [mas]             & [mas/yr]                       & [mas/yr]           \\
\noalign{\smallskip}
\hline
\noalign{\smallskip}
J022422.21+000313.4  & 2500824388329728256  &                     & 5.567  $\pm$ 0.305               & -7.361  $\pm$ 0.224   \\
J080833.77+180221.5  & 668963357040208512   &                     & 6.871  $\pm$ 0.076               & -12.839 $\pm$ 0.051   \\
J082802.03+404008.9  & 914500864915422848   &                     & 0.846  $\pm$ 0.157               & -4.770  $\pm$ 0.130   \\
J084556.85+135211.3  & 609252591386688256   &                     & -5.289 $\pm$ 0.108               & -3.786  $\pm$ 0.080   \\
J090252.98+073533.9  & 584194373496664064   &                     & 9.242   $\pm$ 0.090              & -14.175  $\pm$ 0.073  \\
J091512.06+191114.6  & 635864754212152448   &                     & -8.729  $\pm$ 0.347              & -6.130   $\pm$ 0.250  \\
J094850.47+551631.6  & 1021818521250329600  &                     & -1.929  $\pm$ 0.189              & -9.568   $\pm$ 0.170  \\
J102057.16+013751.2  & 3832278788553971968  &                     & -4.856  $\pm$ 0.100              & -3.718   $\pm$ 0.123  \\
J102439.42+383917.9  & 755712414993199744   &                     & -11.732 $\pm$ 0.082              & -4.190   $\pm$ 0.060  \\
J103810.94+253204.8  & 724806143863825664   &                     & -10.883 $\pm$ 0.069              & -12.708  $\pm$ 0.055  \\
J120352.23+235343.3  & 4002664328779500928  &                     & -28.161 $\pm$ 0.094              & -4.783   $\pm$ 0.062  \\
J120521.47+224702.1  & 4000291720125290496  &                     & -6.939  $\pm$ 0.166              & -1.053   $\pm$ 0.103  \\
J121703.14+454539.3  & 1539014020563855232  &                     & -23.256 $\pm$ 0.067              & -7.289   $\pm$ 0.101  \\
J123137.56+074621.7  & 3902459207006091520  &                     & -9.311  $\pm$ 0.113              & -6.266   $\pm$ 0.082  \\
J123428.30+262757.9  & 3960827945702129792  &                     & 1.743   $\pm$ 0.227              & -9.738   $\pm$ 0.215  \\
J123953.53+062853.0  & 3709894073412554880  &                     & -12.625 $\pm$ 0.148              & -4.454   $\pm$ 0.101  \\
J124248.89+133632.4  & 3929230528435328640  &                     & -7.372  $\pm$ 0.144              & -8.569   $\pm$ 0.131  \\
J124310.59+343358.5  & 1515296828902620928  &                     & -17.121 $\pm$ 0.081              & -12.881  $\pm$ 0.125  \\
J124819.08+035003.1  & 3703904449460695040  &                     & 8.663   $\pm$ 0.101              & -23.259  $\pm$ 0.096  \\
J130543.96+115840.8  & 3737057611255721472  & 0.395  $\pm$ 0.059  & -35.033 $\pm$ 0.075              & -7.858   $\pm$ 0.070  \\
J133135.42+020919.8  & 3711881543758246144  & 0.434  $\pm$ 0.081  & -30.052 $\pm$ 0.088              & 0.405    $\pm$ 0.049  \\
J133417.09+173850.8  & 3745825838529051008  &                     & 6.215   $\pm$ 0.309              & -2.387   $\pm$ 0.197  \\
J135651.26+155810.4  & 1243203435156394624 	&                     & -3.397  $\pm$ 0.472              & -1.648   $\pm$ 0.297  \\
J140532.34+410626.1  & 1497921109214351872  &                     & -1.698  $\pm$ 0.175              & -7.596   $\pm$ 0.194  \\
J143127.88+014416.3  & 3655878434393609728  &                     & -0.607  $\pm$ 0.305              & -5.963   $\pm$ 0.284  \\
J143258.00+011857.3  & 3655054139974647424  &                     & -17.900 $\pm$ 0.161              & -5.513   $\pm$ 0.127  \\
J144209.90+105733.9  & 1177896670713897984  &                     & 7.541   $\pm$ 0.127              & -12.969  $\pm$ 0.122  \\
J145141.40+090645.3  & 1174034021941552000  &                     & 2.887   $\pm$ 0.147              & -10.667  $\pm$ 0.122  \\
J145930.70+175846.1  & 1188517849959404672  &                     & -12.796 $\pm$ 0.285              & 2.046    $\pm$ 0.334  \\
J150222.35+320220.9  & 1288498160259323392  &                     & 2.566   $\pm$ 0.068              & -20.205  $\pm$ 0.096  \\
J151248.61+042205.6  & 1155663858405475584  &                     & -4.173  $\pm$ 0.299              & -15.155  $\pm$ 0.336  \\
J153419.42+372557.2  & 1375718878637792000  &                     & -6.900  $\pm$ 0.170              & -5.917   $\pm$ 0.223  \\
J154958.29+043820.1  & 4426050350708989568  &                     & -15.577 $\pm$ 0.156              & -4.084   $\pm$ 0.140  \\
J161143.29+554044.4  & 1429687032498740992  &                     & -0.065  $\pm$ 0.119              & -0.893   $\pm$ 0.127  \\
J163213.05+205124.0  & 1297264978785132416  &                     & -12.207 $\pm$ 0.085              & -3.901   $\pm$ 0.094  \\
J164419.44+452326.7  & 1406119962829248896  &                     & -1.945  $\pm$ 0.084              & -21.227  $\pm$ 0.101  \\
J164853.25+121702.9  & 4448696338872866432  &                     & -8.575  $\pm$ 0.163              & -10.120  $\pm$ 0.134  \\
J165924.75+273244.3  & 1306852479661771648  &                     & -3.028  $\pm$ 0.302              & -5.423   $\pm$ 0.338  \\
J170256.37+241757.7  & 4571704133509719680  &                     & -2.407  $\pm$ 0.150              & -2.401   $\pm$ 0.192  \\
J171533.85+365214.8  & 1339942973590043264  &                     & -6.109  $\pm$ 0.070              & -8.604   $\pm$ 0.070  \\
J172736.02+361706.3  & 1337048230055229824  &                     & -3.011  $\pm$ 0.152              & -3.275   $\pm$ 0.187  \\
J174211.74+643009.8  & 1440880786719564544  &                     & -3.470  $\pm$ 0.047              & -3.825   $\pm$ 0.049  \\
J180313.45+234000.2  & 4577607243942355456  & 0.322  $\pm$ 0.045  & 2.386   $\pm$ 0.033              & -4.277   $\pm$ 0.048  \\
J184832.51+181540.0  & 4512244400108060288  &                     & -10.769 $\pm$ 0.109              & -12.656  $\pm$ 0.129  \\
J204358.55-065025.8  & 6907270760248588416  &                     & 9.235   $\pm$ 0.180              & -15.037  $\pm$ 0.155  \\
J205030.39-061957.8  & 6913188263109953920  &                     & -0.099  $\pm$ 0.150              & -4.582   $\pm$ 0.104  \\
J210907.28+103640.6  & 1744880526139228032  &                     & -2.830  $\pm$ 0.258              & -6.173   $\pm$ 0.169  \\
J212300.31+043453.0  & 2693134400866685440  &                     &  7.453  $\pm$ 0.288              & -0.700   $\pm$ 0.242  \\
J212449.23+061956.5  & 1738660520141260416  &                     & -1.493  $\pm$ 0.142              & -10.164  $\pm$ 0.121  \\
J215648.71+003620.7  & 2681388764823777664  &                     & 0.807   $\pm$ 0.148              & -3.455   $\pm$ 0.140  \\
J220759.09+204506.0  & 1781407608084436352  &                     & 13.420  $\pm$ 0.071              & 4.231    $\pm$ 0.087  \\
J221728.34+121642.6  & 2727702466623387648  & 0.355  $\pm$ 0.063  & -8.755  $\pm$ 0.065              & -24.883  $\pm$ 0.075  \\
J222515.34-011156.8  & 2629912069952582656  &                     & 1.768   $\pm$ 0.224              & -1.753   $\pm$ 0.217  \\
\noalign{\smallskip}                                                                                                    
\hline\hline                                         
\end{tabular}
\tablefoot{Only parallaxes with uncertainties better than $20\%$ are listed.}
\end{table*}

\begin{table}[!ht]
\centering
\caption{\label{tab:spec_overview} Spectroscopic data overview. }
\resizebox{\textwidth}{!}{
\begin{tabular}{lllllllllll}
\hline\hline
\noalign{\smallskip}
Name                        &  SDSS  & BOSS  &  ISIS  & LAMOST & XSHO  & FORS1 & EFOSC2 & DBSP & TWIN & LRIS  \\
\noalign{\smallskip}
\hline
\noalign{\smallskip}
J022422.21+000313.5         &   5    & 15    &        &        &           &       &        &      &      &       \\
J080833.76+180221.8         &   1    & 2     &        &  1     &           &       &        &      &      &       \\
J082802.03+404008.9         &   1    & 2     &        &  1     &           &       &        &      &      &       \\
J084556.85+135211.3         &   1    &       &        &        &           &       &        &      &      &       \\
J090252.98+073533.9         &   1    &       &        &        &           &       &        &      &      &       \\
J091512.06+191114.6         &   1    & 1     &        &        &           &       &        &      &      &       \\
J094850.47+551631.7         &   1    & 1     &        &  1     &           &       &        &      &      &       \\
J102057.16+013751.2         &   1    &       &        &        &           & 2     &        &      &      &       \\ 
J102439.43+383917.9         &   1    & 1     &  1     &        &           &       &        &      &      &       \\
J103810.94+253204.8         &   1    & 1     &  1     &        &           &       &        &      &      &       \\
J120352.23+235343.3         &   1    & 2     &  1     &        &           &       &        &      &      &       \\
J120521.48+224702.2         &   1    & 1     &        &        &           &       &        & 1    &      &       \\
J121703.14+454539.3         &        & 1     &        &        &           &       &        &      &      &       \\
J123137.56+074621.7         &        & 1     &        &        & 5         &       &        &      &      &       \\
J123428.30+262757.9         &   1    & 1     &        &        &           &       &        &      &      &       \\
J123953.52+062853.0         &        & 1     &        &        &           &       &        &      &      &       \\
J124248.89+133632.6         &   1    & 1     &        &  1     &           &       &        &      &      &       \\
J124310.59+343358.5         &   1    & 1     &  1     &        &           &       &        &      &      &       \\
J124819.08+035003.1         &   1    & 1     &  1     &  1     &           &       &        &      &      &       \\
J130543.96+115840.8         &   1    & 1     &  1     &  4     &           &       &        &      &      &       \\
J133135.41+020919.8         &        & 1     &        &        &           &       &        &      &      &       \\
J133417.09+173850.8         &        & 1     &        &        &           &       &        &      &      & 1     \\
J135651.26+155810.4         &   1    &       &        &        &           &       &        &      &      &       \\
J140532.34+410626.1         &   1    & 1     &        &        &           &       &        &      &      &       \\
J143127.88+014416.2         &        & 1     &        &        &           &       &        &      &      &       \\
J143258.00+011857.3         &        & 1     &        &        &           &       &        &      &      &       \\
J144209.90+105733.9         &   1    & 1     &        &        &           &       &        &      &      &       \\
J145141.40+090645.3         &        & 1     &  1     &        &           &       &        & 1    &      &       \\
J145930.70+175846.1         &        & 1     &        &        &           &       &        &      &      &       \\
J150222.35+320220.9         &        & 1     &        &        &           &       &        &      &      &       \\
J151248.61+042205.6         &        & 1     &  1     &        &           &       &        &      &      & 1     \\
J153419.42+372557.2         &   1    & 1     &        &        &           &       &        &      &      &       \\
J154958.29+043820.1         &   1    & 1     &        &        &           &       &        &      &      &       \\
J161143.29+554044.4         &        & 1     &        &        &           &       &        &      &      &       \\
J163213.05+205124.0         &   1    & 1     &        &        & 10        &  2    &        &      &      &       \\
J164419.44+452326.7         &   1    & 1     &        &        &           &       &        &      & 3    &       \\
J164853.26+121703.0         &   1    &       &        &        &           &       &        &      &      &       \\
J165924.75+273244.3         &   1    &       &        &        &           &       &        &      &      &       \\
J170256.38+241757.9         &   1    &       &        &        &           &       &        &      &      &       \\
J171533.85+365214.8         &        & 1     &        &        &           &       &        &      &      &       \\
J172736.02+361706.3         &        & 1     &        &        &           &       &        &      &      &       \\
J174211.74+643009.8         &   1    &       &  1     &        &           &       &        &      &      &       \\
J180313.45+234000.1         &   1    &       &        &        &           &       &  1     &      &      &       \\
J184832.52+181540.0         &   1    &       &        &        &           &       &        &      & 1    &       \\
J204358.55-065025.8         &   2    &       &  1     &        &           &       &        &      &      &       \\
J205030.39-061957.8         &   1    &       &        &        &           & 2     &        &      &      &       \\
J210907.28+103640.6         &   1    &       &        &        &           &       &        &      &      &       \\
J212300.31+043453.0         &        & 1     &        &        &           &       &        &      &      &       \\
J212449.22+061956.4         &        & 1     &  1     &        &           &       &        &      &      &       \\
J215648.71+003620.7         &   1    & 1     &        &        &           &       &        &      &      &       \\
J220759.09+204506.0         &   1    & 1     &        &        &           &       &        &      &      &       \\
J221728.38+121643.8         &        & 1     &  1     &        &           &       &        &      &      &       \\
J222515.34-011156.8         &   1    & 1     &        &        &           &       &        &      &      &       \\
\noalign{\smallskip}                                                                                                    
\hline\hline                                         
\end{tabular}
}
\end{table}

\begin{table}
%\begin{table*}
\centering
\caption{\label{tab:photo_surveys} Photometric surveys used for the SED analysis}
\begin{tabular}{l}
\hline\hline
\noalign{\smallskip}
\bf{UV}:\\
Revised catalog of GALEX UV sources: Bianchi et al. (\cite{bianchi17})\\                                          %GALEX        II/335/galex_ais
\bf{Optical}:\\
Gaia Early Date Release 3: Riello et al. (\cite{riello21}) \\                                                                     %Gaia        I/350/gaiaedr3
DECam Local Volume Exploration Survey (DELVE) DR2: Drlica-Wagner et al. (\cite{delve}) \\ %Drlica-Wagner et al. (2021ApJS..256....2D)                            %DELVE       DELVE_DR2
The Dark Energy Survey (DES) Data Release 2: Abbott et al. (\cite{abbott21}) \\  %2021ApJS..255...20A                                %DES         DES_DR2
Pan-STARRS Data Release 2: Chambers et al. (\cite{chambers17})\\ %2020ApJS..251....6M                                                 %PS1         PS1_DR2
SDSS Photometric Catalogue Release 12: Alam et al. (\cite{alam15})\\                                           %SDSS        V/147/sdss12
SkyMapper Southern Survey DR2: Onken et al. (\cite{onken19})\\                                          %            SkyMapper Skymapper_DR2
Javalambre Photometric Local Universe Survey Data Release DR3: López-Sanjuan et al. (\cite{lopez24})\\ %2024A&A...683A..29L %
AAVSO Photometric All Sky Survey (APASS) DR9: Henden et al. (\cite{henden15})\\                                 %Johnson       II/336/apass9
Beijing-Arizona-Taiwan-Connecticut (BATC) Large Field Multi-Color Sky Survey: Xu \& Zhaoji (\cite{xu05})\\ %Xu, Z., & Zhaoji, J. 2005, 2005yCat 2262 0X   % BATC        II/262/batc
\bf{IR}:\\
UKIDSS-DR9 LAS, GCS and DXS Surveys: Lawrence et al. (\cite{lawrence12}) \\                                     %UKIDSS        II/319/las9 and II/319/dxs9
VISTA Hemisphere Survey DR5: McMahon et al. (\cite{mcmahon21}) \\ % Paper (2013Msngr.154...35M)                 % VISTA        VHS_DR6
VISTA Kilo-degree Infrared Galaxy Public Survey (VIKING) DR4: Kuijken et al. (\cite{kuijken2019}) \\ %2019A&A...625A...2K
The CatWISE2020 Catalog: Marocco et al. (\cite{marocco21}) \\ %2021ApJS..253....8M                                %  WISE       II/365/catwise
The band-merged unWISE Catalog: Schlafly et al.  (\cite{schlafly19}) \\                                           %   WISE       II/363/unwise
\hline
\end{tabular}
\end{table}

\end{onecolumn}

\end{appendix}              
                            
\end{document}